\documentclass{IEEEtran}
\usepackage{rotating}
\usepackage[boxruled]{algorithm2e}
\usepackage{enumerate}
\usepackage{cite}
\usepackage{flushend}
\usepackage{graphicx}
\usepackage{psfrag}
\usepackage{subfigure}
\usepackage{url}
\usepackage{amsmath}
\usepackage{array}
\usepackage{amssymb}
\usepackage{amsfonts}
\newtheorem{proposition}{Proposition}
\newtheorem{lemma}{Lemma}

\newtheorem{corollary}{Corollary}
\newtheorem{remark}{Remark}
\newtheorem{definition}{Definition}
\newtheorem{theorem}{Theorem}
\newtheorem{example}{Example}

\title{A Matroidal Framework for Network-Error Correcting Codes}
\begin{document}
\author{
\authorblockN{K.~Prasad and B.~Sundar Rajan,}
\authorblockA{Dept. of ECE, IISc, Bangalore 560012, India.\\
Email: \{prasadk5,bsrajan\}@ece.iisc.ernet.in\\}
}
\date{\today}
\maketitle
\thispagestyle{empty}	
\let\thefootnote\relax\footnotetext
{
Parts of the content of this work was presented at ISIT 2012 held at Cambridge, Massachusetts, USA, during July 1 - 6, 2012 and at ISITA 2012 held at Honolulu, Hawaii, USA, during October 28-31, 2012.
}
\begin{abstract}
Matroidal networks were introduced by Dougherty et al. and have been well studied in the recent past. It was shown that a network has a scalar linear network coding solution if and only if it is matroidal associated with a representable matroid. A particularly interesting feature of this development is the ability to construct (scalar and vector) linearly solvable networks using certain classes of matroids. Furthermore, it was shown through the connection between network coding and matroid theory that linear network coding is not always sufficient for general network coding scenarios. The current work attempts to establish a connection between matroid theory and network-error correcting and detecting codes.  In a similar vein to the theory connecting matroids and network coding, we abstract the essential aspects of network-error detecting codes to arrive at the definition of a \textit{matroidal error detecting network} (and similarly, a \textit{matroidal error correcting network} abstracting from network-error correcting codes). An acyclic network (with arbitrary sink demands) is then shown to possess a scalar linear error detecting (correcting) network code if and only if it is a matroidal error detecting (correcting) network associated with a representable matroid. Therefore, constructing such network-error correcting and detecting codes implies the construction of  certain representable matroids that satisfy some special conditions, and vice versa. We then present algorithms which enable the construction of matroidal error detecting and correcting networks with a specified capability of network-error correction. Using these construction algorithms, a large class of hitherto unknown scalar linearly solvable networks with multisource multicast and multiple-unicast network-error correcting codes is made available for theoretical use and practical implementation, with parameters such as number of information symbols, number of sinks, number of coding nodes, error correcting capability, etc. being arbitrary but for computing power (for the execution of the algorithms). The complexity of the construction of these networks is shown to be comparable to the complexity of existing algorithms that design multicast scalar linear network-error correcting codes. Finally we also show that linear network coding is not sufficient for the general network-error correction (detection) problem with arbitrary demands. In particular, for the same number of network-errors, we show a network for which there is a nonlinear network-error detecting code satisfying the demands at the sinks, while there are no linear network-error detecting codes that do the same.
\end{abstract}
\section{Introduction}
\label{sec1}
Network coding, introduced in \cite{ACLY}, is a technique to increase the rate of  information transmission through a network by coding different information flows present in the network. One of the chief problems in network coding is to find whether a given network with a set of sources and sink demands is solvable using a scalar linear network code. Much work has been done on the existence and construction of scalar linear network coding techniques in several papers including \cite{CLY,KoM,JSCEEJT}.

Matroids are discrete objects which abstract the notions of linear dependence among vectors. They arise naturally in several discrete structures including graphs and matrices. The relationship between network coding and matroid theory was first introduced in \cite{DFZ}. The authors of \cite{DFZ} showed that the scalar linear solvability of a network with a given set of demands was related to the existence of a \textit{representable matroid} (matroids which arise from matrices over fields) satisfying certain properties. This connection was further developed and strengthened in \cite{DFZ2,KiM,SHL,LiS,RSG}. Using the techniques of \cite{DFZ,KiM,RSG}, it is known that several network instances which are scalar (or vector) linearly solvable can be constructed using representable matroids and their generalisations. Using the equivalence between networks and matroids, it was shown in \cite{DFZ3} that linear network codes are not always sufficient for solving network coding problems where the sinks have arbitrary demands (i.e., not necessarily multicast). An explicit network was demonstrated which had a nonlinear network coding solution but no linear network coding solutions. 

Linear network-error correcting codes were introduced in \cite{YeC1,YeC2} as special kinds of linear network codes which could correct errors that occurred in the edges of the network. Linear network-error detection codes are simply linear network-error correction codes where the sinks are able to decode their demands in the presence of errors at edges known to the sinks. Together with the subsequent works \cite{Zha,Mat,YaY}, the bounds and constructions similar to classical block coding techniques were carried over to the context of linear network-error correction and detection. As network-error correcting (detecting) codes are essentially special kinds of network codes, the issues of network coding especially with respect to existence and construction have their equivalent counterparts in network-error correction (detection). Network-error correction was extended to case of non-multicast in \cite{VHEKE}. In \cite{KTT}, linear network-error correction schemes were found to be incapable of satisfying the demands for networks with node adversaries rather than edge adversaries. Nonlinear error correction schemes are also found to perform better than linear error correction in networks with unequal edge capacities \cite{KHEA}. 

In the current work, we present the connection between matroids and network-error correcting and detecting codes. The results of this work may be considered as the network-error correction and detection counterparts of some of the results of \cite{DFZ,KiM,DFZ3}. The organisation and the chief contributions of our work are as follows. 
\begin{itemize}
\item After reviewing linear network-error correcting and detecting codes in Section \ref{sec2} and matroid theory in Section \ref{sec3}, in Section \ref{sec4} we define the notion of a \textit{matroidal error detecting network} associated with a particular matroid. Using this definition, we show that an acyclic network has a scalar linear network-error detecting code (satisfying general demands) if and only if there exists a representable matroid  $\cal M$ such that the given network is a matroidal error detecting network associated with $\cal M.$ Therefore, networks with scalar linear network-error detecting codes are shown to be analogous to representable matroids satisfying a certain set of properties. Because of the equivalence between network-error detection and network-error correction, all these results have their counterparts for network-error correcting codes also.
\item In Section \ref{sec5}, we give algorithms which construct multisource multicast and multiple-unicast matroidal error correcting networks associated with general matroids (not necessarily representable) satisfying the required properties. If the matroids associated with such networks are representable over finite fields, then these networks are obtained along with their corresponding scalar linear network-error correcting codes. Therefore, our results generate a large class of hitherto unknown networks which have scalar linear network-error correcting codes, a few of which are shown in this paper by implementing the representable matroids version of our algorithms in MATLAB. Though the implementation the nonrepresentable matroids version of our algorithm is difficult, we do give a small result as a first step in this direction in Subsection \ref{mecnnonrepresentable}. The complexity of the construction of multicast and multiple-unicast networks associated with representable matroids is shown to be comparable to the complexity of existing algorithms that design multicast scalar linear network-error correcting codes for given networks in Section \ref{seccomplexity}.
\item Based on the results from \cite{DFZ3}, in Section \ref{secinsufficiency}, we prove the insufficiency of linear network coding for the network-error detection problem on networks with general demands (i.e., not necessarily multicast). In particular, we demonstrate a network (adapted from the network used in \cite{DFZ3} to demonstrate the insufficiency of linear network coding for the general network coding problem) for which there exists a nonlinear single edge network-error detecting code that satisfies the sink demands, while there are no linear network-error detecting codes that do the same. 
\item In Subsection \ref{insuffnetwork}, we show that this network, for which linear network-error detection is insufficient, is a matroidal error detecting network with respect to a nonrepresentable matroid. Thus our definition of matroidal error detecting networks is not limited to networks with linear network-error detecting schemes alone, instead has a wider scope, accommodating nonlinear error detection schemes also. 
\end{itemize}

Though algorithms for constructing network-error correcting codes are known for given single source multicast networks \cite{YeC1,YeC2,YaY}, there is no general characterisation of networks and demands for which  scalar linear network-error correction codes can be designed. The authors believe that the algorithm given in this paper could provide useful insights in this regard. Furthermore, it could also prove useful in the design of practical network topologies in which network coding and network-error correction (detection) have advantages over routing and classical error correction (detection). We also highlight that though there are many papers in network coding literature which discuss network coding for multiple-unicast networks, the results obtained in our paper are some of the first in network-error correction literature which talk about network-error correction codes for multiple-unicast networks. 

\textit{Notations: }The following notations will be followed throughout the paper. The disjoint union of any two sets $A$ and $B$ is denoted by $A\uplus B.$ For a finite set $A,$ the power set of $A$ is denoted by $2^A.$ A finite field is denoted by the symbol $\mathbb{F}.$ For some positive integer $k,$ the identity matrix of size $k$ over $\mathbb{F}$ is denoted by $I_k.$ The rank of a matrix $A$ over $\mathbb{F}$ is denoted by $rank(A),$ and its transpose is denoted by $A^T.$ The $\mathbb{F}$-vector space spanned by the columns of a matrix $A$ over $\mathbb{F}$ is denoted by $\langle A\rangle.$ The set of columns of $A$ is denoted by $cols(A).$ The support set of a vector $\boldsymbol{x}$ and its Hamming weight are denoted by $supp(\boldsymbol{x})$ and  $w_H(\boldsymbol{x})$ respectively. The symbol $\boldsymbol{0}$ represents an all zero vector or matrix of appropriate size indicated explicitly or known according to the context. 
For some matrix $A,$ we denote by $A^l$ the $l^{th}$ column of $A,$ and for a subset ${\cal L}$ of the column indices of $A,$ we denote by $A^{\cal L}$ the submatrix of $A$ with columns indexed by $\cal L.$ Likewise, we denote by $A_{j}$ the $j^{th}$ row of $A,$ and by $A_{\cal J}$ the submatrix of $A$ with rows given by the subset ${\cal J}$ of the row indices.
\section{Network-Error Correcting and Detecting Codes}
\label{sec2}
As in \cite{KoM,YeC1}, we model the directed acyclic network as a directed acyclic multigraph (one with parallel edges) ${\cal G}({\cal V},{\cal E})$ where ${\cal V}$ is the set of vertices of $\cal G$ representing the nodes in the network and ${\cal E}$ is the set of edges representing the links in the network. An ancestral ordering is assumed on $\cal E$ as the network is acyclic. Each edge is assumed to carry at most one finite field symbol at any given time instant. A non-empty subset ${\cal S} \subseteq {\cal V}$, called the set of sources, generates the information that is meant for the sinks in the network, represented by another non-empty subset ${\cal T} \subseteq {\cal V}$. Each sink demands a particular subset of the information symbols generated by the sources. Any node in the network can be a source and a sink simultaneously, however not generating and demanding the same information.
Let $n_{s_i}$ be the number of information symbols (from some finite field $\mathbb{F}$) generated at source $s_i.$  Let ${\mu} = \left\{1,2,...,\sum_{i=1}^{|{\cal S}|} n_{s_i}=n\right\}$ denote the ordered index set of messages (each corresponding to a particular information symbol) generated at all the sources. 
For each edge $e\in {\cal E},$ we denote by $tail(e)$ the node from which $e$ is outgoing, and by $head(e)$ the node to which $e$ is incoming. Also, for each node $v\in {\cal V},$ let $In(v)$ denote the union of the messages (a subset of $\mu$) generated by $v$ and the set of incoming edges at $v.$ Similarly, let $Out(v)$ denote the union of the subset of messages demanded by $v$ and the set of outgoing edges from $v.$ Further, for any $e\in {\cal E},$ we denote by $In(e)$ the set $In(tail(e)).$

A network code on $\cal G$ is a collection of functions, one associated with each node of the network mapping the incoming symbols at that node to its outgoing symbols. When these functions are scalar linear, the network code is said to be a scalar linear network code. To be precise, a scalar linear network code is an assignment to the following matrices.
\begin{itemize} 
\item A matrix $A_{s_i}$ of size $n_{s_i} \times |{\cal E}|,$ for each source $s_i \in {\cal S}, i=1,2,...,|{\cal S}|,$ denoting the linear combinations taken by the sources mapping information symbols to the network, with non-zero entries (from $\mathbb{F}$) only in those columns which index the outgoing edges from $s_i.$
\item A matrix $K$ of size $|{\cal E}| \times |{\cal E}|$ which indicates the linear combinations taken by the nodes in the network to map incoming symbols to outgoing symbols. For $i<j,$ the $(i,j)^{th}$ element of $K,$ $K_{i,j},$ is an element from $\mathbb{F}$ representing the network coding coefficient between edge $e_i$ and $e_j.$ Naturally $K_{i,j}$ can be non-zero only if $e_j$ is at the downstream of $e_i.$
\end{itemize}
Also, to each sink $t \in {\cal T},$ we associate a matrix $B_{t}$ of size $|{\cal E}| \times n_{t},$  where $n_{t}$ is the number of incoming edges at $t.$ Corresponding to the $n_t$ rows that index these incoming edges, we fix the $n_t \times n_t$ submatrix of $B_{t}$ as an identity submatrix. The other entries of $B_t$ are fixed as zeroes.

For $i=1,2,...,|{\cal S}|,$ let $\boldsymbol{x_{s_i}} \in \mathbb{F}^{n_{s_i}}$ be the row vector representing the information symbols at source $s_i$. Let $\boldsymbol{F}=\left(I_{|{\cal E}|}-K\right)^{-1}$ and $A_{s_i}\boldsymbol{F}B_{t} = \boldsymbol{F_{s_i,t}}.$
Let ${\cal A}$ be the $n \times |{\cal E}|$ row-wise concatenated matrix
\begin{equation}
\label{formofA}
\left(
\begin{array}{c}
A_{s_1}\\
A_{s_2}\\
.\\
.\\
A_{s_{|{\cal S}|}}
\end{array}
\right).
\end{equation}
The columns of ${\cal A}\boldsymbol{F}$ are known as the \textit{global encoding vectors} corresponding to the edges of the network, indicating  the particular linear combinations of the information symbols which flow in the edges. We assume that no edge is assigned an all zero global encoding vector, for then it can simply be removed from the network and a smaller graph can be assumed. The global encoding vector corresponding to the $n$  messages are fixed to be the $n$ standard basis vectors over $\mathbb{F},$ the concerned field. A network code can also be specified completely by specifying global encoding vectors for all edges in the network, provided that they are valid assignments, i.e., global encoding vectors of outgoing edges are linear combinations of those of the incoming edges.

Let  $\boldsymbol{x}=\left(\boldsymbol{x_{s_1}}~~~\boldsymbol{x_{s_2}}~~ ... ~~\boldsymbol{x_{s_{|{\cal S}|}}}\right)$ be the vector of all information symbols. Let $\boldsymbol{{\cal D}_{t}} \subseteq {\mu}$ denote the set of demands at sink $t,$ and let $\boldsymbol{x_{s_{{\cal D}_t}}}$ denote the subvector of the super-vector $\boldsymbol{x}$ corresponding to the information symbols indexed by $\boldsymbol{{\cal D}_{t}}.$ 


An edge is said to be in error if its input symbol (from $tail(e)$) and output symbol (to $head(e)$), both from $\mathbb{F}$, are not the same. We call this as a \textit{network-error}. We model the network-error as an additive error from $\mathbb{F}$. A \textit{network-error vector} is a $|{\cal E}|$ length row vector over $\mathbb{F}$, whose components indicate the additive errors on the corresponding edges. The case of multicast network-error correction, where a single source multicasts all its symbols to all sinks in the presence of errors, has been discussed in several papers (see for example, \cite{YeC1,YeC2,YaY}) all being equivalent in some sense. 

Now we briefly review the results for network-error correcting and detecting codes in the case of arbitrary number of sources and sinks with arbitrary demands. Let $\boldsymbol{z}$ be the network-error vector corresponding to a particular instance of communication in the network. Let $\boldsymbol{F_{{\cal S},t}}$ be the matrix ${\cal A}\boldsymbol{F}B_t.$ Let $\boldsymbol{F}B_{t}=\boldsymbol{F_{t}}.$ Then a sink $t$ receives the $n_t$ length vector 
\begin{equation}
\label{networkinoutrelationship}
\boldsymbol{y_{t}}=\boldsymbol{x}\boldsymbol{F_{{\cal S},t}}+\boldsymbol{z}\boldsymbol{F_{t}}.
\end{equation}
One way to interpret the input-output relationship shown by (\ref{networkinoutrelationship}) is to think of the network as a finite state machine whose states are the symbols flowing on the edges. The matrix $\boldsymbol{F_{{\cal S},t}}$ then describes the transfer matrix of this state machine between the sources and sink $t$. Some of the states of this network could be in error (i.e. the network-errors at the edges), which is captured by the network-error vector $\boldsymbol{z}.$ These errors are also reflected at the sink outputs, in their appropriate linear combinations, given by the matrix $\boldsymbol{F_{t}}.$ For more details the reader is referred to \cite{YaY}.

A network code which enables every sink to successfully recover the desired information symbols in the presence of any network-errors in any set of edges of cardinality at most $\alpha$ is said to be a $\alpha$-\textit{network-error correcting code}. A network code which enables the sink demands to be recovered in the presence of errors in at most $\beta$ edges which are \textit{known} to the sinks, is called a $\beta$-\textit{network-error detecting code}. 

It is not difficult to see that a scalar linear network code is a scalar linear $\alpha$-network-error correcting code if and only if the following condition holds at each sink $t\in{\cal T}$. 
\begin{align}
\nonumber
\boldsymbol{y_{t}}& = \boldsymbol{x}\boldsymbol{F_{{\cal S},t}} +\boldsymbol{z}\boldsymbol{F_{t}} \neq \boldsymbol{0} \in \mathbb{F}^{n_t},\\
\label{eqn2}
&\forall ~  \boldsymbol{x}\in \mathbb{F}^{n}: \boldsymbol{x_{s_{{\cal D}_{t}}}} \neq \boldsymbol{0},~\forall~ \boldsymbol{z}\in \mathbb{F}^{|{\cal E}|}: w_H(\boldsymbol{z}) \leq 2\alpha.
\end{align}
Similarly, for a $\beta$-network-error detecting code, we must have the following condition holding true for all sinks. 
\begin{align}
\nonumber
\boldsymbol{y_{t}}& = \boldsymbol{x}\boldsymbol{F_{{\cal S},t}} +\boldsymbol{z}\boldsymbol{F_{t}} \neq \boldsymbol{0} \in \mathbb{F}^{n_t},\\
\label{eqn2a}
&\forall ~  \boldsymbol{x}\in \mathbb{F}^{n}: \boldsymbol{x_{s_{{\cal D}_{t}}}} \neq \boldsymbol{0},~\forall~ \boldsymbol{z}\in \mathbb{F}^{|{\cal E}|}: w_H(\boldsymbol{z}) \leq \beta.
\end{align}
The proof that (\ref{eqn2}) indeed implies a $\alpha$-network-error correcting code follows from the fact that we can always demonstrate a pair of information vectors $\boldsymbol{x}$ and $\boldsymbol{x'}$ with $\boldsymbol{x_{s_{{\cal D}_{t}}}} \neq\boldsymbol{x'_{s_{{\cal D}_{t}}}}$ and a corresponding pair of error vectors $\boldsymbol{z}$ and $\boldsymbol{z'}$ with $w_H(\boldsymbol{z}) \leq \alpha$ and $w_H(\boldsymbol{z'})\leq \alpha$ such that the corresponding outputs $\boldsymbol{y_{t}}$ and $\boldsymbol{y'_{t}}$ are equal, if and only if the sink $t$ is not able to distinguish between $\boldsymbol{x_{s_{{\cal D}_{t}}}}$ and $\boldsymbol{x'_{s_{{\cal D}_{t}}}}$ in the presence of errors. 
A similar argument can be given for (\ref{eqn2a}). 

Thus, by (\ref{eqn2}) and (\ref{eqn2a}), it is clear that a $\beta$-network-error detecting code is also a $\lfloor \frac{\beta}{2}\rfloor$-network-error correcting code, while an $\alpha$-network-error correcting code is also a $2\alpha$-network-error detecting code.

The \textit{error pattern} corresponding to a network-error vector $\boldsymbol{z}$ is defined as its support set $supp(\boldsymbol{z}),$ which we shall also alternatively refer to using the corresponding subset of $\cal E.$ Let $\boldsymbol{F_{supp(\boldsymbol{z}),{t}}}$ denote the submatrix of $\boldsymbol{F_{t}}$ consisting of those rows of $\boldsymbol{F_{t}}$ which are indexed by $supp(\boldsymbol{z}).$ The condition (\ref{eqn2}) can then be rewritten as 
\begin{align}
\nonumber
\boldsymbol{y_{t}}= & \left(\boldsymbol{x}~~\boldsymbol{\bar{z}}\right)
\left(
\begin{array}{c}
\boldsymbol{F_{{\cal S},t}} \\
\boldsymbol{F_{supp(\boldsymbol{z}),{t}}}
\end{array}
\right) \neq \boldsymbol{0} ,~\forall ~ \boldsymbol{x}\in \mathbb{F}^{n}: \boldsymbol{x_{s_{{\cal D}_{t}}}} \neq \boldsymbol{0},\\
\label{eqn3}
&~\forall~ \boldsymbol{\bar{z}}\in \mathbb{F}^{2\alpha},~\forall~supp(\boldsymbol{z})\in \left\{{\cal F} \subseteq {\cal E}:|{\cal F}|=2\alpha \right\}.
\end{align}
Similarly condition (\ref{eqn2a}) becomes
\begin{align}
\nonumber
\boldsymbol{y_{t}}= & \left(\boldsymbol{x}~~\boldsymbol{\bar{z}}\right)
\left(
\begin{array}{c}
\boldsymbol{F_{{\cal S},t}} \\
\boldsymbol{F_{supp(\boldsymbol{z}),{t}}}
\end{array}
\right) \neq \boldsymbol{0} ,~\forall ~ \boldsymbol{x}\in \mathbb{F}^{n}: \boldsymbol{x_{s_{{\cal D}_{t}}}} \neq \boldsymbol{0},\\
\label{eqn3a}
&~\forall~ \boldsymbol{\bar{z}}\in \mathbb{F}^{\beta},~\forall~supp(\boldsymbol{z})\in \left\{{\cal F} \subseteq {\cal E}:|{\cal F}|=\beta \right\}.
\end{align}

For the special case of a single source multicast, the condition (\ref{eqn3}) becomes 
\begin{align}
\nonumber
\boldsymbol{y_{t}}= & \left(\boldsymbol{x}~~\boldsymbol{\bar{z}}\right)
\left(
\begin{array}{c}
\boldsymbol{F_{s,t}} \\
\boldsymbol{F_{supp(\boldsymbol{z}),{t}}}
\end{array}
\right) \neq \boldsymbol{0} \in \mathbb{F}^{n_{t}},~\forall ~  \boldsymbol{x} \neq \boldsymbol{0},\\
\label{eqnmulticastcondition}
&~\forall~ \boldsymbol{\bar{z}}\in \mathbb{F}^{2\alpha},~\forall~supp(\boldsymbol{z})\in \left\{{\cal F} \subseteq {\cal E}:|{\cal F}|=2\alpha \right\},
\end{align}
which is known from \cite{YeC1,YeC2,Zha,YaY}. Some of these papers also discuss the case of unequal error correcting capabilities at different sinks, but in our paper we only consider $\alpha$-network-error correction at all sinks uniformly. The extension to the unequal error capabilities is natural and therefore omitted.

For the multiple-unicast case, where each source has only one symbol to unicast to some sink and each sink has only one information symbol to receive from some source, the condition (\ref{eqn2}) becomes 
\begin{align}
\nonumber
\boldsymbol{y_{t}}= & x_{s_{\boldsymbol{{\cal D}_{t}}}}\boldsymbol{F_{s_{{\cal D}_{t}},t}} + \left(\sum_{i=1,i\neq \boldsymbol{{\cal D}_{t}}}^{|{\cal S}|}x_{s_i}\boldsymbol{F_{s_i,t}}+\boldsymbol{z}\boldsymbol{F_{t}}\right)  \neq \boldsymbol{0},\\
\label{multipleunicastcondition}
&~\forall ~  \boldsymbol{x}: x_{s_{\boldsymbol{{\cal D}_{t}}}} \neq 0,~\forall~ \boldsymbol{z}\in \left\{\boldsymbol{z}\in \mathbb{F}^{|{\cal E}|}: w_H(\boldsymbol{z}) \leq 2\alpha\right\},
\end{align}
where the first term above represents the signal part of the received vector and the second term denotes the interference plus noise part. Note that $ x_{s_{\boldsymbol{{\cal D}_{t}}}}$ denotes the demanded information symbol at sink $t,$ while $x_{s_i}$ denotes the information symbol generated at source $s_i.$  Equations similar to (\ref{eqnmulticastcondition}) and (\ref{multipleunicastcondition}) can be obtained for $\beta$-network-error detecting codes also, by simply replacing $2\alpha$ by $\beta.$

\subsection{A technical lemma}
We now present a technical lemma, which will be used in Section \ref{sec4}. The result of the lemma can be inferred from the results of \cite{VHEKE}, but we give it here for the sake of completeness. 
\begin{lemma}
\label{lemmadecoding} 
Let $I_{\boldsymbol{{\cal D}_{t}}}$ denote the $(n+\beta) \times |\boldsymbol{{\cal D}_{t}}|$ matrix with a $|\boldsymbol{{\cal D}_{t}}|\times |\boldsymbol{{\cal D}_{t}}|$ identity submatrix in $|\boldsymbol{{\cal D}_{t}}|$ of the first $n$ rows corresponding to the demands $\boldsymbol{{\cal D}_{t}}$ at sink $t,$ and with all other elements being zero. For some $supp(\boldsymbol{z})\in \left\{{\cal F} \subseteq {\cal E}:|{\cal F}|=\beta \right\},$ the condition 
\begin{align}
\label{eqn4}
\left(\boldsymbol{x}~~\boldsymbol{\bar{z}}\right)
\left(
\begin{array}{c}
\boldsymbol{F_{{\cal S},t}} \\
\boldsymbol{F_{supp(\boldsymbol{z}),{t}}}
\end{array}
\right) \neq \boldsymbol{0},~\forall\boldsymbol{x}\in\mathbb{F}^n:\boldsymbol{x_{s_{{\cal D}_{t}}}} \neq \boldsymbol{0},~\forall\boldsymbol{\bar{z}}\in \mathbb{F}^{\beta}
\end{align}
holds if and only if the following condition holds
\begin{equation}
\label{eqnlemma}
cols(I_{\boldsymbol{{\cal D}_{t}}}) \subseteq \left\langle\left(
\begin{array}{c}
\boldsymbol{F_{{\cal S},t}} \\
\boldsymbol{F_{supp(\boldsymbol{z}),{t}}}
\end{array}
\right)\right\rangle.
\end{equation}
Therefore a given network code is $\beta$-network-error detecting (or $\lfloor\frac{\beta}{2}\rfloor$-network-error correcting) if and only if the condition (\ref{eqnlemma}) 
holds for all $supp(\boldsymbol{z})\in \left\{{\cal F} \subseteq {\cal E}:|{\cal F}|=\beta \right\}$ at all sinks $t\in{\cal T}.$
\end{lemma}
\begin{IEEEproof}
We first prove the \textit{If} part. Since $cols(I_{\boldsymbol{{\cal D}_{t}}})$ is in the subspace $\left\langle\left(
\begin{array}{c}
\boldsymbol{F_{{\cal S},t}} \\
\boldsymbol{F_{supp(\boldsymbol{z}),{t}}}
\end{array}
\right)\right\rangle,$ linear combinations of the columns of $\left(
\begin{array}{c}
\boldsymbol{F_{{\cal S},t}} \\
\boldsymbol{F_{supp(\boldsymbol{z}),{t}}}
\end{array}
\right)$ should generate the columns of $I_{\boldsymbol{{\cal D}_{t}}}.$ Thus, we must have for some matrix $X$ of size $n_{t}\times |\boldsymbol{{\cal D}_{t}}|,$ 
\[
\left(
\begin{array}{c}
\boldsymbol{F_{{\cal S},t}} \\
\boldsymbol{F_{supp(\boldsymbol{z}),{t}}}
\end{array}
\right)X = I_{\boldsymbol{{\cal D}_{t}}}.
\]
Now suppose for some $\left(\boldsymbol{x}~~\boldsymbol{\bar{z}}\right)$ with $\boldsymbol{x_{s_{{\cal D}_{t}}}} \neq \boldsymbol{0}$ and some $~\boldsymbol{\bar{z}}\in \mathbb{F}^{\beta}$ we have 
\[
\left(\boldsymbol{x}~~\boldsymbol{\bar{z}}\right)
\left(
\begin{array}{c}
\boldsymbol{F_{{\cal S},t}} \\
\boldsymbol{F_{supp(\boldsymbol{z}),{t}}}
\end{array}
\right) = \boldsymbol{0}.
\] 

Multiplying both sides by $X,$ we then have $\boldsymbol{x_{s_{{\cal D}_{t}}}} = \boldsymbol{0},$ a contradiction. This proves the If part.

Now we prove the \textit{only if} part. Let $\boldsymbol{F_{{\cal S},t,{\cal D}_{t}}}$ denote the submatrix of $\boldsymbol{F_{{\cal S},t}}$ consisting of the $|\boldsymbol{{\cal D}_{t}}|$ rows corresponding to the symbols demanded by $t.$ Let $\boldsymbol{F_{{\cal S},t,\overline{{\cal D}_{t}}}}$ denote the submatrix of $\boldsymbol{F_{{\cal S},t}}$ with rows other than those in $\boldsymbol{F_{{\cal S},t,{\cal D}_{t}}}.$ Then because (\ref{eqn4}) holds, we must have 
\begin{align*}
rank&\left(
\begin{array}{c}
\boldsymbol{F_{{\cal S},t}} \\
\boldsymbol{F_{supp(z),{t}}}
\end{array}
\right)\\
&= rank(\boldsymbol{F_{{\cal S},t,{\cal D}_{t}}}) + rank
\left(
\begin{array}{c}
\boldsymbol{F_{{\cal S},t,\overline{{\cal D}_{t}}}} \\
\boldsymbol{F_{supp(z),{t}}}
\end{array}
\right).
\end{align*}
The above equation follows because (\ref{eqn4}) requires that the rows of $\boldsymbol{F_{{\cal S},t,{\cal D}_{t}}}$ and $\left(
\begin{array}{c}
\boldsymbol{F_{{\cal S},t,\overline{{\cal D}_{t}}}} \\
\boldsymbol{F_{supp(z),{t}}}
\end{array}
\right)$ be linearly independent. Thus,
\begin{align}
\label{eqnno33}
rank\left(
\begin{array}{c}
\boldsymbol{F_{{\cal S},t}} \\
\boldsymbol{F_{supp(z),{t}}}
\end{array}
\right)= |\boldsymbol{{\cal D}_{t}}| + rank
\left(
\begin{array}{c}
\boldsymbol{F_{{\cal S},t,\overline{{\cal D}_{t}}}} \\
\boldsymbol{F_{supp(z),{t}}}
\end{array}
\right).
\end{align}
Let the concatenated matrix
\[
\left(
\begin{array}{cc}
\boldsymbol{F_{{\cal S},t}} \vspace{-0.2cm}& \\ 
\vspace{-0.2cm} & I_{\boldsymbol{{\cal D}_{t}}}\\
\boldsymbol{F_{supp(\boldsymbol{z}),{t}}} & \\
\end{array} 
\right)
\]
be denoted by $Y.$ Again, it is easy to see that
\begin{align*}
&rank(Y)\\ 
&= 
rank
\left(
\begin{array}{cc}
\boldsymbol{F_{{\cal S},t,{\cal D}_{t}}} & I_{|\boldsymbol{{\cal D}_{t}}|}
\end{array} 
\right)
 + 
rank
\left(
\begin{array}{cc}
\boldsymbol{F_{{\cal S},t,\overline{{\cal D}_{t}}}} \\
\boldsymbol{F_{supp(\boldsymbol{z}),{t}}}
\end{array} 
\right)\\
&= |\boldsymbol{{\cal D}_{t}}| 
+
rank
\left(
\begin{array}{cc}
\boldsymbol{F_{{\cal S},t,\overline{{\cal D}_{t}}}} \\
\boldsymbol{F_{supp(\boldsymbol{z}),{t}}}
\end{array} 
\right) \\ 
&= rank\left(
\begin{array}{c}
\boldsymbol{F_{{\cal S},t}} \\
\boldsymbol{F_{supp(z),{t}}}
\end{array}
\right),
\end{align*}
where the last equality follows from (\ref{eqnno33}). This proves the only if part. Together with (\ref{eqn3a}), the lemma is proved.
\end{IEEEproof}
\section{Matroids}
\label{sec3}
In this section, we provide some basic definitions and results from matroid theory that will be used throughout this paper. For more details, the reader is referred to \cite{Oxl}.

\begin{definition}
\label{matroiddefnindp}
Let $E$ be a finite set. A \textit{matroid} $\cal{M}$ on $E$ is an ordered pair $(E,\cal{I}),$ where the set $\cal{I}$ is a collection  of subsets of $E$ satisfying the following three conditions
\begin{enumerate}
\item[\textbf{I1}] $\phi \in \cal{I}.$ 
\item[\textbf{I2}] If $X \in \cal{I}$ and $X' \subseteq X,$ then $X'  \in \cal{I}.$
\item[\textbf{I3}] If $X_1$ and $X_2$ are in $\cal{I}$ and $|X_1|<|X_2|,$ then there is an element $e$ of $X_2-X_1$ such that $X_1 \cup e\in \cal{I}.$
\end{enumerate}
The set $E$ is called the \textit{ground set} of the matroid and is also referred to as $E(\cal{M}).$ The members of set $\cal{I}$ (also referred to as ${\cal I}(\cal{M})$) are called the \textit{independent sets} of $\cal{M}.$ A maximal independent subset of $E$ is called a \textit{basis} of $\cal M$, and the set of all bases of ${\cal M}$ is denoted by ${\cal B}({\cal M}).$ The set ${\cal I}({\cal M})$ is then obtained as ${\cal I}({\cal M})=\left\{X\subseteq B: B \in {\cal B}({\cal M})\right\}.$ A subset of $E$ which is not in $\cal{I}$ is called a \textit{dependent set}. 
A minimal dependent set of $E$ (any of whose proper subsets is in $\cal{I}$) is called a \textit{circuit} and the set of circuits of $E$ is denoted by $\cal{C}$ or $\cal{C}(\cal{M}).$ With ${\cal M},$ a function called the \textit{rank} function is associated, whose domain is the power set $2^E$ and codomain is the set of non-negative integers. The rank of any $X \subseteq E$ in $\cal{M},$ denoted by $r_{\cal{M}}(X)$, is defined as the maximum cardinality of a subset of $X$ that is a member of $\cal{I}(\cal{M}).$ We denote $r_{\cal{M}}\left(E({\cal M})\right)=r({\cal M}).$
\end{definition}

The set of circuits of a matroid $\cal M$ satisfy the property that if $C_1,C_2 \in {\cal C}({\cal M}),$ and $e\in C_1 \cap C_2,$ then there exists a circuit $C_3 \subseteq \left(C_1\cup C_2\right) - e.$ This is known as the \textit{circuit-elimination axiom}.

Besides using the independent sets, a matroid on $E$ can defined by several other ways, including by specifying the set of circuits, the set of bases or the rank function. We now give the definition of a matroid based on the properties satisfied by the rank function for our use in Section \ref{secinsufficiency}.
\begin{definition}
\label{matroiddefnrank}
Let $E$ be a finite set. A function $r:2^E\rightarrow \mathbb{Z}^+\cup\left\{0\right\}$ is the \textit{rank} function of a matroid on $E$ if and only if $r$ satisfies the following conditions. 
\begin{enumerate}
\item[\textbf{R1}] If $X\subseteq E,$ then $0\leq r(X) \leq |X|.$
\item[\textbf{R2}] If $X\subseteq Y \subseteq E,$ then $r(X)\leq r(Y).$
\item[\textbf{R3}] If $X$ and $Y$ are subsets of $E,$ then 
\[
r(X\cup Y)+r(X\cap Y) \leq r(X)+r(Y).
\]
\end{enumerate}
\end{definition}
\begin{definition}
Two matroids ${\cal M}_1$ and ${\cal M}_2$  are said to be \textit{isomorphic}, denoted as ${\cal M}_1 \widetilde{=} {\cal M}_2,$ if there is a bijection $\varphi$ from $E({\cal M}_1)$  to $E({\cal M}_2)$ such that, for all $X \subseteq E({\cal M}_1),$ $\varphi(X)$ is independent in ${\cal M}_2$ if and only if $X$ is independent in ${\cal M}_1.$
\end{definition}
\begin{definition}
The \textit{vector matroid} associated with a matrix $A$ over some field $\mathbb{F}$, denoted by ${\cal M}[A],$ is defined as the ordered pair $(E,\cal{I})$ where $E$ consists of the set of column labels of $A,$ and $\cal{I}$ consists of all the subsets of $E$ which index columns that are linearly independent over $\mathbb{F}$. An arbitrary matroid $\cal M$ is said to be $\mathbb{F}$-\textit{representable} if it is isomorphic to a vector matroid associated with some matrix $A$ over some field $\mathbb{F}.$ The matrix $A$ is then said to be a \textit{representation} of $\cal M.$ The rank function of a representable matroid $\cal{M}$, given by $r_{\cal{M}}(X),X\subseteq E$ is therefore equal to the rank of the submatrix of columns corresponding to $X$ in the matrix $A$ to which the matroid is associated. A matroid which is not representable over any finite field is called a \textit{nonrepresentable}
 matroid.
\end{definition}
\begin{example}
\label{exm1}
Let 
$A =
\left(
\begin{array}{cccc} 
1 & 0 & 0 & 1 \\
0 & 1 & 0 & 1
\end{array}
\right)
$ with elements from $\mathbb{F}_2.$ Then the matroid ${\cal M}[A]$ over the set $E=\left\{1,2,3,4\right\}$ of column indices of $A$ is defined by 
\[
{\cal I}({\cal M}) = \left\{\left\{1\right\},\left\{2\right\},\left\{4\right\},\left\{1,2\right\},\left\{1,4\right\},\left\{2,4\right\}\right\}.
\]    
\end{example}
\begin{definition}
Let $E = \left\{1,2,...,m\right\}$ for some positive integer $m.$ For some non-negative integer $k \leq m,$ let ${\cal I} = \left\{I \subseteq E: |I|\leq k\right\}.$ The set $\cal I$ satisfies the axioms of independent sets of a matroid on $E,$ referred to as the \textit{uniform matroid} ${\cal U}_{k,m}.$
\end{definition}
\begin{remark}
The vector matroid of a generator matrix of an MDS code of length $m$ and with number of information symbols $k$ is isomorphic to the uniform matroid $U_{k,m}.$
\end{remark}
\begin{definition}
Let $\{{\cal M}_i:i=1,2,..,m\}$ be a collection of matroids defined on the disjoint groundsets $\left\{E_i:i=1,2,..,m\right\}$ respectively. The \textit{direct sum} of the matroids, denoted by $\boxplus_{i=1}^m{\cal M}_i,$ over the groundset $\uplus_{i=1}^m E_i$ is the matroid with the independent sets as follows.
\[
{\cal I}=\left\{\uplus I_i: I_i\in{\cal I}({\cal M}_i)\right\}.
\]
\end{definition} 
\begin{lemma}[\cite{Oxl}]
\label{lemmamatroidequivalence}
Let ${\cal M}={\cal M}[A],$ $A$ being a matrix over some field $\mathbb{F}.$ The matroid $\cal M$ remains unchanged if any of the following operations are performed on $A$
\begin{itemize}
\item Interchange two rows.
\item Multiply a row by a non-zero member of $\mathbb{F}$.
\item Replace a row by the sum of that row and another.
\item Adjoin or delete a zero row.
\item Multiply a column by a non-zero member of $\mathbb{F}$.
\end{itemize}
\end{lemma}

By the row operations of Lemma \ref{lemmamatroidequivalence}, it is clear that any $\mathbb{F}$-representable matroid can be uniquely expressed as the vector matroid of a matrix of the form $\left(I_{r({\cal M})}~~~A_{r({\cal M})\times (|E({\cal M})|-r({\cal M}))}\right),$ with elements from $\mathbb{F}.$

\begin{definition}
Let $\cal{M}$ be the matroid $(E,{\cal I})$ and suppose that $X\subseteq E.$ Let ${\cal I}|X=\left\{I\subseteq X: I \in {\cal I}\right\}.$ Then the ordered pair $(X,{\cal I}|X)$ is a matroid and is called the \textit{restriction} of $\cal{M}$ to $X$ or the \textit{deletion} of $E-X$ from $\cal{M}.$ It is denoted as ${\cal M}|X$ or ${\cal M}\backslash(E-X).$ It follows that the circuits of ${\cal M}|X$ are given by ${\cal C}({\cal M}|X)=\left\{C\subseteq X : C\in {\cal C}(\cal M)\right\}.$
\end{definition}

The restriction of a $\mathbb{F}$-representable matroid is also $\mathbb{F}$-representable. The restriction of a vector matroid ${\cal M}[A]$ to a subset $T$ of the column indices of $A$ is also obtained as the vector matroid of a matrix $A'$ where $A'$ is obtained from $A$ by considering only those columns of $A$ indexed by $T.$
 \begin{example}
Let ${\cal M}={\cal M}[A]$ be the matroid from Example \ref{exm1}. Let $T=\left\{1,2,3\right\}\subseteq E({\cal M}).$ The matroid ${\cal M}|T$ is given by ${\cal I}({\cal M}|T) = \left\{\left\{1\right\},\left\{2\right\},\left\{1,2\right\}\right\} = {\cal I}({\cal M}[A']),$ where $A' = 
\left(\begin{array}{ccc}
1 & 0 & 0 \\
0 & 1 & 0 
\end{array}\right).$
\end{example}
\begin{definition}
Let $\cal M$ be a matroid and ${\cal B}^*({\cal M})$ be $\left\{E({\cal M})-B:B\in {\cal B}({\cal M})\right\}$. Then the set ${\cal B}^*({\cal M})$ forms the set of bases of a matroid on $E({\cal M}),$ defined as the \textit{dual matroid} of $\cal M,$ denoted as ${\cal M}^*.$ Clearly $({\cal M}^*)^* = {\cal M}.$ We also have  
\[
r_{{\cal M}^*}(X)=|X|-r({\cal M})+r_{\cal M}(E({\cal M})-X),
\]
for any $X\subseteq E({\cal M}).$
\end{definition}
\begin{example}
\label{exmdual}
The dual matroid of the matroid ${\cal M}[A]$ given in Example \ref{exm1} is given by the vector matroid ${\cal M}[A']$ corresponding to the matrix
$A'=
\left(\begin{array}{cccc}
0 & 0 & 1 & 0  \\
1 & 1 & 0 & 1
\end{array}\right).
$
\end{example}
\begin{definition}
\label{contraction}
Let $\cal M$ be a matroid on $E$ and $T \subseteq E.$ The \textit{contraction} of $T$ from $\cal M,$ denoted as ${\cal M}/T,$ is given by the matroid $({\cal M}^*\backslash T)^*$ with $E-T$ as its ground set. The set of independent sets of ${\cal M}/T$ is as follows. 
\begin{equation}
\label{independentcontraction}
{\cal I}({\cal M}/T) =\left\{I\subseteq E-T : I \cup B_T \in {\cal I}({\cal M}) \right\} \\
\end{equation}
where $B_T$ is some basis of ${\cal M}|T.$ The set of circuits of ${\cal M}/T$ consists of the minimal non-empty members of $\left\{C-T:C\in {\cal C}({\cal M})\right\}.$
\end{definition}

In Section \ref{sec4}, we show that for a network to be a matroidal error detecting (or correcting) network associated with a matroid $\cal M,$ the circuits of $\cal M$ have to satisfy certain conditions. Thus the concept of circuits of a matroid is the gateway for our results concerning matroidal error detecting (correcting) networks. This is in contrast with the theory of matroidal networks developed in \cite{DFZ,KiM}, where any arbitrary matroid can give rise to a corresponding matroidal network. 

\begin{example}
Let $\cal M$ be the matroid with ground set $E=\left\{a,b,c,d,e\right\}$ and with set of bases
${\cal B}$ being the set of all subsets of $E$ of size four. We wish to find ${\cal M}/\left\{d,e\right\}.$ It can be seen that the dual matroid ${\cal M}^*$ has the set of all singletons of $E$ as its set of bases ${\cal B}^*.$ Then, the matroid ${\cal M}^*\backslash\left\{d,e\right\}$ has the ground set $E'=\left\{a,b,c\right\}$ and the set of bases 
\[
{\cal B}' = \left\{\{a\},\{b\},\{c\}\right\}.
\]
The dual matroid of ${\cal M}^*\backslash\left\{d,e\right\}$ is the matroid ${\cal M}/\left\{d,e\right\}$ with the ground set $\left\{a,b,c\right\}$ and the set of bases 
\[
{\cal B}'' = \left\{\{a,b\},\{a,c\},\{b,c\}\right\}.
\]
\end{example}

\begin{remark}\cite{Oxl}
\label{remarkcontraction}
The contraction of a $\mathbb{F}$-representable matroid is also $\mathbb{F}$-representable. Let ${\cal M}[A]$ be the vector matroid associated with a matrix $A$ over $\mathbb{F}.$ Let $e$ be the index of a non-zero column of $A.$ Suppose using the elementary row operations listed in Lemma \ref{lemmamatroidequivalence}, we transform $A$ to obtain a matrix $A'$ which has a single non-zero entry in column $e.$ Let $A''$ denote the matrix which is obtained by deleting the row and column containing the only non-zero entry of column $e.$  Then 
\[
{\cal M}[A]/\left\{e\right\} = \left({\cal M}[A]^*\backslash\left\{e\right\}\right)^* = {\cal M}[A''],
\] 
where ${\cal M}[A]^*$ is the dual matroid of ${\cal M}[A].$
\end{remark}
\begin{example}
Let ${\cal M}={\cal M}[A]$ be the matroid from Example \ref{exm1}. We want to find ${\cal M}[A]/\left\{4\right\}.$ We first obtain ${\cal M}[A]/\left\{4\right\}$ in a straightforward manner according to the definition of contraction. The dual matroid of ${\cal M}[A]$ is the vector matroid corresponding to the matrix
\[
A_d=
\left(
\begin{array}{cccc}
0 & 0 & 1 & 1 \\
1 & 1 & 0 & 1
\end{array}
\right).
\]
Now ${\cal M}[A_d]\backslash\left\{4\right\}$ is the vector matroid corresponding to the matrix 
\[
A'_d=
\left(
\begin{array}{ccc}
0 & 0 & 1  \\
1 & 1 & 0 
\end{array}
\right).
\]
According to the definition of contraction, ${\cal M}[A'_d]^*={\cal M}[A]/\left\{4\right\}.$ The set of bases of  ${\cal M}[A'_d]^*$ is $\left\{\{1\},\{2\}\right\}.$ Thus the matroid 
\[
{\cal M}[A]/\left\{4\right\} = \left(E=\left\{1,2,3\right\},{\cal I}=\left\{\phi,\{1\},\{2\}\right\}\right).
\]  
We can also obtain ${\cal M}[A]/\left\{4\right\}$ using the technique shown in Remark \ref{remarkcontraction}. Towards that end, using row operations on $A,$ we obtain the matrix 
\[
A' =
\left(
\begin{array}{cccc}
1 & 0 & 0 & 1 \\
1 & 1 & 0 & 0
\end{array}
\right).
\]
By removing the row corresponding to the only non-zero entry in the $4^{th}$ column of $A'$ and the $4^{th}$ column itself, we obtain the matrix $A''=(1~ 1~ 0).$ It is easily verified that ${\cal M}[A'_d]^* = {\cal M}[A''].$ 
\end{example}
\begin{definition}Let $\cal M$ be a matroid on $E$ and $X$ be a subset of $E$. The \textit{closure} of $X$ is then defined to be the set $cl_{\cal M}(X)=\left\{x\in E:r_{\cal M}(X\cup x)=r_{\cal M}(X)\right\}.$ If $X=cl_{\cal M}(X),$ then $X$ is said to be a \textit{flat} of $\cal M.$ A flat $H$ such that $r_{{\cal M}}(H) = r({\cal M})-1$ is called a \textit{hyperplane} of $\cal M.$ Moreover, $X \subset E$ is a hyperplane of $\cal M$ if and only if $E-X$ is a circuit of ${\cal M}^*.$
\end{definition}
\begin{example}
Consider the matroid ${\cal M}[A]$ of Example \ref{exm1}. Let $X=\left\{1\right\},$ then $cl_{\cal M}(X)=\left\{1,3\right\}$ is a flat. Moreover it is also a hyperplane of $\cal M.$ Also, it can be easily verified that the set
\[
E(\{{\cal M}[A]\})-\left\{1,3\right\} = \left\{2,4\right\}
\]
is a circuit of the dual matroid ${\cal M}[A]^*,$ given in Example \ref{exmdual}.
\end{example}
\begin{definition}
Let $\cal N$ be a matroid on $E.$ If for some $e\in E,$ $\left\{e,f\right\}\in {\cal C}({\cal N})$ for some $f \in E,$ then the matroid $\cal N$ is said to be a \textit{parallel extension} of ${\cal M}={\cal N}\backslash \{e\},$  and is denoted by ${\cal M}+^p_f e.$ The element $e$ is said to be \textit{added in parallel} with element $f.$ 
Also, a parallel extension ${\cal N}^*$ of ${\cal M}^*$ is said to be a \textit{series extension} of ${\cal M},$ in which case ${\cal M}={\cal N}/\{e\}$ and $\cal N$ is denoted by ${\cal M}+^s_f e.$ The element $e$ is then said to be \textit{added in series} with element $f.$
\end{definition}

The following two lemmas summarise equalities which can be proved easily from the definitions of the series and parallel matroids and the duality relations between them. We state them here without proof so that we may use them later in Section \ref{sec5}.
\begin{lemma}
\label{remparallel}
Let $f\in E({\cal M})$ such that $\{f\}\notin{\cal C}({\cal M}).$ In a parallel extension ${\cal N}={\cal M}+^p_f e$ of $\cal M.$ The following statements are true.
\begin{align}
\label{eqn201}
&r_{\cal N}(X)=r_{\cal M}(X), \forall X\subseteq E({\cal M}).\\
\label{eqn202}
&r_{\cal N}(X-f+e)=r_{\cal M}(X), \forall X\subseteq E({\cal M})~\text{with}~f\in X.\\
\label{eqn203}
&r({\cal N})=r({\cal M}).\\
\label{eqn204}
&{\cal M}={\cal N}\backslash\{e\}.
\end{align}
\end{lemma}
\begin{lemma}
\label{remseries}
Let $f\in E({\cal M})$ such that $\{f\}\notin{\cal C}({\cal M}).$ In a series extension ${\cal N}={\cal M}+^s_f e$ of $\cal M,$ The following statements are then true.
\begin{align}
\label{eqn301}
&{\cal B}({\cal N})=\{B\cup\{e\}:B \in {\cal B}({\cal M})\}.\\
\label{eqn302}
&r_{\cal M}(X)=r_{\cal N}(X), \forall X\subseteq E({\cal M})~\text{such that}~f\notin X.\\
\label{eqn303}
&{\cal M}={\cal N}/\{e\}.
\end{align}
\end{lemma}

We now present two lemmas, which will be useful for describing the construction of matroidal error detecting (correcting) networks in Section \ref{sec5}. They also serve as examples for parallel and series extensions of a matroid. To the best of our knowledge they are not explicitly found in existing matroid literature. Therefore, we prove them here for the sake of completeness.
\begin{lemma}
\label{parallelrepresentation}
Let $A$ be an $n\times N$ matrix over $\mathbb{F}.$ For some $1\leq i \leq n,$ let ${A^i}$ be a non-zero column of $A.$ Let $B$ be the $n\times (N+1)$ matrix
\[
\left( 
\begin{array}{ccccc}
{A^1} & {A^2} & ... & {A^{N}} & {A^{i}}
\end{array}
\right).
\]
Then, ${\cal M}[B]={\cal M}[A]+^p_i \left\{N+1\right\},$ i.e., ${\cal M}[B]$ is a parallel extension of the vector matroid associated with $A.$
\end{lemma}
\begin{IEEEproof}
Clearly ${\cal M}[B]\backslash \left\{N+1\right\} = {\cal M}[A] = {\cal M}.$ Moreover, in ${\cal M}[B],$ the $\left(N+1\right)^{th}$ column of $B$ is equal to the $i^{th}$ column, thus $\left\{i,N+1\right\} \in {\cal C}({{\cal M}[B]}).$ Thus, by definition, ${\cal M}[B] = {\cal M}[A]+^p_i \left\{N+1\right\},$ the parallel extension of ${\cal M}[A]$ at $i.$ This proves the lemma.
\end{IEEEproof}
\begin{lemma} 
\label{seriesrepresentation}
Let 
$
A = \left( 
\begin{array}{cccc}
{A^1} & {A^2} & ... & {A^N}  
\end{array}
\right)
$
be an $n\times N$  matrix over $\mathbb{F},$ where ${A^j}$ denotes the $j^{th}$ column of $A.$ For some $1\leq i \leq n,$  let ${A^i}$ be a non-zero column of $A$ such that $A^i \in \left\langle\left(A^{\left\{1,...,N\right\}-i}\right)\right\rangle$ . Let $B$ be the  $(n+1)\times (N+1)$ matrix
\[
\left( 
\begin{array}{ccccccccc}
{A^1} & {A^2} & ...& {A^{i-1}} & {A^{i}} & {A^{i+1}} & ... & {A^{N}} & \boldsymbol{0} \\
0			 &   0		&	... & 0 & 	1 	 & 0 & .... & 0 & 1 
\end{array}
\right),
\]
where $\boldsymbol{0} \in \mathbb{F}^n.$ Then the vector matroid associated with $B,$  ${\cal M}[B],$ is a series extension of the vector matroid associated with $A,$ ${\cal M}[A]$ at $i,$ i.e., ${\cal M}[B]={\cal M}[A]+^s_i \left\{N+1\right\}.$
\end{lemma}
\begin{IEEEproof}
Because $A^i \in \left\langle\left(A^{\left\{1,...,N\right\}-i}\right)\right\rangle,$ we must have $B^i \in \left\langle\left(B^{\left\{1,...,N,N+1\right\}-i}\right)\right\rangle.$ Also from the form of $B,$ we have $B^i \notin \left\langle\left(B^{\left\{1,...,N\right\}-i}\right)\right\rangle.$ Thus, $\left\{1,2,..,N+1\right\}-\left\{i,N+1\right\}$ of columns forms a hyperplane of ${\cal M}[B].$ Therefore, $\left\{i,N+1\right\}$ is a circuit in ${\cal M}[B]^*.$ Also, as ${\cal M}[B]/\left\{N+1\right\}={\cal M}[A],$ we must have ${\cal M}[B]^*\backslash\left\{N+1\right\}={\cal M}[A]^*.$ Thus ${\cal M}[B]^*$ is a parallel extension of ${\cal M}[A]^*,$ i.e., ${\cal M}[B]^*={\cal M}[A]^*+_i^p\left\{N+1\right\}.$ Hence ${\cal M}[B]={\cal M}[A]+_i^s\left\{N+1\right\},$ i.e., ${\cal M}[B]$ is a series extension of ${\cal M}[A].$ This proves the lemma. 
\end{IEEEproof}
\begin{definition}
If a matroid $\cal M$ is obtained from a matroid $\cal N$ by deleting a non-empty subset $T$ of $E({\cal N}),$ then $\cal N$ is called an \textit{extension} of $\cal M.$  In particular, if $|T|=1$, then ${\cal N}$ is said to be a \textit{single-element extension} of ${\cal M}.$
\end{definition}
\begin{definition}
\label{singleelement}
Let $\cal K$ be a set of flats of $\cal M$ satisfying the following conditions.
\begin{itemize}
\item If $F\in {\cal K}$ and $F'$ is a flat of $\cal M$ containing $F,$ then $F' \in{\cal K}.$
\item If $F_1,F_2\in{\cal K}$ are such that $r_{\cal M}(F_1)+r_{\cal M}(F_2)=r_{\cal M}(F_1\cup F_2)+r_{\cal M}(F_1\cap F_2),$ then $F_1 \cap F_2 \in {\cal K}.$ 
\end{itemize}
Any set $\cal K$ of flats of $\cal M$ which satisfies the above conditions is called a \textit{modular cut} of $\cal M.$ There is a one-one correspondence between the set of all modular cuts of a matroid and the set of all single-element extensions of a matroid. We denote the single-element extension ${\cal N}$ corresponding to the modular cut $\cal K$ as ${\cal M}+_{_{\cal K}} e,$ where $e$ is the new element that is added. Also, the set ${\cal K}$ consists precisely of those flats of $\cal M$ such that for each $F\in {\cal K},$ we have $r_{\cal N}(F\cup e) = r_{\cal N}(F).$ 
\end{definition}

\begin{example}
\label{examplesingle}
Let ${\cal M}$ be the vector matroid of the matrix over $\mathbb{F}_2$
\[
B=\left(
\begin{array}{ccccc}
1&0&0&0&1\\
0&1&0&0&1\\
0&0&1&0&1
\end{array}
\right).
\]
Consider the flats $F_1=\{3,4,5\}$ and $F_2=\{1,2,3,4,5\}.$ Note that the flats $F_1$ and $F_2$ form a modular cut $\cal K$ satisfying the conditions in Definition \ref{singleelement}. Thus there exists a single-element extension of $\cal M$ which corresponds to this modular cut. Let ${\cal M}'$ be this matroid. It can be verified that ${\cal M}'$ is the vector matroid of the matrix over $\mathbb{F}_3$ 
\[
B'=\left(
\begin{array}{cccccc}
1&0&0&0&1&1\\
0&1&0&0&1&1\\
0&0&1&0&1&2
\end{array}
\right).
\]
However, ${\cal M}'$ does not have a representation over the field $\mathbb{F}_2.$
\end{example}
\begin{definition}
Let $\cal M$ be a matroid. For a flat $F$ in the set of flats of $\cal M,$ let ${\cal K}_{F}$ denote the set of all flats of $\cal M$ which contain $F.$ Then ${\cal K}_{F}$ can be easily verified to be a modular cut of $\cal M$ and is defined as the \textit{principal modular cut} of $\cal M$ \textit{generated by the flat} $F.$ The single-element extension of $\cal M$ corresponding to this principal modular cut is then defined as the \textit{principal extension of} $\cal M$ \textit{generated by the flat} $F,$ and is denoted by ${\cal M}+_{_{{\cal K}_F}} e,$ where $e$ is the new element added. 
\end{definition}
\begin{example}
The single-element extension shown in Example \ref{examplesingle} is a principal extension of the matroid ${\cal M}$ generated by the flat $F_1.$ The principal modular cut corresponding to this extension is then ${\cal K}.$
\end{example}
\section{Matroidal Error Correcting and Detecting Networks}
\label{sec4}
In this section, we define \textit{matroidal error correcting and detecting networks} and establish the link between matroids and network-error correcting and detecting codes. The contents of this section are logical extensions of the concept of the matroidal networks defined in \cite{DFZ} which gave the connection between matroids and network codes. The definition of a matroidal network is as follows. 
\begin{definition}[\cite{DFZ}]
\label{matroidalnetworkdefinition}
Let ${\cal G}({\cal V},{\cal E})$ be a network with a message set $\mu.$ Let ${\cal M}=(E,{\cal I})$ be a matroid. The network ${\cal G}$ is said to be a \textit{matroidal network} associated with ${\cal M}$ if there exists a function 
$f:\mu\cup{\cal E}\rightarrow E$ such that the following conditions are satisfied.
\begin{enumerate}
\item $f$ is one-one on $\mu.$
\item $f(\mu) = \cup_{m \in \mu}f(m)  \in{\cal I}.$
\item $r_{\cal M}(f(In(v)))=r_{\cal M}(f(In(v)\cup Out(v))),$ $\forall v\in {\cal V}.$
\end{enumerate}
\end{definition}

Suppose $\cal M$ is a representable matroid. Then the first two conditions of Definition \ref{matroidalnetworkdefinition} can be interpreted as associating independent global encoding vectors with the information symbols. The last condition will then ensure that flow conservation holds throughout the network, and also that the sinks are able to decode the demanded information symbols. Thus Definition \ref{matroidalnetworkdefinition} can be looked at as the matroidal generalization of a scalar linear network code, which is confirmed by the following theorem proved in parts in \cite{DFZ} and \cite{KiM}.
\begin{theorem}
\label{matroidalnetworkthm}
A network $\cal G$ is matroidal in association with a representable matroid if and only if it has a scalar linear network coding solution.
\end{theorem}

Let ${\cal G}({\cal V},{\cal E})$ be an acyclic network with a collection of sources $\cal S$ with message set $\mu$ (with $n$ elements) and sinks $\cal T,$ and a given topological order on $\cal E.$ Let $\beta < |{\cal E}|$ be a non-negative integer, and $\mathfrak{F}=\left\{{\cal F} \subseteq {\cal E}:|{\cal F}| = \beta \right\}$ be the collection of error patterns of size $\beta.$ Let $\cal M$ be a matroid over a ground set $E$ with $n + 2|{\cal E}|$ elements, and with $r({\cal M})= n+|{\cal E}|.$ We now define \textit{matroidal error detecting and correcting networks} by extending the definition of matroidal networks of \cite{DFZ} for the case of networks where errors occur. 
\begin{definition}
\label{matroidalerrornetworkdefinition}
The network ${\cal G}$ is said to be a \textit{matroidal} $\beta$-\textit{error detecting network} associated with $\cal M,$ if there exists a function $f:\mu\cup{\cal E} \rightarrow E({\cal M})$ such that the following conditions are satisfied.
\begin{enumerate}[(A)]
\item
\textbf{Independent inputs condition}: $f$ is one-one on $\mu,$ where $f(\mu) = \cup_{m \in \mu}f(m) \in {\cal I}({\cal M})$.
\item	\textbf{Flow conservation condition}: For some basis $B$ of ${\cal M}$ obtained by extending $f(\mu)$ (where  $B-f(\mu)=\left\{b_{n+1},...,b_{n+|{\cal E}|}\right\}$ is ordered according to the given topological order on $\cal E$), the following conditions should hold for all $e_i \in {\cal E}.$
\begin{enumerate}[(B1)]
\item $f(e_i) \notin cl_{\cal M}(B-f(\mu))$
\item $r_{\cal M}\left(f\left(In(e_i)\right)\cup f(e_i)\cup b_{n+i}\right)$ 
\vspace {-0.2cm} 
\begin{align*}
& = r_{\cal M}\left(f\left(In(e_i)\right)\cup b_{n+i}\right)\\
& = r_{\cal M}\left(f\left(In(e_i)\right)\right) + r_{\cal M}(b_{n+i}) \\
& = r_{\cal M}\left(f\left(In(e_i)\right)\right) + 1.
\end{align*}
\end{enumerate}
\item \textbf{Successful decoding condition}: For each error pattern ${\cal F}=\left\{e_{i_1},e_{i_2},...,e_{i_{\beta}}\right\}\in\mathfrak{F},$ let $B_{\overline{{\cal F}}} = B-f(\mu)-\left\{b_{n+i_1},b_{n+i_2},...,b_{n+i_{\beta}}\right\}.$ Let ${\cal M}_{\cal F}$ be the $n+\beta+|{\cal E}|$ element matroid ${\cal M}/B_{\overline{\cal F}}.$ 
Then, at every sink $t\in {\cal T},$ for each ${\cal F} \in \mathfrak{F},$ we must have
\[
r_{{\cal M}_{\cal F}}\left(f\left(In_{\cal E}(t)\right)\cup f\left(\boldsymbol{{\cal D}_t}\right)\right)=r_{{\cal M}_{\cal F}}\left(f\left(In_{\cal E}(t)\right)\right),
\]
where $In_{\cal E}(t) \subseteq In(t)$ denotes the set of incoming edges at sink $t$ and $\boldsymbol{{\cal D}_t}$ is the set of demands at $t.$
\end{enumerate}
\end{definition}
\begin{definition}
\label{matroidalerrorcorrectionnetworkdefinition}
The network ${\cal G}$ is said to be a \textit{matroidal} $\alpha$-\textit{error correcting network} associated with a matroid $\cal M,$ if it is a  matroidal $2\alpha$-\textit{error detecting network} associated with $\cal M.$
\end{definition}
\begin{remark}
As with Definition \ref{matroidalnetworkdefinition}, Definitions \ref{matroidalerrornetworkdefinition} and \ref{matroidalerrorcorrectionnetworkdefinition} can be viewed as the matroidal abstractions of a scalar linear network-error detecting and correcting codes (Theorem \ref{matroidalerrornetworkthm} will present the formal statement and proof of this abstraction). If $\cal M$ is a representable matroid, then as in Definition \ref{matroidalnetworkdefinition}, Condition (A) is equivalent to saying that the global encoding vectors corresponding to the information symbols are linearly independent. Condition (B1) is equivalent to saying that the symbol flowing on any edge in the network is a \textit{non-zero} linear combination of the information symbols, added with a (not necessarily non-zero) linear combination of the network-errors in the network. Such a condition is not a restriction, because if an edge carries an all-zero linear combination of the input symbols, then such an edge can simply be removed from the network. Condition (B2) is equivalent to a modified flow conservation condition in networks with errors, implying that the symbol flowing through any edge $e$ in the network is a linear combination of the incoming symbols at $In(e)$ and the network-error in that particular edge. Condition (C) ensures that the sinks can decode their demands. Although our definitions are abstracted from scalar linear network-error detecting and correcting codes, we will show in Section \ref{secinsufficiency} that it applies to nonlinear schemes also.
\end{remark}
\begin{remark}
The Condition (C) of Definition \ref{matroidalerrornetworkdefinition} requires that $f(x), \forall x\in\mu\cup{\cal E}$ exist in $E({\cal M}_{\cal F})$ in the first place. However, this is ensured by Condition (B1). To see this, first we note that $f(\mu)\subset E({\cal M}_{\cal F})$ because these elements are in $B$ and are not contracted out of $\cal M.$ Now consider the set $f(e)\cup(B-f(\mu))$ for any $e\in {\cal E},$ which is independent in $\cal M$ because of Condition (B1). By (\ref{independentcontraction}) in the definition of the contraction of a matroid, we have that $f(e)$ exists and is also not dependent in ${\cal M}_{\cal F}.$ Therefore, $f(x)$ is well defined in ${\cal M}_{\cal F}$ also. 
\end{remark}
\begin{remark}
Although Definition \ref{matroidalnetworkdefinition} and Definition \ref{matroidalerrornetworkdefinition} in the case of no network-errors do not immediately appear to agree, it can be shown that a network is a matroidal network associated with some matroid $\cal M,$ if and only if it is a matroidal error detecting network with  $\beta=0,$ with respect to another matroid derived using extensions of $\cal M.$ This can be inferred easily from the remainder of this paper, therefore we leave it without an explicit proof.
\end{remark}

We now present the main result of this paper which is the counterpart of the results from \cite{DFZ,KiM} which relate networks with scalar linearly solvable network codes to representable matroids. 
\begin{theorem}
\label{matroidalerrornetworkthm}
Let ${\cal G}({\cal V},{\cal E})$ be an acyclic communication network with sources $\cal S$ and sinks $\cal T.$ The network ${\cal G}$ is a \textit{matroidal} $\beta$-\textit{error detecting network} associated with a  $\mathbb{F}$-representable matroid if and only if it has a scalar linear network-error detecting code over $\mathbb{F}$ that can correct network-errors at any $\beta$ edges which are known to the sinks.
\end{theorem}
\begin{IEEEproof}
\textit{If part:}
Suppose there exists a scalar linear $\beta$-network-error detecting code over $\mathbb{F}$ for $\cal G$ with the matrices $A_{s_i} (i=1,2,...,|{\cal S}|), \boldsymbol{F}$ and $B_{t},~t\in{\cal T},$ as defined in Section \ref{sec2}, according to the given topological ordering on $\cal E$. Let $\cal A$ be the matrix as in (\ref{formofA}).

Let $\cal X$ be the row-wise concatenated matrix 
$
\left( 
\begin{array}{c}
{\cal A}\boldsymbol{F} \\
\boldsymbol{F}
\end{array}
\right)
$
of size $(n+|{\cal E}|)\times |{\cal E}|,$ and  $\cal Y$ be the column-wise concatenated matrix 
$
\left(I_{n+|{\cal E}|}~~~{\cal X}\right).
$
Also, let ${\cal M}={\cal M}[{\cal Y}],$ the vector matroid associated with $\cal Y,$ with $E({\cal M})$ being the set of column indices of $\cal Y.$ Let 
$
f:{\cal E}\cup\mu\rightarrow E({\cal M})
$
be the function defined as follows. 
\begin{align*}
& f(m_i) = i,~~m_i \in \mu, i=1,2,...,n. \\
& f(e_i) = n+|{\cal E}|+i,~\forall~e_i \in {\cal E}~\text{in the given ordering}. 
\end{align*}
We shall consider the basis for $\cal M$ as $B=\left\{1,2,...,n+|{\cal E}|\right\},$  i.e., the first $n+|{\cal E}|$ columns of ${\cal Y}$. This basis will be used repeatedly in the proof. We shall now prove that the matroid $\cal M$ and function $f$ satisfy the conditions of Definition \ref{matroidalerrornetworkdefinition}. Towards this end, first we see that Condition (A) holds by the definition of function $f.$ 

We first prove that Condition (B1) holds. We have that ${\cal Y}^{n+|{\cal E}|+i} \notin \left\langle({\cal Y}^{B-f(\mu)})\right\rangle,$ because no edge is assigned a zero-global encoding vector, i.e., no column of ${\cal A}\boldsymbol{F}$ is zero. Thus Condition (B1) holds. 

To show Condition (B2), first note that because the given set of coding coefficients for the network is a (valid) network code, $\boldsymbol{F}$ is such that 
\begin{equation}
\label{eqn5}
\boldsymbol{F}^{j} = \left(\sum_{\scriptsize \begin{array}{c} e_i \in {\cal E}:\\ tail(e_j)=head(e_i) \end{array}}\hspace{-1cm}K_{i,j}\boldsymbol{F}^{i}\right)+\boldsymbol{1_j},
\end{equation}
where $\boldsymbol{1_j}$ is a column vector in $\mathbb{F}^{|{\cal E}|}$ with all zeros except for the $j^{th}$ entry which is $1 \in \mathbb{F}.$  Also, (\ref{eqn5}) implies that 
\vspace{-0.07cm}
\begin{align}
\nonumber
\left({\cal A}\boldsymbol{F}\right)^{j} & = {\cal A}\boldsymbol{F}^{j} \\
\nonumber
& = {\cal A}\left(\sum_{\scriptsize \begin{array}{c} e_i \in {\cal E}:\\ tail(e_j)=head(e_i) \end{array}}\hspace{-1cm}K_{i,j}\boldsymbol{F}^{i}\right)+{\cal A}\boldsymbol{1_{j}}\\
\label{eqn6}
&= \left(\sum_{\scriptsize \begin{array}{c} e_i \in {\cal E}:\\ tail(e_j)=head(e_i) \end{array}}\hspace{-1cm}K_{i,j}\left({\cal A}\boldsymbol{F}\right)^{i}\right)+{\cal A}^j.
\end{align}
Thus, combining (\ref{eqn5}) and (\ref{eqn6}), we have 
\begin{align*}
{\cal X}^j &= {\cal Y}^{n+|{\cal E}|+ j} \\
& = \left(\sum_{\scriptsize \begin{array}{c} e_i \in {\cal E}:\\ tail(e_j)=head(e_i) \end{array}}\hspace{-1cm}K_{i,j}{\cal X}^i\right) + {\cal Y}^{f(\mu)}{\cal A}^j + {\cal Y}^{n+j} \\ 
& = \left(\sum_{\scriptsize \begin{array}{c} e_i \in {\cal E}:\\ tail(e_j)=head(e_i) \end{array}}\hspace{-1cm}K_{i,j}{\cal Y}^{n+|{\cal E}|+ i}\right) + {\cal Y}^{f(\mu)}{\cal A}^j + {\cal Y}^{n+j},\\
\end{align*}
where ${\cal Y}^{n+j}$ corresponds to $b_{n+j} \in B-f(\mu)$ and the non-zero coefficients of ${\cal A}^j$ can occur only in those positions corresponding to the set of messages generated at $tail(e_j),$ if any, which is a subset of $In(tail(e_j))=In(e_j)$. Also, for any $e_i \in {\cal E}$ with $tail(e_j)=head(e_i),$ the vector ${\cal Y}^{n+|{\cal E}|+ i}$  is some column of the matrix ${\cal Y}^{f(In(e_j))}.$ Thus 
\begin{equation}
\label{eqnno20}
{\cal Y}^{n+|{\cal E}|+j} \in \left\langle\left({\cal Y}^{f(In(e_j))\cup b_{n+j}}\right)\right\rangle.
\end{equation}
We also note that the $(n+j)^{th}$ row of ${\cal Y}^{n+j}$ contains $1$ (indicating the error at the edge $e_j$) while the $(n+j)^{th}$ row of ${\cal Y}^{f(In(e_j))}$ is all-zero because of the topological ordering in the acyclic network (as symbols flowing in any edge can have contribution only from upstream errors). Therefore ${\cal Y}^{n+|{\cal E}|+j} \notin \left\langle{\cal Y}^{f(In(e_j))}\right\rangle.$ Along with (\ref{eqnno20}), this proves that Condition (B2) holds.

Now we prove that Condition (C) also holds. Let $I({\cal F})=\left\{i_1,i_2,...,i_{\beta}\right\}$ be the index set following the topological ordering corresponding to an arbitrary error pattern ${\cal F}\in\mathfrak{F}$ and let the set $\left\{n+i_1,n+i_2,...,n+i_{\beta}\right\}$ be denoted as $n+I({\cal F}).$ First we note that by definition, ${\cal M}_{\cal F}$ is the vector matroid of the matrix 
\begin{equation}
\label{eqnno21}
{\cal Z} = {\cal Y}_{f(\mu)\cup (n+I({\cal F}))} = \left(I_{n+\beta}~~ {\cal X}_{f(\mu)\cup (n+I({\cal F}))}\right),
\end{equation}
where 
$
{\cal X}_{f(\mu)\cup (n+I({\cal F}))} = \left( 
\begin{array}{c}
{\cal A}\boldsymbol{F} \\
\boldsymbol{F}_{I({\cal F})}
\end{array}
\right).
$
Now for a sink $t\in{\cal T},$ 
\[
{\cal Z}^{f(In_{\cal E}(t))}={\cal X}_{f(\mu)\cup (n+I({\cal F}))}^{f(In_{\cal E}(t))}=
\left( 
\begin{array}{c}
{\cal A}\boldsymbol{F}^{f(In_{\cal E}(t))} \\
\boldsymbol{F}_{I({\cal F})}^{f(In_{\cal E}(t))}
\end{array}
\right).
\]
But according to Section \ref{sec2}, we have, ${\cal A}\boldsymbol{F}^{f(In_{\cal E}(t))} = \boldsymbol{F_{{\cal S},t}},$ and $\boldsymbol{F}_{I({\cal F})}^{f(In_{\cal E}(t))} = \boldsymbol{F_{supp(\boldsymbol{z}),t}},$ where $supp(\boldsymbol{z})={\cal F}.$ By Lemma \ref{lemmadecoding}, as the given network code is $\beta$-network-error detecting, we must have 
\[
cols(I_{\boldsymbol{{\cal D}_{t}}}) \subseteq \left\langle\left( 
{\cal Z}^{f(In_{\cal E}(t))}\right)\right\rangle,
\]
where $\boldsymbol{{\cal D}_{t}}\subseteq \mu$ is the set of demands at $t$. But then $I_{\boldsymbol{{\cal D}_{t}}}={\cal Z}^{f(\boldsymbol{{\cal D}_{t}})}$ by (\ref{eqnno21}). This proves Condition (C) for sink $t.$ The choice of error pattern and sink being arbitrary, this proves the If part of the theorem. 

\textit{Only If part:}
Let $\cal M$ be the given $\mathbb{F}$-representable matroid, along with the function $f,$ and basis $B=f(\mu)\uplus\left\{b_{n+1},b_{n+2},...,b_{n+|{\cal E}|}\right\}$ that satisfy the given set of conditions. Let 
${\cal Y}=(I_{n+{|{\cal E}|}}~~{\cal X})$ be a representation of $\cal M$ over $\mathbb{F},$ such that $B=\left\{1,2,...,n+|{\cal E}|\right\}.$ First we prove the following claim.

\textit{Claim:}
There exists an $n\times |{\cal E}|$ matrix ${\cal A},$  and a $|{\cal E}|\times |{\cal E}|$ matrix $\boldsymbol{F}$ of the form $\boldsymbol{F}=(I_{|{\cal E}|}-K)^{-1}$ for some strictly upper-triangular matrix $K$, such that 
\begin{equation}
\label{formofX}
{\cal X}=\left( 
\begin{array}{c}
{\cal A}\boldsymbol{F} \\
\boldsymbol{F}
\end{array}
\right).
\end{equation}

\textit{Proof of claim}:

Consider an edge $e_j\in {\cal E}.$ Let $\mu_{tail(e_j)}$ denote indices of the set of messages generated at $tail(e_j).$ As Condition (B2) holds, ${\cal Y}^{f(e_j)}$ is such that
\begin{align}
\nonumber
&{\cal Y}^{f(e_j)} \\
\label{eqn7}
&= \hspace{-0.4cm}\sum_{\scriptsize \begin{array}{c} e_i \in {\cal E}:\\ tail(e_j)=head(e_i) \end{array}}\hspace{-1cm}a'_{i,j}{\cal Y}^{f(e_i)} +\hspace{-0.3cm} \sum_{m_i \in {\mu}_{tail(e_j)}}\hspace{-0.5cm}c'_{i,j}{\cal Y}^{f(m_i)}+a'_{j,j}{\cal Y}^{n+j},
\end{align}
for some $a'_{i,j}$ and $c'_{i,j}$ in $\mathbb{F}.$ Note that if $e_j$ is such that $In(e_j)\subseteq \mu,$ then by (\ref{eqn7}), ${\cal Y}^{f(e_j)}$ is just a linear combination of ${\cal Y}^{{\mu}_{tail(e_j)}}$ and ${\cal Y}^{n+j}.$ Following the ancestral ordering for $j,$ it can be seen that for any edge $e_j,$ ${\cal Y}^{f(e_j)}$ is a linear combination of ${\cal Y}^{\left\{1,2,...,n+j\right\}}$ and ${\cal Y}^{\mu}.$ Thus we have,
%
%
%
\[
{\cal Y}^{f(e_j)} = \sum_{\scriptsize e_i \in {\cal E}: i \leq j}a_{i,j}{\cal Y}^{n+i}  +\hspace{-0.2cm} \sum_{m_i \in {\mu}}c_{i,j}{\cal Y}^{f(m_i)}.
\]
%
As Condition (B1) holds, we must have at least one $c_{i,j}\neq 0, \forall i=1,2,...,n$ and because of Condition (B2), we must have $a_{j,j} = a'_{j,j}\neq 0.$ This structure of ${\cal Y}^{f(e_j)}$ also implies that ${\cal Y}^{f(e_j)}\neq {\cal Y}^b,$ for any $b\in B.$ Moreover, we also see that ${\cal Y}^{f(e_i)}\neq {\cal Y}^{f(e_j)},$ for any distinct pair $e_i,e_j$ of edges in $\cal E.$  Arranging all the ${\cal Y}^{f(e_i)}$s in the given topological order (i.e., with $f(e_j)=n+|{\cal E}|+j$), we get ${\cal Y}^{f({\cal E})} = {\cal X},$ and
\[
{\cal X} = 
\left(\begin{array}{c} 
J_{n\times|{\cal E}|} \\ L_{|{\cal E}|\times|{\cal E}|}
\end{array}\right),
\]
where $J$ comprises of the elements $c_{i,j}, 1\leq i \leq n,1\leq j \leq |{\cal E}|$ and $L$ is the matrix
\[
\scriptsize
L = \left(\begin{array}{ccccc} 
a_{1,1} & a_{1,2} & .& .& a_{1,|{\cal E}|}\\
0 & a_{2,2} & .& .& a_{2,|{\cal E}|}\\
. & 0 & .& .& .\\
. & .  & .& .& .\\
. & .  & .& .& .\\
0 & 0  & .& 0 & a_{|{\cal E}|,|{\cal E}|}
\end{array}\right).
\]

By Lemma \ref{lemmamatroidequivalence}, the matroid $\cal M$ does not change if some row or some column of ${\cal Y}=(I_{n+{|{\cal E}|}}~~{\cal X})$ is multiplied by a non-zero element of $\mathbb{F}.$ Let ${\cal Y}'$ be the matrix obtained from ${\cal Y}$ by multiplying the rows $\left\{n+1,n+2,...,n+|{\cal E}|\right\}$ by the elements $\left\{a_{1,1}^{-1},a_{2,2}^{-1},...,a_{|{\cal E}|,|{\cal E}|}^{-1}\right\}$ respectively, and then multiplying the columns $\left\{n+1,n+2,...,n+|{\cal E}|\right\}$ by $\left\{a_{1,1},a_{2,2},...,a_{|{\cal E}|,|{\cal E}|}\right\}$ respectively. The matrix ${\cal Y}'$ is then of the form  $(I_{n+{|{\cal E}|}}~~{\cal X}'),$ where $
{\cal X}' = 
\left(\begin{array}{c} 
J \\ L'_{|{\cal E}|\times|{\cal E}|}
\end{array}\right),
$
$L'$ being the upper-triangular matrix obtained from $L,$ i.e.,
\[
\scriptsize
L' = \left(\begin{array}{ccccc} 
1 & a_{1,2}a_{1,1}^{-1} & . & .& a_{1,|{\cal E}|}a_{1,1}^{-1}\\
0 & 1 & . & .& a_{2,|{\cal E}|}a_{2,2}^{-1}\\
. & 0 & . & .& .\\
. & . & . & .& .\\
. & . & . & .& .\\
0 & 0 & . & 0& 1
\end{array}\right).
\]
As $\cal M$ is the vector matroid of ${\cal Y}'$ also, without loss of generality we assume that ${\cal Y}={\cal Y}',$ with $a_{1,1}=a_{2,2}=...=a_{|{\cal E}|,|{\cal E}|}=1.$

Now let $H$ be the $n\times |{\cal E}|$ matrix whose columns are populated as follows. For all $j=1,2,...,|{\cal E}|,$ 
\[
H^j=J^j - \hspace{-0.2cm}\sum_{\scriptsize \begin{array}{c} e_i \in {\cal E}:\\ tail(e_j)=head(e_i) \end{array}}\hspace{-1cm}a'_{i,j}J^i= \sum_{m_i \in {\mu}_{tail(e_j)}}\hspace{-0.5cm}c'_{i,j}{\cal Y}^{f(m_i)}_{f(\mu)}.
\]
We shall now show that $J^j=HL^j,~\forall~j=1,2,...,|{\cal E}|.$ Clearly for any edge $e_j$ such that $In(e_j)\subset \mu,$ (such edges exist because of acyclicity of $\cal G$), we have $J^j=HL^j,$ as $L^j$ is the basis vector which picks the $j^{th}$ column of $H,$ which is equal to $J^j.$ We now use induction on $j$ (according to the topological order) to show that $J^j=HL^j,~\forall~j=1,2,...,|{\cal E}|$. Now assume that for some $e_j,$ all $e_i\in In(e_j)$ are such that $J^i=HL^i.$ By (\ref{eqn7}), we have
\begin{align*}
J^j &= \hspace{-0.2cm}\sum_{\scriptsize \begin{array}{c} e_i \in {\cal E}:\\ tail(e_j)=head(e_i) \end{array}}\hspace{-1cm}a'_{i,j}J^i \hspace{-0.1cm} + \sum_{m_i \in {\mu}_{tail(e_j)}}\hspace{-0.5cm}c'_{i,j}{\cal Y}^{f(m_i)}_{f(\mu)}\\
&= \hspace{-0.15cm}\sum_{\scriptsize \begin{array}{c} e_i \in {\cal E}:\\ tail(e_j)=head(e_i) \end{array}}\hspace{-1cm}a'_{i,j}HL^i + H^j \\
& = H\left(\hspace{-0.15cm}\sum_{\scriptsize \begin{array}{c} e_i \in {\cal E}:\\ tail(e_j)=head(e_i) \end{array}}\hspace{-1cm}a'_{i,j}L^i+\boldsymbol{1_j}\right)\\
&= HL^j,
\end{align*}
where the second equality above follows from the induction assumption and the definition of $H^j,$ $\boldsymbol{1_j}$ is a column vector of length $|{\cal E}|$ with all zeros except for the $1$ at $j^{th}$ position, and the last equality follows from (\ref{eqn7}). Thus we have $J^j=HL^j.$ Continuing the induction on $j,$ we have that $J^j=HL^j,~\forall~j=1,2,..,|{\cal E}|.$ Therefore, we have 
$
{\cal X}=\left( 
\begin{array}{c}
HL \\
L
\end{array}
\right).
$
Thus, with ${\cal A}=H,$ and $\boldsymbol{F}=L,$ we have that $\cal X$ is of the form as in (\ref{formofX}). This proves the claim. 

We finally show that there is a scalar linear $\beta$-network-error detecting code for ${\cal G}.$  Let the matrices $A_{s_i},i=1,2,...,|{\cal S}|$ be obtained according to (\ref{formofA}) with $H={\cal A},$ and let the network coding matrix $K = I-L^{-1}.$ Then, the columns of the matrix $HL$ denote the global encoding vectors of the edges of $\cal E$ in the given topological order. Clearly this is a valid network code for $\cal G,$ by the structure of the matrices $H$ and $L.$ 

For some arbitrary error pattern, ${\cal F}\in\mathfrak{F},$ ${\cal M}_{\cal F}$ (as in Condition (C)) is clearly the vector matroid of the matrix 
\[
{\cal Z} = {\cal Y}_{f(\mu)\cup (n+I({\cal F}))} = \left(I_{n+\beta}~~ {\cal X}_{f(\mu)\cup (n+I({\cal F}))}\right),
\]
where $I({\cal F})=\left\{i_1,i_2,...,i_{\beta}\right\}$ is the index set corresponding to $\cal F,$ and
$
{\cal X}_{f(\mu)\cup (n+I({\cal F}))} = \left( 
\begin{array}{c}
HL \\
L_{I({\cal F})}
\end{array}
\right).
$
Now for a sink $t\in{\cal T},$ 
\[
{\cal Z}^{f(In_{\cal E}(t))}={\cal X}_{f(\mu)\cup (n+I({\cal F}))}^{f(In_{\cal E}(t))}=
\left( 
\begin{array}{c}
HL^{f(In_{\cal E}(t))} \\
L_{I({\cal F})}^{f(In_{\cal E}(t))}
\end{array}
\right).
\]

By Condition (C), we have $cols({\cal Z}^{f(\boldsymbol{{\cal D}_{t}})}) \subseteq \left\langle({\cal Z}^{f(In_{\cal E}(t))})\right\rangle.$ But we have by the notations of Section \ref{sec2}, for $supp(\boldsymbol{z})={\cal F}$
\begin{align*}
&{\cal Z}^{f(\boldsymbol{{\cal D}_{t}})} = I_{\boldsymbol{{\cal D}_{t}}}\\
&HL^{f(In_{\cal E}(t))} = \boldsymbol{F_{{\cal S},t}}\\ 
&L^{f(In_{\cal E}(t))}_{I({\cal F})}=\boldsymbol{F_{supp(\boldsymbol{z}),t}}.
\end{align*}
Thus, $cols(I_{\boldsymbol{{\cal D}_{t}}})\subseteq \left\langle\left(
\begin{array}{c}
\boldsymbol{F_{{\cal S},t}} \\
\boldsymbol{F_{supp(\boldsymbol{z}),{t}}}
\end{array}
\right)\right\rangle.$ As the choice of sink and error pattern was arbitrary, using Lemma \ref{lemmadecoding} it is seen that the network code given by the column vectors of $HL$ is $\beta$-network-error detecting. This completes the proof of the theorem.
\end{IEEEproof}

Theorem \ref{matroidalerrornetworkthm} has the following corollary which is easy to prove.
\begin{corollary}
\label{matroidalerrorcorrectingnetworkthm}
Let ${\cal G}({\cal V},{\cal E})$ be an acyclic communication network with sources $\cal S$ and sinks $\cal T.$ The network ${\cal G}$ is a \textit{matroidal} $\alpha$-\textit{error correcting network} associated with a  $\mathbb{F}$-representable matroid if and only if it has a scalar linear network-error correcting code over $\mathbb{F}$ that can correct network-errors at any $\alpha$ edges in the network.
\end{corollary}
\section{Constructions of Multisource Multicast and Multiple-Unicast Error Correcting networks}
\label{sec5}
In the theory of matroidal networks developed in \cite{DFZ,KiM}, we could start with any matroid and obtain a network which is matroidal with respect to that matroid. In particular, if we start with a representable matroid, we always obtain a network which has a scalar linear network code. On the other hand, to obtain matroidal error detecting (correcting) networks, the matroid has to satisfy the conditions of Definition \ref{matroidalerrornetworkdefinition}, in particular Condition (C) which puts restrictions on the choice of the matroid according to the nature of its contractions. If we are looking for networks with scalar linear network-error correcting codes, such matroids should be representable. Thus, unlike \cite{DFZ,KiM}, it is not straightforward how to obtain or construct such matroids (representable or otherwise). In this section, we propose algorithms for constructing such matroids (not necessarily representable) along with their corresponding networks (in particular multisource multicast and multiple-unicast), such that these networks are matroidal error correcting networks associated with the constructed matroids. The matroidal $\alpha$-error correcting networks constructed by our algorithms naturally are also matroidal $2\alpha$-error detecting networks. The construction of matroidal $\beta$-error detecting networks (for general $\beta$) can be done in a similar fashion, and therefore we omit it. 

Each such matroidal error correcting network is obtained by constructing a series of networks and a corresponding series of matroids associated with which the networks are matroidal error correcting. The series of networks are constructed using two types of nodes defined as follows.
\begin{itemize}
\item \textit Nodes which have a single incoming edge from a coding node and multiple outgoing edges to other coding nodes or sinks are known as \textit{forwarding nodes}. We denote the set of all forwarding nodes as ${\cal V}_{fwd}.$ 
\item Nodes which combine information from several incoming edges from the forwarding nodes and transmit the coded information to their corresponding forwarding nodes are known as \textit{coding nodes}.
\end{itemize}
If the series of matroids constructed are representable matroids, then the networks constructed are obtained along with scalar linear network-error correcting codes that satisfy the sink demands successfully.

Let $In({\cal V}_{fwd})$ be the set of all incoming edges of all forwarding nodes ${\cal V}_{fwd}.$ In a network with the property that coding and forwarding nodes alternate in any path from a source to a sink in the network, it is sufficient to consider error patterns that are subsets of $In({\cal V}_{fwd})$ to define the error correcting capability of the network, rather than subsets of all the edges in the network. If errors corresponding to such error patterns are correctable, then in such networks other errors are also correctable, as symbols flowing through edges other than $In({\cal V}_{fwd})$ are only copies of symbols flowing through $In({\cal V}_{fwd}).$ The networks that we design using our algorithms are restricted to have these properties, and therefore it is sufficient to construct a matroid $\cal M$ with ${\cal E}=In({\cal V}_{fwd})$ that satisfies the conditions in Definition \ref{matroidalerrornetworkdefinition}.

The goal of the construction algorithms is to generate a network defined by the following parameters that are to be given as inputs to the algorithms.
\begin{itemize}
\item \textit{Number of sources} ($|{\cal S}|$): The number of sources in the multisource multicast network or in the multiple-unicast network.
\item \textit{Number of information symbols} ($n=\sum_{s_k\in{\cal S}}n_{s_k}$): For multicast, $n_{s_k}$ is the number of information symbols generated by  $s_k,$ while $n$ is the total number of information symbols generated by all sources. For the multiple-unicast case, $n$ represents the number of non-collocated sources present in the network, each generating one information symbol.
\item \textit{Number of correctable network-errors} ($\alpha$): 
This fixes the number of outgoing edges from the source(s). For multicast, the number of outgoing edges from the source $s_k$ is fixed as $N_k = n_{s_k}+2\alpha.$ For multiple-unicast, the number of outgoing edges from each source is fixed as $1+2\alpha.$ These edges and their head nodes are for the sake of clearly presenting our algorithm, and can be absorbed back into the corresponding sources once the algorithm is completed. 
\item \textit{Number of network-coding nodes} ($N_C$): At each iteration in our algorithm, one network-coding node and one forwarding node will be added to the network, and a corresponding matroid constructed associated with which the extended network will be a matroidal error correcting network. The algorithm will run until $N_C$ forwarding nodes have been added. 
\item \textit{Number of multicast sinks} ($|{\cal T}|$): This value indicates the number of sinks to which the information symbols is to be multicast. For the multiple-unicast case, we assume  that the number of sinks is equal to the number of sources (i.e. messages).
\end{itemize}

\subsection{Sketch of Construction and Illustrative Examples}
\label{sketchandillustrexamples}
\begin{sidewaysfigure*}
\centering
\includegraphics[totalheight=6in]{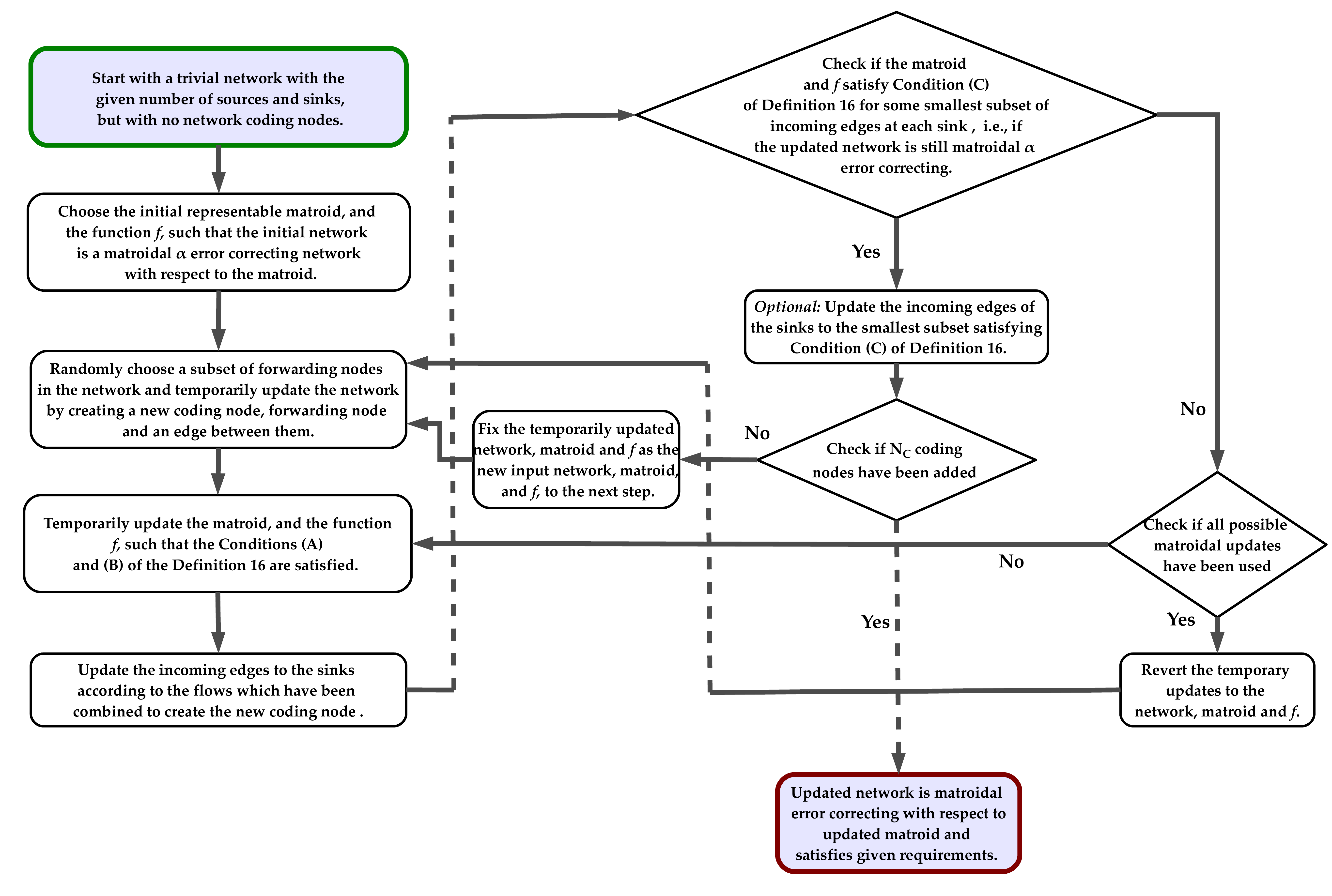}
\caption{Flowchart of the construction of matroidal error correcting networks. Some subpaths are shown dashed as they criss-cross with others.}	
\label{fig:flowchart}	
\end{sidewaysfigure*}

Fig. \ref{fig:flowchart} presents a sketch of our algorithm for constructing acyclic matroidal $\alpha$-error correcting multisource multicast and multiple-unicast networks. The full description of the algorithm for multisource multicast is given in Section \ref{subsec5b} and for multiple-unicast in Section \ref{subsec5c}. We now present a couple of illustrative examples before we give the full description of our algorithm.

\begin{figure*}
\centering
\subfigure[Network with $2$ sources, $3$ information symbols (of which $S_1$ generates two, and $S_2$ generates one), $2$ sinks and $\alpha=1,$ at initial stage of multicast construction]{
\includegraphics[width=2.6in]{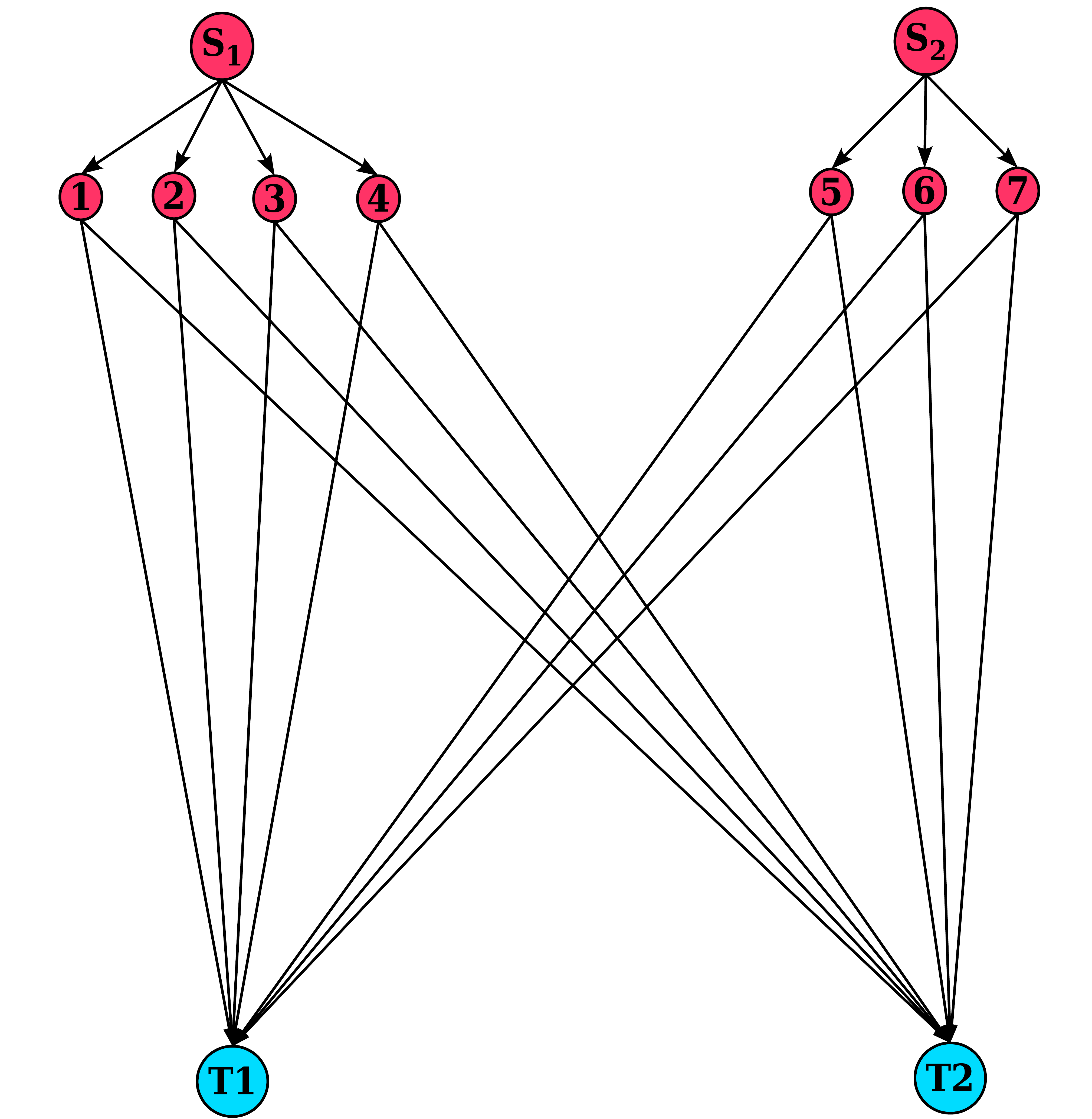}
\label{fig:multicastconstructionexample0}	
}
\subfigure[Multicast network after first iteration]{
\includegraphics[width=2.6in]{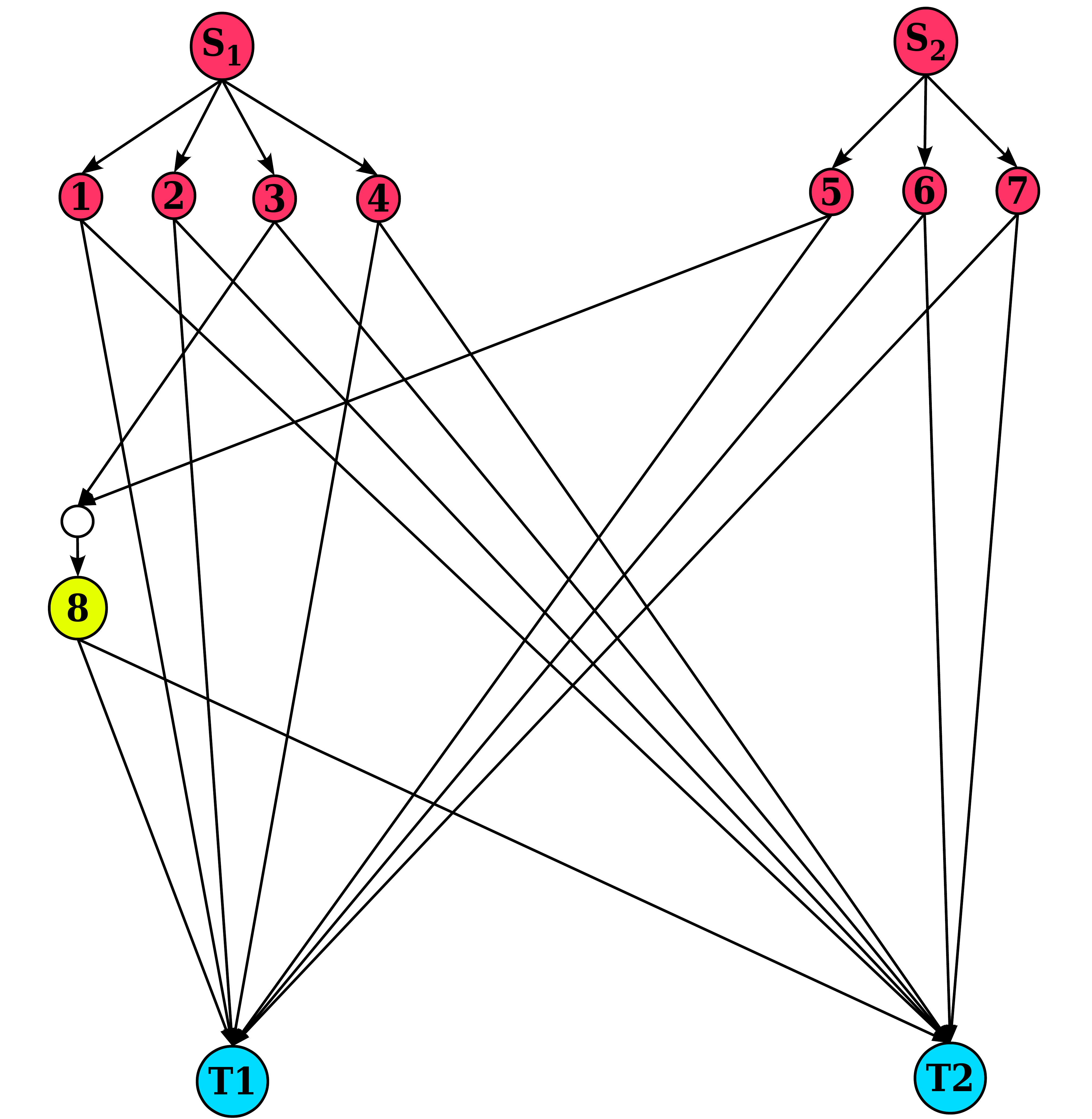}
\label{fig:multicastconstructionexample1}	
}
\subfigure[Multicast network after second iteration]{
\includegraphics[width=2.6in]{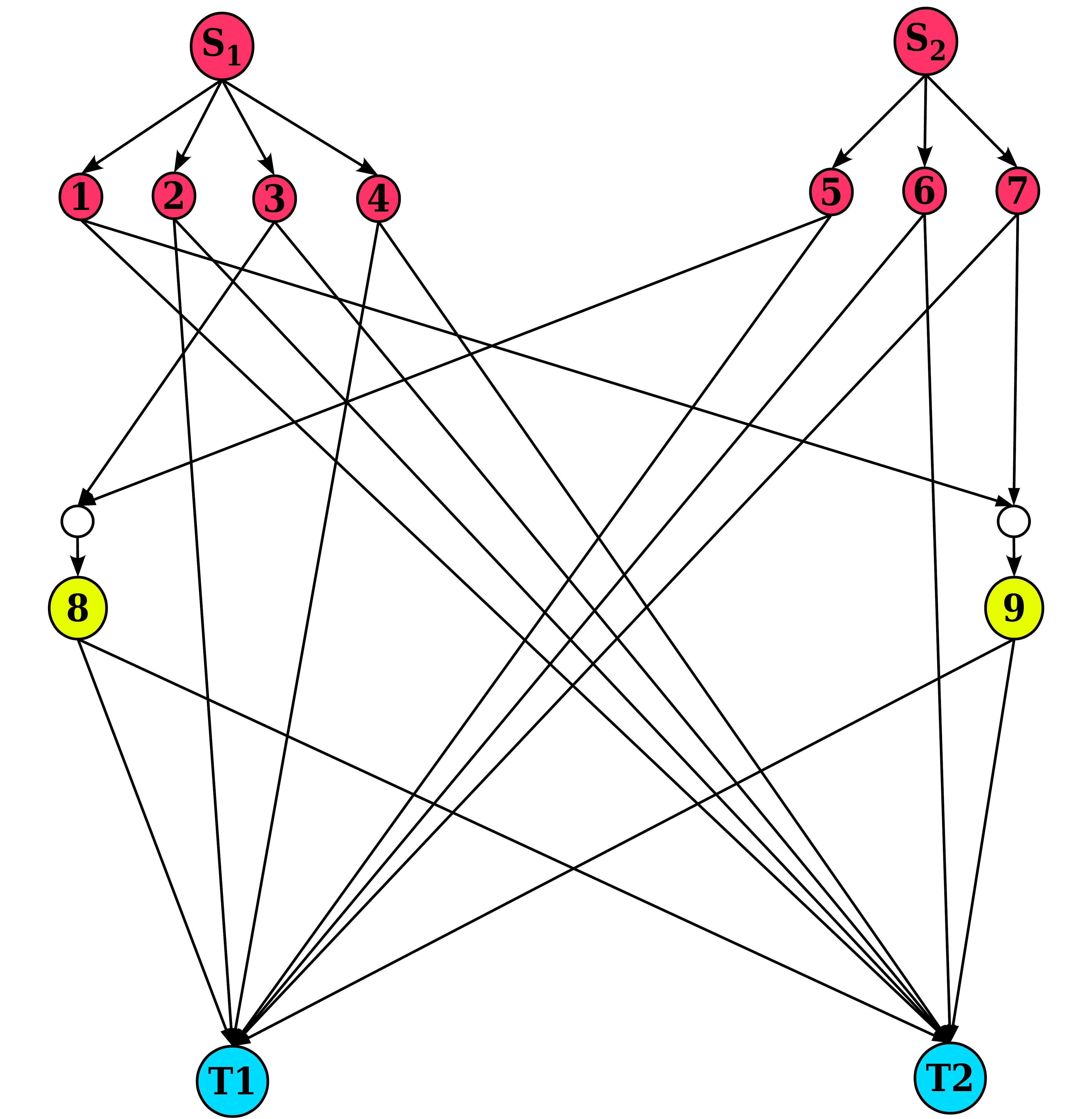}
\label{fig:multicastconstructionexample2}	
}
\subfigure[Multicast network after third iteration. Notice that the number of incoming edges to sink $T_2$ drops from $7$ to $5$. The reason for this is explained in the full description of our algorithm in Subsection \ref{subsec5b}.]{
\includegraphics[width=2.6in]{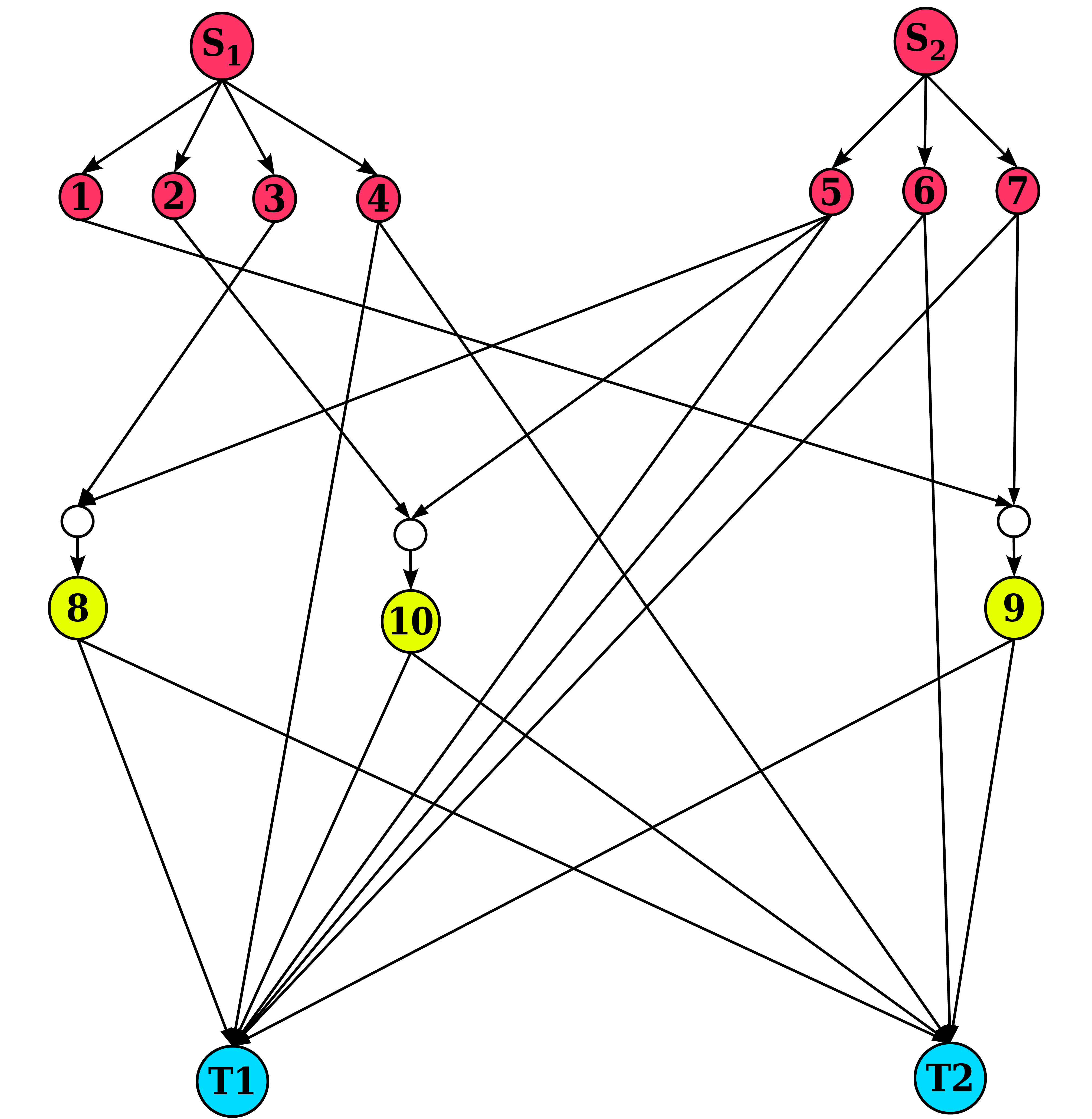}
\label{fig:multicastconstructionexample3}	
}	
\subfigure[The final multicast network with $3$ information symbols and $3$ sinks with single edge network-error correction]{
\includegraphics[width=2.6in]{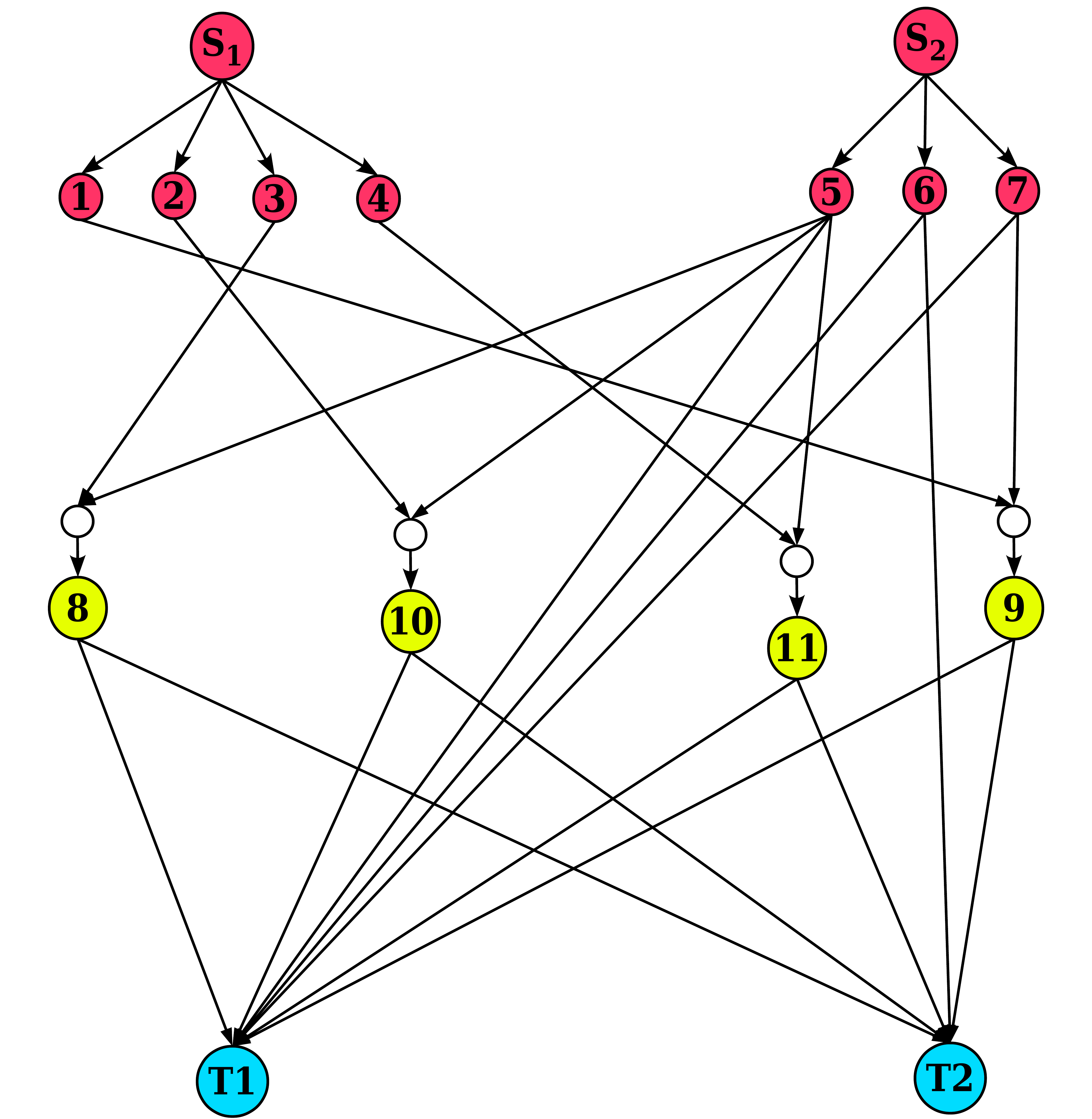}
\label{fig:multicastconstructionexample4}	
}
\caption{The stages of network evolution in the construction of a multicast network with a $1$-error correcting network code. Fig. \ref{fig:multicasteqns} shows the representations of the matroids associated with these networks.}
\label{fig:multicaststagesnws}
\end{figure*}

\begin{example}
\label{multicastex}
Fig. \ref{fig:multicastconstructionexample0}-\ref{fig:multicastconstructionexample4} describe the stages of a two source multicast network with input parameters $n_{s_1}=2, n_{s_2}=1, \alpha = 1, |{\cal T}|=2,$ and $N_C=4,$ as it evolves through the iterations in the construction shown in the sketch. The network shown in Fig. \ref{fig:multicastconstructionexample0} is the initial naive network. A representation of the initial matroid corresponding to this naive network is shown in (\ref{eqnstage0multicast}) in Fig. \ref{fig:multicasteqns} and is obtained from two MDS codes over $\mathbb{F}_8$, one of length $n_{s_1}+2\alpha=4$ implemented at source $s_1$ and another at source $s_2$ with length $n_{s_2}+2\alpha=3.$ Both codes have minimum distance $3.$ Each successive iteration in the construction adds a new coding node to the network, and a new column and row to the matrix representing the matroid. The equations (\ref{eqnstage1multicast})-(\ref{eqnstage4multicast}) shown in Fig. \ref{fig:multicasteqns} indicate the matrices representative of the representable matroids which correspond to the networks shown in Fig. \ref{fig:multicastconstructionexample1}-\ref{fig:multicastconstructionexample4}, respectively.

Let $e_i$ be the incoming edge at forwarding node $i.$ The function $f$ for each corresponding pair of network and matroid is defined as follows. 
\begin{align*}
&f(\mu) = \left\{1,2,3\right\}.\\
&f(e_i) = 3+|In({\cal V}_{fwd})|+i,~\forall~e_i \in In({\cal V}_{fwd}).
\end{align*}
For reasons mentioned in the beginning of this section, it is sufficient to define $f$ for the input indices $\mu$ and the set of edges $In({\cal V}_{fwd}).$ Each network is seen to be matroidal $1$-error correcting with respect to the corresponding matroid along with the function $f.$
\end{example}

\begin{figure*}
\centering
\subfigure[Unicast Network with $3$ information symbols and $\alpha = 1$ at initial stage of multiple-unicast construction]{
\includegraphics[width=2.8in,totalheight=2.8in]{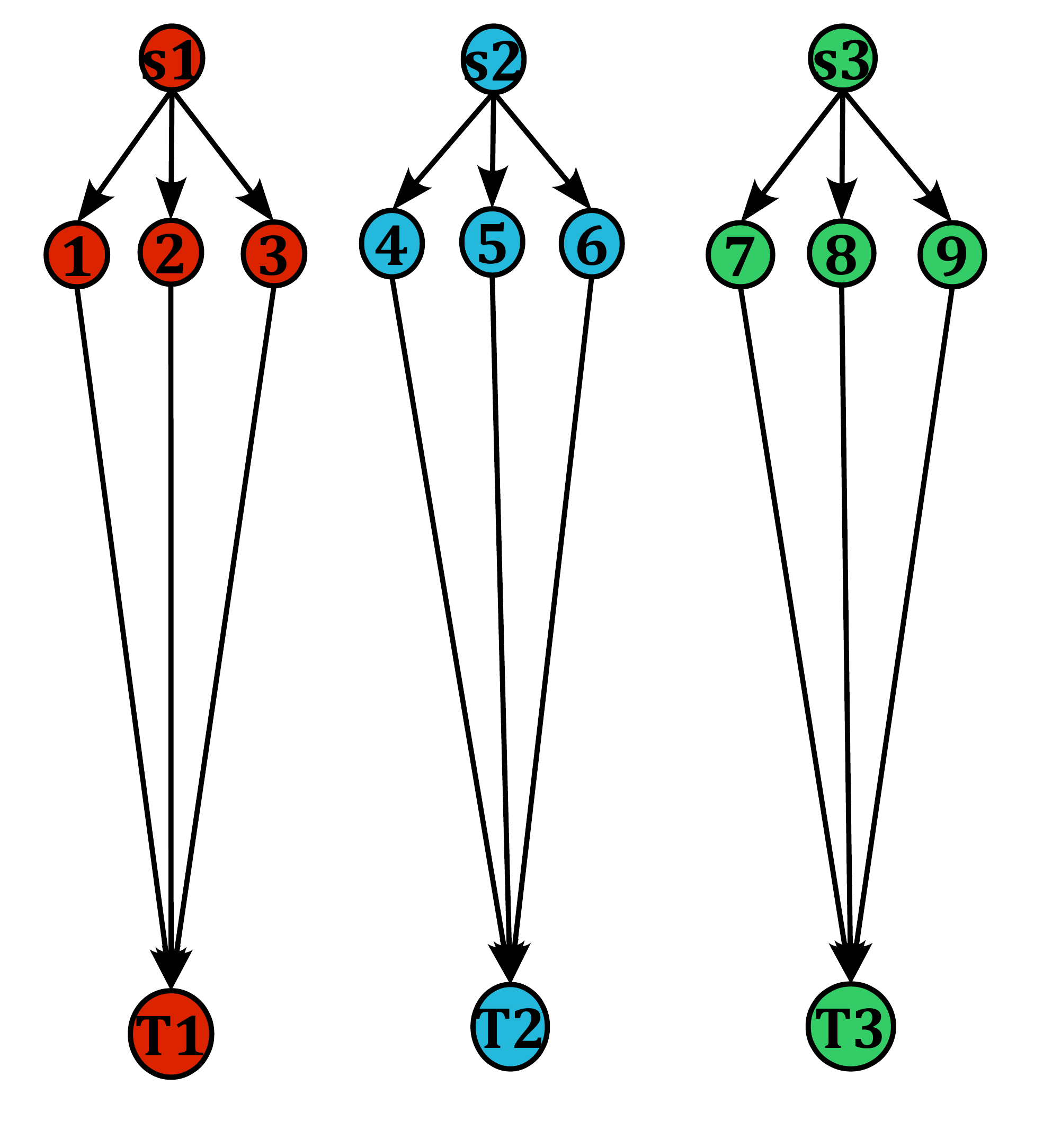}
\label{fig:unicastconstructionexample0}	
}
\subfigure[Multiple-unicast network after first iteration]{
\includegraphics[width=2.8in,totalheight=2.8in]{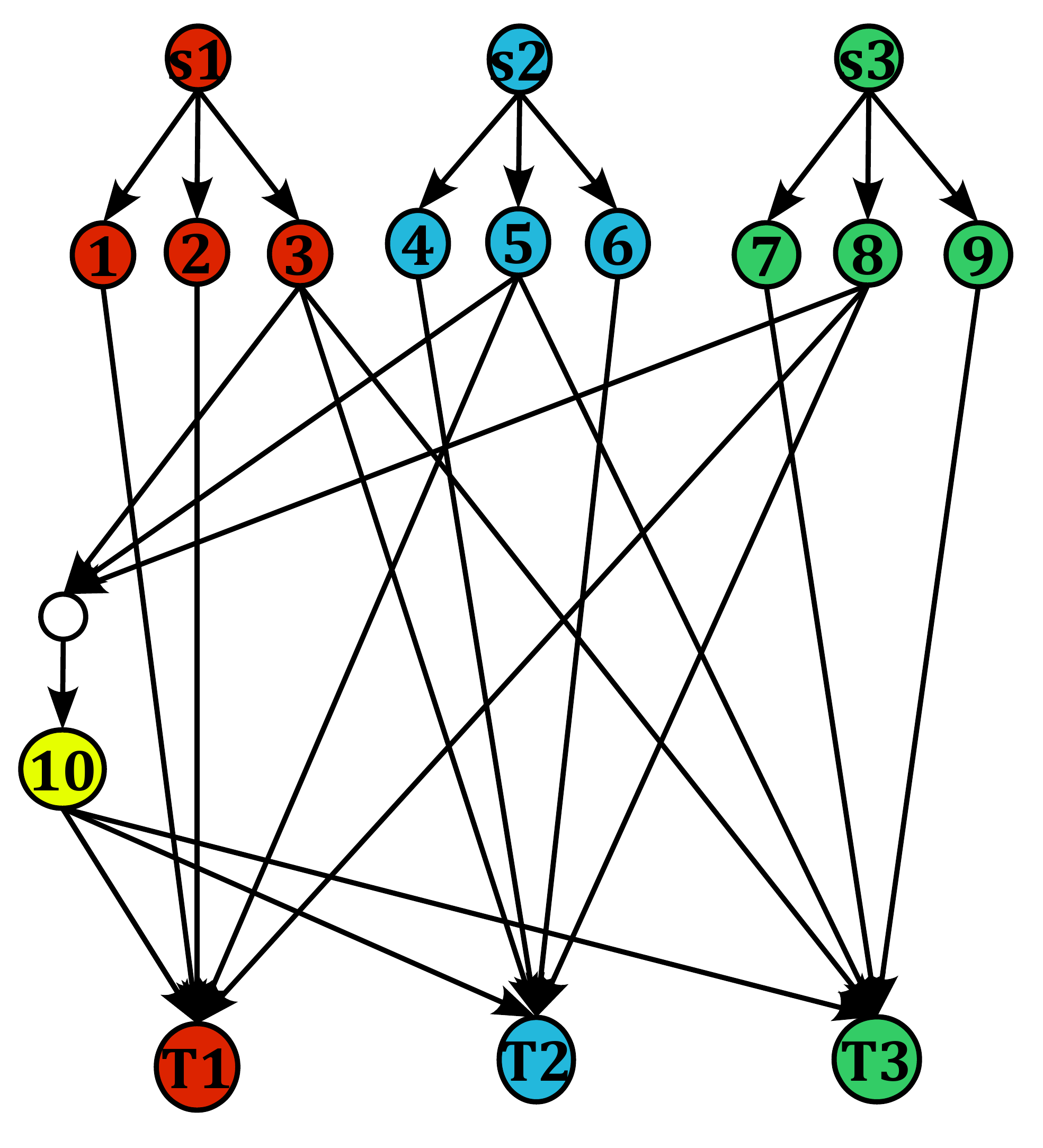}
\label{fig:unicastconstructionexample1}	
}
\subfigure[Multiple-unicast network after second iteration]{
\includegraphics[width=2.8in,totalheight=2.8in]{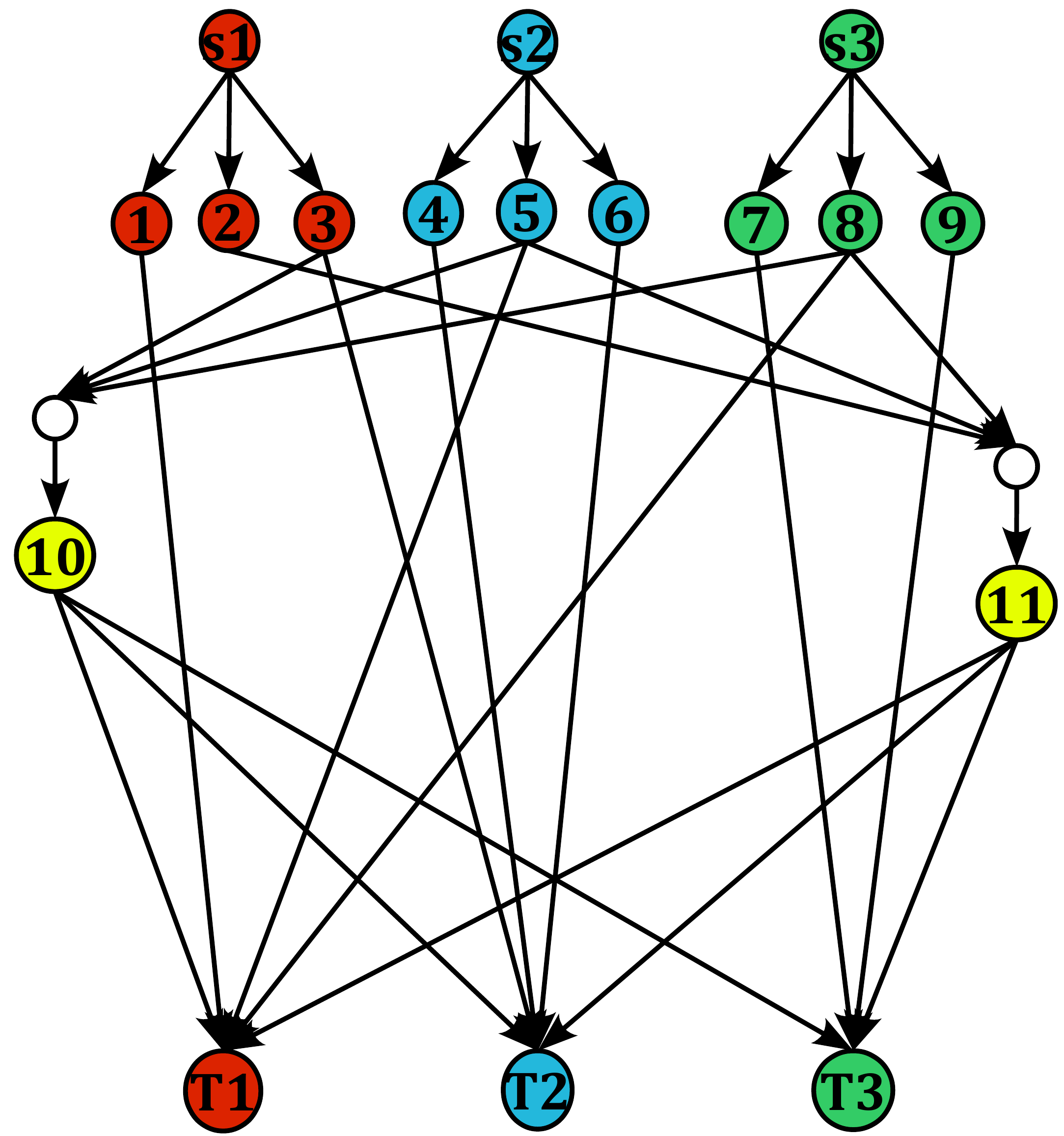}
\label{fig:unicastconstructionexample2}	
}
\subfigure[Multiple-unicast network after third iteration]{
\includegraphics[width=2.8in,totalheight=2.8in]{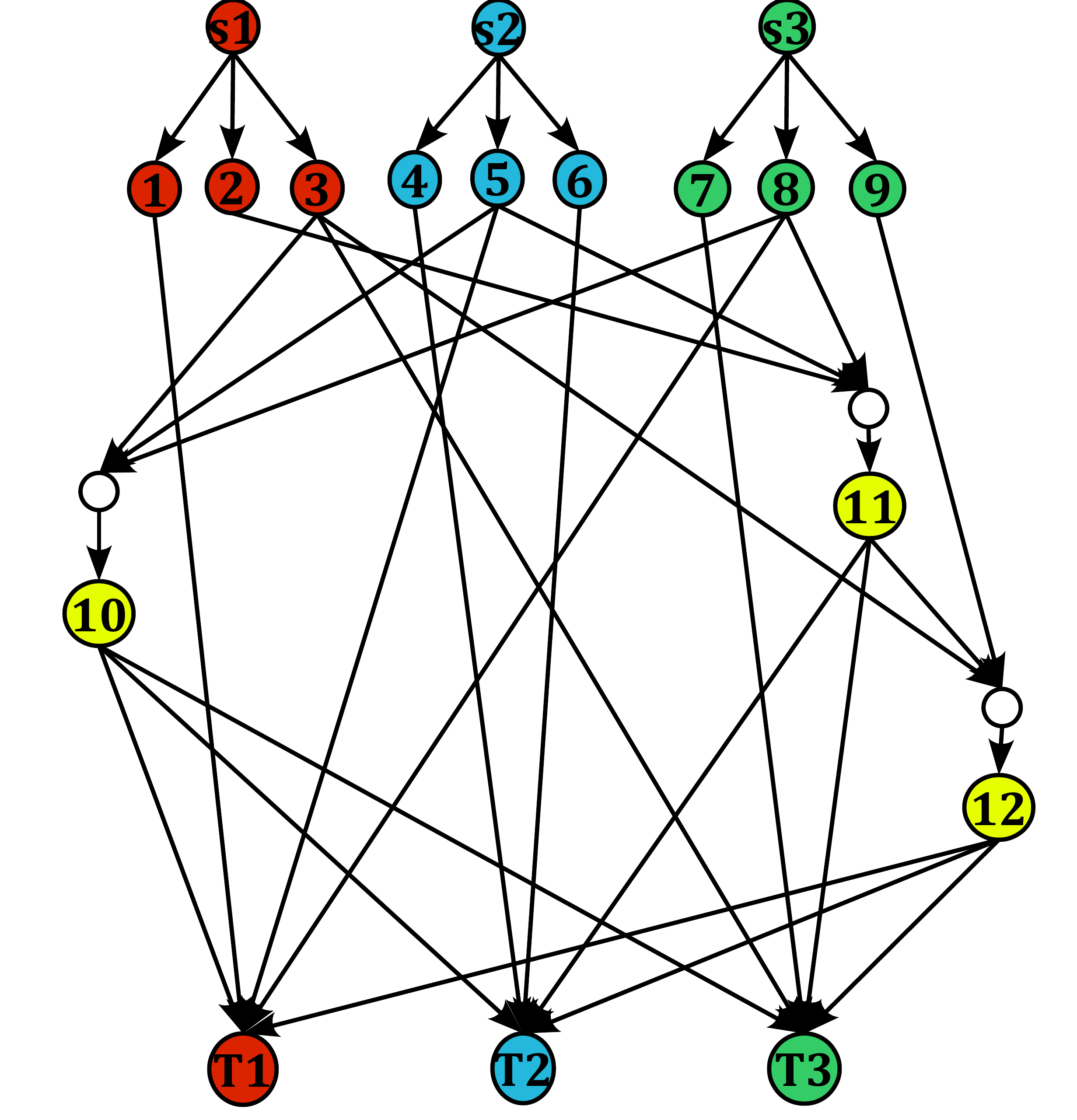}
\label{fig:unicastconstructionexample3}	
}	
\caption{The stages of network evolution in the construction of a multiple-unicast network with a $1$-error correcting network code. The representations of the matroids associated with these networks are shown in Fig. \ref{fig:unicasteqns}.}
~\hrule
\label{fig:unicastnetworkstages}
\end{figure*}

\begin{example}
Fig. \ref{fig:unicastconstructionexample0}-\ref{fig:unicastconstructionexample3} show the stages of the network evolution of a multiple-unicast network with parameters $n=3, \alpha = 1,$ and $N_C=3.$ For $i=1,2,3$, the $k^{th}$ sink demands the information symbol generated by the $k^{th}$ source. The representative matrices of the corresponding matroids are shown in (\ref{eqnstage0unicast})-(\ref{eqnstage3unicast}) in Fig. \ref{fig:unicasteqns}. The initial matroid represented by the matrix in (\ref{eqnstage0unicast}) is obtained from a repetition code of length $3$ and minimum distance $3.$ The function $f$ is defined in the same way as in the multicast example. Again, every network is matroidal $1$-error correcting with the corresponding matroid and function $f.$ 
\end{example}

The example networks shown in this paper which are obtained using our construction algorithms (executed in MATLAB) are matroidal error correcting networks with respect to a representable matroid, i.e., all the example networks have a scalar linear solution. The reason for presenting networks associated only with representable matroids is that obtaining matroidal error correcting networks associated with nonrepresentable matroids seems to be a computationally difficult problem. This is because our algorithms have to repeatedly compute various types of matroid extensions satisfying different kinds of properties. Computations and descriptions of the extensions of nonrepresentable matroids is a computationally intensive task. We further elaborate on the difficulty of obtaining networks associated with representable matroids in Subsection \ref{mecnnonrepresentable}. Using stronger mathematical machinery with respect to nonrepresentable matroids and their minors, the complexity of obtaining associated networks could be reduced and our algorithms can then be used to obtain examples of the same. In Subsection \ref{mecnnonrepresentable}, we present a result which can be considered as a first step towards obtaining matroidal error correcting networks which are associated with nonrepresentable matroids.
\begin{figure*}
~\\
~\\
~\\
~\\
~\\
~\\
~\\
~\\
\begin{minipage}[htbp]{.5\textwidth}
\begin{equation}
\label{eqnstage0multicast}
\left(
\begin{array}{cccccccc}
&1&1&1&1&0&0&0\\
&1&2&4&3&0&0&0\\
&0&0&0&0&1&1&1\\
&&&&&&&\\
I_{10}&1&0&0&0&0&0&0\\
&0&1&0&0&0&0&0\\
&0&0&1&0&0&0&0\\
&0&0&0&1&0&0&0\\
&0&0&0&0&1&0&0\\
&0&0&0&0&0&1&0\\
&0&0&0&0&0&0&1\\
\end{array}
\right)
\end{equation}
\end{minipage}
\begin{minipage}[htbp]{.5\textwidth}
\begin{equation}
\label{eqnstage1multicast}
\left(
\begin{array}{ccccccccc}
&1&1&1&1&0&0&0&1\\
&1&2&4&3&0&0&0&4\\
&0&0&0&0&1&1&1&6\\
&&&&&&&&\\
&1&0&0&0&0&0&0&0\\
I_{11}&0&1&0&0&0&0&0&0\\
&0&0&1&0&0&0&0&1\\
&0&0&0&1&0&0&0&0\\
&0&0&0&0&1&0&0&6\\
&0&0&0&0&0&1&0&0\\
&0&0&0&0&0&0&1&0\\
&0&0&0&0&0&0&0&1\\
\end{array}
\right)
\end{equation}
\end{minipage}
\begin{minipage}[htbp]{.5\textwidth}
\begin{equation}
\label{eqnstage2multicast}
\left(
\begin{array}{cccccccccc}
&1&1&1&1&0&0&0&1&1\\
&1&2&4&3&0&0&0&4&1\\
&0&0&0&0&1&1&1&6&5\\
&&&&&&&&&\\
&1&0&0&0&0&0&0&0&1\\
&0&1&0&0&0&0&0&0&0\\
I_{12}&0&0&1&0&0&0&0&1&0\\
&0&0&0&1&0&0&0&0&0\\
&0&0&0&0&1&0&0&6&0\\
&0&0&0&0&0&1&0&0&0\\
&0&0&0&0&0&0&1&0&5\\
&0&0&0&0&0&0&0&1&0\\
&0&0&0&0&0&0&0&0&1\\
\end{array}
\right)
\end{equation}
\end{minipage}
\begin{minipage}[htbp]{.5\textwidth}
\begin{equation}
\label{eqnstage3multicast}
\left(
\begin{array}{ccccccccccc}
&1&1&1&1&0&0&0&1&1&1\\
&1&2&4&3&0&0&0&4&1&2\\
&0&0&0&0&1&1&1&6&5&1\\
&&&&&&&&&&\\
&1&0&0&0&0&0&0&0&1&0\\
&0&1&0&0&0&0&0&0&0&1\\
I_{13}&0&0&1&0&0&0&0&1&0&0\\
&0&0&0&1&0&0&0&0&0&0\\
&0&0&0&0&1&0&0&6&0&1\\
&0&0&0&0&0&1&0&0&0&0\\
&0&0&0&0&0&0&1&0&5&0\\
&0&0&0&0&0&0&0&1&0&0\\
&0&0&0&0&0&0&0&0&1&0\\
&0&0&0&0&0&0&0&0&0&1\\
\end{array}
\right)
\end{equation}
\end{minipage}

\begin{minipage}[htbp]{\textwidth}
\begin{equation}
\label{eqnstage4multicast}
\left(
\begin{array}{cccccccccccc}
&1&1&1&1&0&0&0&1&1&1&1\\
&1&2&4&3&0&0&0&4&1&2&3\\
&0&0&0&0&1&1&1&6&5&1&1\\
&&&&&&&&&&&\\
&1&0&0&0&0&0&0&0&1&0&0\\
&0&1&0&0&0&0&0&0&0&1&0\\
I_{14}&0&0&1&0&0&0&0&1&0&0&0\\
&0&0&0&1&0&0&0&0&0&0&1\\
&0&0&0&0&1&0&0&6&0&1&1\\
&0&0&0&0&0&1&0&0&0&0&0\\
&0&0&0&0&0&0&1&0&5&0&0\\
&0&0&0&0&0&0&0&1&0&0&0\\
&0&0&0&0&0&0&0&0&1&0&0\\
&0&0&0&0&0&0&0&0&0&1&0\\
&0&0&0&0&0&0&0&0&0&0&1\\
\end{array}
\right)
\end{equation}
\end{minipage}
~\\
~\\
~\\
~\\
\caption{The stages of evolution in the representable matroid in the construction of a $2$-source multicast network (shown in Fig. \ref{fig:multicaststagesnws}) with a $1$-error correcting network code. All matrices are over $\mathbb{F}_8$ (with modulo polynomial $x^3+x+1$) and the entries are the decimal equivalents of the polynomial representations of elements from $\mathbb{F}_8.$ 
}
~\\
~\\
~\\
~\\
\label{fig:multicasteqns}	
\end{figure*}
\begin{figure*}

\begin{minipage}[htbp]{.5\textwidth}
\begin{equation}
\label{eqnstage0unicast}
\left(
\begin{array}{cccccccccc}
&1&1&1&0&0&0&0&0&0\\
&0&0&0&1&1&1&0&0&0\\
&0&0&0&0&0&0&1&1&1\\
&&&&&&&&&\\
&1&0&0&0&0&0&0&0&0\\
&0&1&0&0&0&0&0&0&0\\
&0&0&1&0&0&0&0&0&0\\
I_{12} &0&0&0&1&0&0&0&0&0\\
&0&0&0&0&1&0&0&0&0\\
&0&0&0&0&0&1&0&0&0\\
&0&0&0&0&0&0&1&0&0\\
&0&0&0&0&0&0&0&1&0\\
&0&0&0&0&0&0&0&0&1\\
\end{array}
\right)
\end{equation}
\end{minipage}
\begin{minipage}[htbp]{.5\textwidth}
\begin{equation}
\label{eqnstage1unicast}
\left(
\begin{array}{ccccccccccc}
&1&1&1&0&0&0&0&0&0&1\\
&0&0&0&1&1&1&0&0&0&4\\
&0&0&0&0&0&0&1&1&1&4\\
&&&&&&&&&&\\
&1&0&0&0&0&0&0&0&0&0\\
&0&1&0&0&0&0&0&0&0&0\\
&0&0&1&0&0&0&0&0&0&1\\
I_{13} &0&0&0&1&0&0&0&0&0&0\\
&0&0&0&0&1&0&0&0&0&4\\
&0&0&0&0&0&1&0&0&0&0\\
&0&0&0&0&0&0&1&0&0&0\\
&0&0&0&0&0&0&0&1&0&4\\
&0&0&0&0&0&0&0&0&1&0\\
&0&0&0&0&0&0&0&0&0&1\\
\end{array}
\right)
\end{equation}
\end{minipage}
\begin{minipage}[htbp]{.5\textwidth}
\begin{equation}
\label{eqnstage2unicast}
\left(
\begin{array}{cccccccccccc}
&1&1&1&0&0&0&0&0&0&1&1\\
&0&0&0&1&1&1&0&0&0&4&4\\
&0&0&0&0&0&0&1&1&1&4&3\\
&&&&&&&&&&&\\
&1&0&0&0&0&0&0&0&0&0&0\\
&0&1&0&0&0&0&0&0&0&0&1\\
&0&0&1&0&0&0&0&0&0&1&0\\
I_{14} &0&0&0&1&0&0&0&0&0&0&0\\
&0&0&0&0&1&0&0&0&0&4&4\\
&0&0&0&0&0&1&0&0&0&0&0\\
&0&0&0&0&0&0&1&0&0&0&0\\
&0&0&0&0&0&0&0&1&0&4&3\\
&0&0&0&0&0&0&0&0&1&0&0\\
&0&0&0&0&0&0&0&0&0&1&0\\
&0&0&0&0&0&0&0&0&0&0&1\\
\end{array}
\right)
\end{equation}
\end{minipage}
\begin{minipage}[htbp]{.5\textwidth}
\begin{equation}
\label{eqnstage3unicast}
\left(
\begin{array}{ccccccccccccc}
&1&1&1&0&0&0&0&0&0&1&1&2\\
&0&0&0&1&1&1&0&0&0&4&4&7\\
&0&0&0&0&0&0&1&1&1&4&3&1\\
&&&&&&&&&&&&\\
&1&0&0&0&0&0&0&0&0&0&0&0\\
&0&1&0&0&0&0&0&0&0&0&1&3\\
&0&0&1&0&0&0&0&0&0&1&0&1\\
I_{15} &0&0&0&1&0&0&0&0&0&0&0&0\\
&0&0&0&0&1&0&0&0&0&4&4&7\\
&0&0&0&0&0&1&0&0&0&0&0&0\\
&0&0&0&0&0&0&1&0&0&0&0&0\\
&0&0&0&0&0&0&0&1&0&4&3&5\\
&0&0&0&0&0&0&0&0&1&0&0&4\\
&0&0&0&0&0&0&0&0&0&1&0&0\\
&0&0&0&0&0&0&0&0&0&0&1&3\\
&0&0&0&0&0&0&0&0&0&0&0&1\\
\end{array}
\right)
\end{equation}
\end{minipage}
\caption{The stages of evolution in the representable matroid in the construction of a multiple-unicast network (shown in Fig. \ref{fig:unicastnetworkstages}) with a $1$-error correcting network code. All matrices are over $\mathbb{F}_8$ (with modulo polynomial $x^3+x+1$) and the entries are the decimal equivalents of the polynomial representations of elements from $\mathbb{F}_8.$}
\hrule
\label{fig:unicasteqns}	
\end{figure*}
\subsection{Multisource Multicast Construction}
\label{subsec5b}
We now give the full description of our construction for the case of multisource multicast. The construction generates a multisource multicast network with the given parameters $|{\cal S}|, \left\{n_{s}:s\in {\cal S}\right\}, \alpha,N_C$, and $|{\cal T}|,$ along with a matroid (not necessarily representable) with respect to which the network is matroidal $\alpha$-error correcting.  For the sake of the completeness of the description of our construction algorithm, we present a simple lemma.
\begin{lemma}
\label{seriesextensionlemma2}
Let $\cal N$ be a series extension of the matroid ${\cal M} = {\cal N}/e_2$ at $e_1,$ i.e., ${\cal N}={\cal M}+_{e_1}^s e_2.$ Let $C$ be a circuit of $\cal M$ containing $e_1,$ then $C\cup e_2$ is a circuit of $\cal N.$
\end{lemma}
\begin{IEEEproof}
As $C \in {\cal C}({\cal M}),$ $E({\cal M})-C$ is a hyperplane of ${\cal M}^*$ not containing $e_1.$ To prove $C\cup e_2 \in {\cal C}({\cal N}),$ we prove that $E({\cal N})-C\cup e_2 = E({\cal M})-C$ is a hyperplane (obviously not containing $e_1$ or $e_2$) in ${\cal N}^*$ also.

Note that  ${\cal N}^*$ is a parallel extension of ${\cal M}^*.$ In a parallel extension ${\cal N}^*$ of ${\cal M}^*,$ the rank of any subset $X\subseteq E({\cal M}^*)$ does not change in the extension. Therefore $r_{{\cal N}^*}(E({\cal M})-C)=r_{{\cal M}^*}(E({\cal M})-C)=r_{{\cal M}^*}-1=r_{{\cal N}^*}-1.$ 

Now all that we have to prove is that $E({\cal M})-C$ is a flat in ${\cal N}^*$ also. Suppose not, then we must have that $cl_{{\cal N}^*}(E({\cal M})-C) = E({\cal N}^*).$ Thus, as $e_1 \notin (E({\cal M})-C),$ there should be a circuit $C'$ such that $C' \subseteq \left(E({\cal M})-C\right)\cup e_1,$ with $e_1\in C'.$ But then this means $C'\in {\cal C}({\cal M}^*)$ also, which implies that $e_1\in cl_{{\cal M}^*}(E({\cal M})-C) = E({\cal M})-C.$ But this is not the case. Hence  $E({\cal M})-C$ is a flat, and hence a hyperplane, in ${\cal N}^*.$ Therefore $C\cup e_2 = (E({\cal N})-(E({\cal M})-C)) \in {\cal C}({\cal N}).$ This proves the lemma.
\end{IEEEproof}

We now present our construction as an elaboration of the algorithm sketch shown in Fig. \ref{fig:flowchart}. The details of the functionality of the algorithm sketch, such as the method of updating the incoming edges to the sinks, the method of updating the matroid, field size issues which govern the possibility of adding new coding nodes and representability of matroidal extensions, etc., can be inferred through the description of our algorithm and the discussion that follows. The construction is based on matroids which need not always be representable. However, at all the appropriate junctures, the equivalent scenario for representable matroids is given as remarks. Throughout the remainder of this section we will assume that a matroid remains unchanged when its elements are reordered according to some permutation, as this implies only a relabeling of the matroid elements.
~\\
~\\
\textit{\textbf{Step 1: Initializing the network:}}~\\
The network ${\cal G}$ is initialized  by creating the collection of source nodes ${\cal S}$ and a collection of sink nodes ${\cal T}.$ 

Corresponding to each source $s_k \in {\cal S},$ create a set of $N_{s_k}=n_{s_k}+2\alpha$ forwarding nodes, each with one incoming edge from $s_k.$ Let the collection of these incoming edges be $e_1,...,e_{N},$  where $N=\sum_{s_k}N_{s_k}$ is the total number of forwarding nodes added.

\begin{figure}[htbp]
\centering
\includegraphics[totalheight=2.3in]{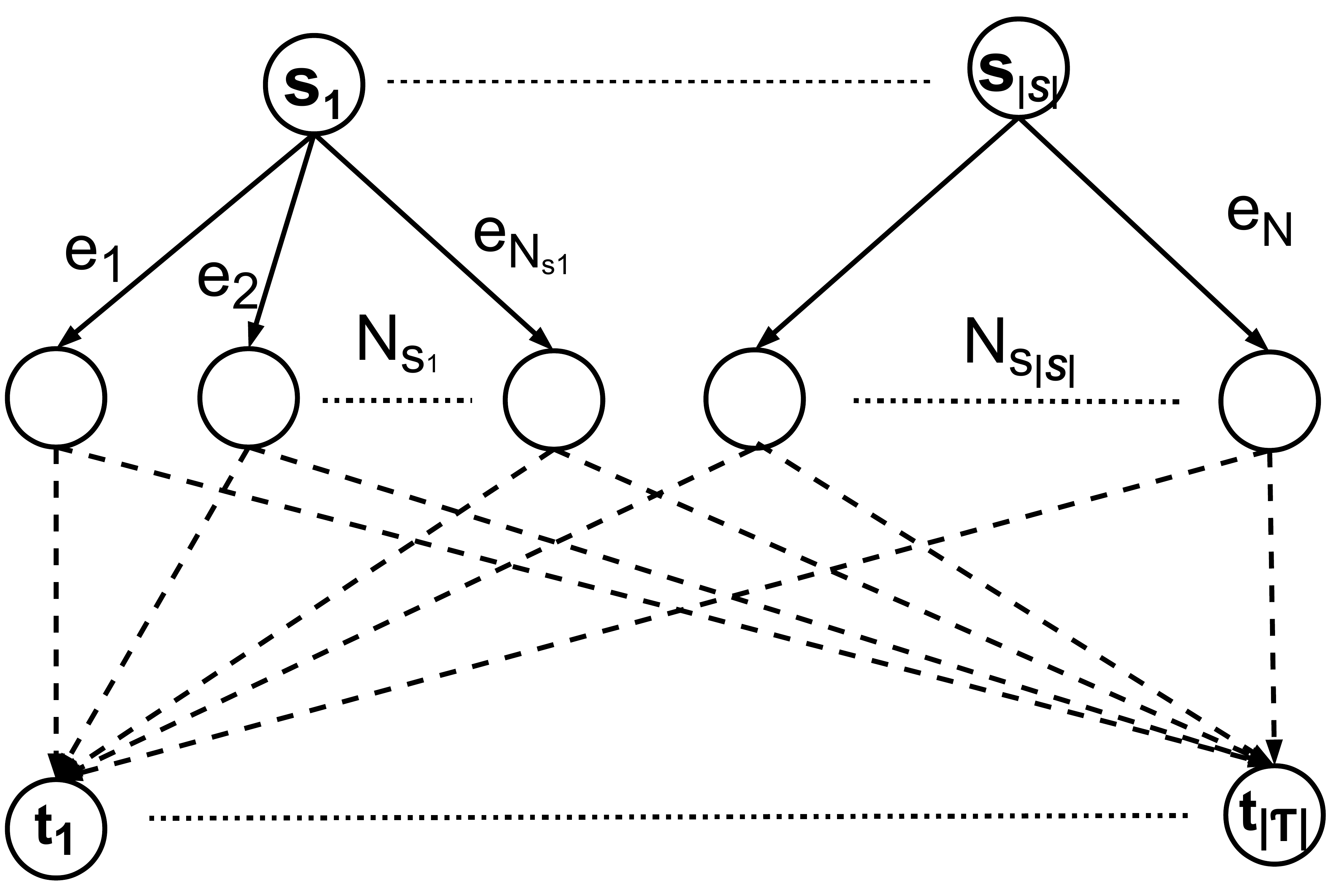}
\caption{The initialization of the multisource multicast network}	
\label{fig:multicastinit}	
\end{figure}

For each sink $t,$ create $N$ temporary incoming edges $In(t)$ originating from the $N$ forwarding nodes. Because it is sufficient to consider error patterns on the incoming edges at the forwarding nodes, we abuse our notation to say that $In(t)=\left\{e_1,...,e_N\right\}={\cal E},~\forall~t\in{\cal T}.$ This initialized network is represented in Fig. \ref{fig:multicastinit}.
~\\~\\
\textit{\textbf{Step 2: Initializing the matroid}}\\

We now obtain a matroid $\cal M$ such that the network $\cal G$ is a matroidal $\alpha$-error correcting network with respect to this matroid $\cal M.$ Towards that end, we consider the direct sum  
\[
{\cal U}=\boxplus_{k=1}^{|{\cal S}|}{\cal U}_{n_{s_k},N_{s_k}},
\]
where ${\cal U}_{n_{s_k},N_{s_k}}$ is the uniform matroid of rank $n_{s_k}$ with the groundset with $N_{s_k}$ elements given as follows. 
\[
E({\cal U}_{n_{s_k},N_{s_k}})=\left\{u_1^{k},u_2^{k},...,u_{N_{s_k}}^{k}\right\}.
\]
The matroid ${\cal U}$ has rank $n=\sum_{k=1}^{|{\cal S}|}n_{s_k}.$ Let the ground set of this matroid be 
\begin{equation}
\label{no18eqn}
E({\cal U})=\{u_1,u_2,...,u_N\}=\uplus_{k=1}^{|{\cal S}|}\{u_1^{k},u_2^{k},...,u_{N_{s_k}}^{k}\},
\end{equation}
where 
\[
\{u_1,u_2,...,u_n\}=\uplus_{k=1}^{|{\cal S}|}\{u_1^k,u_2^k,...,u_{n_{s_k}}^k\}
\]
is a basis for $\cal U.$ 
\begin{remark}
If an MDS code of length $N_{s_k}$ and with $n_{s_k}$ information symbols exists, then  ${\cal U}_{n_{s_k},N_{s_k}}$ corresponds to the vector matroid of a generator matrix of an $N_{s_k}$-length MDS code which has minimum distance $2\alpha+1$. If such an MDS code exists, let this generator matrix be the $n_{s_k}\times N_{s_k}$ matrix of the form $U_{s_k} = \left(I_{n_{s_k}}~~~A_{s_k}\right).$ If such MDS codes exist for each source, then a representation of the matroid ${\cal U}$ is given as 
\[
\left(\begin{matrix}
U_{s_1} &  \boldsymbol{0}  & \ldots & \boldsymbol{0}\\
\boldsymbol{0}  &  U_{s_2} & \ldots & \boldsymbol{0}\\
\vdots & \vdots & \ddots & \vdots\\
\boldsymbol{0}  &   \boldsymbol{0}      &\ldots & U_{s_{|{\cal S}|}}
\end{matrix}
\right).
\]
Rearranging the columns of the above representation, we have the alternative representation for ${\cal U}$ which we shall use in the description of our algorithm.
\begin{equation}
\label{MDSrep}
U=\left(I_n~~~A\right),
\end{equation}
where 
\[
A=\left(\begin{matrix}
A_{s_1} &  \boldsymbol{0}  & \ldots & \boldsymbol{0}\\
\boldsymbol{0}  &  A_{s_2} & \ldots & \boldsymbol{0}\\
\vdots & \vdots & \ddots & \vdots\\
\boldsymbol{0}  &   \boldsymbol{0}      &\ldots & A_{s_{|{\cal S}|}}
\end{matrix}
\right).
\]
\end{remark}

Corresponding to the elements $u_i,i=1,2,...,n,$ we add the elements $u_i^p,i=1,2,...,n$ respectively in parallel. By definition of a parallel extension, it can be seen that the order in which these elements are added does not matter. Let the resultant matroid be  ${\cal U}_p.$ The set
\begin{align*}
E({\cal U}_p)=\{u_1^p, u_2^p,...,u_n^p, u_1,u_2,...,u_N\}
\end{align*}
is the ground set of ${\cal U}_p$ such that $\{u_i^p,u_i\}, \forall i=1,2,..,n$ are circuits in ${\cal U}_p.$ By repeatedly using (\ref{eqn202}) for the succession of parallel extensions, it can be seen that the set $\{u_1^p, u_2^p,...,u_n^p\}$ forms a basis of ${\cal U}_p$. 
\begin{remark}
If ${\cal U}$ is representable, by Lemma \ref{parallelrepresentation} a representation of the matroid ${\cal U}_p$ is then the matrix $U' = \left(I_n~~~I_n~~~A\right).$  
\end{remark}

Corresponding to the elements $u_i,i=1,2,...,N,$ we now add the elements $u_i^s,i=1,2,...,N$ respectively in series. Again, the order in which these elements are added does not matter. Let ${\cal U}_{p,s}$ be the resultant matroid. We then have
\begin{align*}
E({\cal U}_{p,s})&=\{u_1^p, u_2^p,...,u_n^p,u^s_1,u^s_2,...,u^s_N,u_1,u_2,...,u_N\}\\
&=\uplus_{k=1}^{|{\cal S}|}\{u_1^{k},u_2^{k},...,u_{N_{s_k}}^{k}\}\cup\{u_1^p,...,u_n^p,u^s_1,...,u^s_N\}.
\end{align*}
By repeatedly using Lemma \ref{seriesextensionlemma2}, we see that all the circuits of  ${\cal U}_{p,s}$ containing $u_i$ will also contain $u_i^s$ for all $i=1,2,..,N.$ In particular, the set of circuits include $\{u_i^p,u_i,u^s_i\}, \forall  i=1,2,..,n.$ Moreover, by repeatedly using (\ref{eqn301}), we also see that the set $\{u_1^p, u_2^p,...,u_n^p,u^s_1,u^s_2,...,u^s_N\}$ forms a basis for ${\cal U}_{p,s}.$


Let $\cal M$ be the matroid ${\cal U}_{p,s}.$ Consider the initialized network ${\cal G}$ with edges ${\cal E}=\left\{e_1,e_2,...,e_{|{\cal E}|}\right\}$ and with $\cal E$ being the $N$ incoming edges (abusing the notation) at all sinks. For $k=0,1,...,|{\cal S}|-1,$ we define $R_k=\sum_{j=1}^{k}N_{s_j},$ where $R_0=0.$ Let 
\[
f:{\cal E}\cup\mu\rightarrow E({\cal M})
\]
be a function such that 
\begin{itemize}
\item $f(e_{R_k+j})=u_j^{k+1},~j=1,2,...,N_{s_k},~k=0,1,...,|{\cal S}|-1.$
\item $f(m_j)=u_j^p, m_j\in\mu,~j=1,2,..,n.$
\end{itemize}
Let 
\[
B=\left\{b_1,b_2,..,b_{n+{|{\cal E}|}}\right\}=\{u_1^p, u_2^p,...,u_n^p,u^s_1,u^s_2,...,u^s_N\},
\]
taken in the following one-one correspondence. 
\begin{align*}
b_i=&u_i^p,& i=&1,2,..,n \\
b_{n+R_k+j}=&u_i^s~(\text{where}~u_i=u_j^{k+1})& j=&1,2,...,N_{s_k},\\
&&k=&0,1,...,|{\cal S}|-1.
\end{align*}
Thus, the basis vector corresponding to the $i^{th}$ input ($1\leq i \leq n$) is $b_i=u_i^p$ and the basis vector corresponding to the error at the edge $e_{R_k+j}$ (for some $k$ and $j$ as above) is $b_{n+R_k+j}=u_i^s$ (for some $i$ such that $u_i=u_j^{k+1}$).
\begin{remark}
Suppose ${\cal U}$ is representable, by Lemma \ref{seriesrepresentation} a representation of the matroid ${\cal U}_{p,s}$ is
\begin{equation}
\label{eqn9}
U'' = 
\left(
\begin{array}{ccccc}
I_n & \boldsymbol{0} & \boldsymbol{0} & I_n & A \\
\boldsymbol{0} & I_n & \boldsymbol{0} & I_n & \boldsymbol{0}\\
\boldsymbol{0} & \boldsymbol{0} & I_{N-n} & \boldsymbol{0} &  I_{N-n}
\end{array}
\right),
\end{equation}
where $N=|{\cal E}|.$ Thus $U''$ is of the form $(I_{n+{|{\cal E}|}}~~~{\cal X}),$ where ${\cal X}$ is the appropriate $(n+{|{\cal E}|})\times {|{\cal E}|}$ matrix in (\ref{eqn9}). It is not difficult to see that with the assignment $f$ to $\mu\cup{\cal E},$ and basis $B,$ the network $\cal G$ is a matroidal $\alpha$-error correcting network in association with the representable matroid $\cal M,$ as $\left(I_n~~~A\right)$ corresponds to a matrix defined as in (\ref{MDSrep}), whose columns correspond to the columns of generator matrices of MDS codes implemented at each source. 
\end{remark}

However, we claim that even when $\cal U$ is not representable, the network $\cal G$ is still a matroidal $\alpha$-error correcting network in association with $\cal M,$ with this assignment $f$ to $\mu\cup{\cal E},$ and with basis $B.$ We now prove this claim by verifying the conditions of Definition \ref{matroidalerrornetworkdefinition} as follows.

\textbf{Condition (A):} Condition (A) is verified as 
\[
f(\mu)=\{u_i^p:~i=1,2,..,n\} \subseteq B
\]
and therefore is independent in ${\cal U}_{p,s}.$

\textbf{Condition (B1):} Suppose for some $e\in {\cal E},$ Condition (B1) is not satisfied, i.e., 
\[
f(e)=u_j^{k+1}=u_i \in cl_{{\cal U}_{p,s}}(B-f(\mu)).
\]
This means that there is a circuit $C_1 \subseteq (B-f(\mu))\cup\{u_i\}$ with $u_i \in C_1.$ Note that in ${{\cal U}_{p,s}},$ the set $C_2=\{u_i^p,u_i,u^s_i\}$ is also a circuit. Thus applying the circuit elimination axiom to the circuits $C_1$ and $C_2$ with $u_i\in C_1\cap C_2,$ we have that there is some circuit 
\[
C_3\subseteq (B-f(\mu))\cup\{u_i^p,u^s_i\} \subseteq B.
\]
However, $B$ is an independent set in ${\cal U}_{p,s}.$ Thus 
\[
f(e)=u_i \notin cl_{{\cal U}_{p,s}}(B-f(\mu)), \forall i=1,2,..,N.
\]
Hence Condition (B1) is satisfied.

\textbf{Condition (B2):} Consider $e_{R_k+j}\in {\cal E}$ such that $f(e_{R_k+j})=u_j^{k+1}=u_i$ (for some $i$). As $\{u_1^p, u_2^p,...,u_n^p\}$ is a basis in ${\cal U}_p,$ we must have some circuit $C_i \subseteq \{u_1^p, u_2^p,...,u_n^p,u_i\},$ with $u_i\in C_i,$ for each $u_i,i=1,2,..,N.$  Therefore, in ${\cal U}_{p,s},$ by Lemma \ref{seriesextensionlemma2}, $C'_i=C_i\cup\{u_i^s\}$ is a circuit. Thus 
\[
u_i \in cl_{{\cal U}_{p,s}}(\{u_1^p, u_2^p,...,u_n^p\}\cup\{u_i^s\}).
\]
In other words, $f(e_{R_k+j})\in cl_{{\cal U}_{p,s}}(f(\mu)\cup\{b_{n+R_k+j}\}).$ As $f(\mu)=f(In(e_{R_k+j})),$ 
\begin{align}
\label{eqnno23}
f(e_{R_k+j})\in cl_{{\cal U}_{p,s}}(f(In(e_{R_k+j}))\cup\{b_{n+R_k+j}\}).
\end{align}
Moreover, 
\begin{align}
\label{eqnno24}
f(e_{R_k+j})=u_i \notin cl_{{\cal U}_{p,s}}(f(\mu))=cl_{{\cal U}_{p,s}}(f(In(e_{R_k+j}))),
\end{align}
where $u_i \notin cl_{{\cal U}_{p,s}}(f(\mu))$ follows from the fact that any circuit containing $u_i$ in ${\cal U}_{p,s}$ must also contain $u_i^s,$ by Lemma \ref{seriesextensionlemma2}. Thus, by (\ref{eqnno23}) and (\ref{eqnno24}), Condition (B2) is satisfied.

\textbf{Condition (C):} Let ${\cal F}=\{e_{R_{k_1}+j_1},...,e_{R_{k_{2\alpha}}+j_{2\alpha}}\}\in\mathfrak{F}$ be an arbitrary error pattern with
\[
B_{\overline{\cal F}}=B-f(\mu)-\{u_{i_1}^s,...,u_{i_{2\alpha}}^s\},
\] 
where $\{u_{i_1}^s,...,u_{i_{2\alpha}}^s\}$ corresponds to the basis vectors of the errors at ${\cal F}.$ The contraction ${\cal M}/B_{\overline{\cal F}}$ then has the ground set 
\[
E({\cal M}/B_{\overline{\cal F}})=\{u_1^p,...,u^p_n,u_1,u_2,..,u_N,u_{i_1}^s,...,u_{i_{2\alpha}}^s\}.
\]
By repeatedly using (\ref{eqn303}), we see that this matroid is precisely the matroid obtained from ${\cal U}_p$ by adding the elements $\{u_{i_1}^s,...,u_{i_{2\alpha}}^s\}$ in series with $\{u_{i_1},...,u_{i_{2\alpha}}\}$ respectively. Now to verify Condition (C), we have to show that 
\begin{equation}
\label{eqn100}
\{u_1^p,...,u^p_n\} \subset cl_{{\cal M}/B_{\overline{\cal F}}}(\{u_1,u_2,..,u_N\}),
\end{equation}
as $f(\mu)=\{u_1^p,...,u^p_n\}$ and $ f(In_{\cal E}(t))=\{u_1,u_2,..,u_N\}, \forall t\in{\cal T}.$ To show (\ref{eqn100}), we consider the set 
\[
U_{\cal F}=\{u_1,...,u_N\}-\{u_{i_1},...,u_{i_{2\alpha}}\}=\uplus_{k=1}^{|{\cal S}|}U_{\cal F}^k,
\]
where $U_{\cal F}^k=\{u_1^{k},u_2^{k},...,u_{N_{s_k}}^{k}\}-\{u_{i_1},...,u_{i_{2\alpha}}\}.$
For each $k,$ the set $U_{\cal F}^k$ contains at least $n_{s_k}$ elements. Thus, $U_{\cal F}^k$  contains a basis of ${\cal U}_{n_{s_k},N_{s_k}}.$ Therefore, $U_{\cal F}$ contains a basis of ${\cal U}.$ This means that $U_{\cal F}$ contains a basis of ${\cal U}_p$ also. This is seen by repeatedly using (\ref{eqn201}), given the fact that $U_{\cal F}$ contains a basis of ${\cal U}.$ Moreover as $u_j \notin (U_{\cal F}\cup f(\mu)), \forall j=i_1,...,i_{2\alpha},$ again by repeatedly using (\ref{eqn302}), we have
\begin{align}
\label{eqn101}
&r_{{\cal M}/B_{\overline{\cal F}}}(U_{\cal F})=r_{{\cal U}_p}(U_{\cal F}) = n,\\
\label{eqn102}
&r_{{\cal M}/B_{\overline{\cal F}}}(U_{\cal F}\cup f(\mu))=r_{{\cal U}_p}(U_{\cal F}\cup f(\mu))=n,
\end{align}
where the final equalities in both (\ref{eqn101}) and (\ref{eqn102}) follow from the fact that $U_{\cal F}$ has a basis of ${\cal U}_p.$ Equations (\ref{eqn101}) and (\ref{eqn102}) together prove (\ref{eqn100}), which proves that Condition (C) also holds.

Thus we have verified all the conditions of Definition \ref{matroidalerrornetworkdefinition}. Therefore the matroid ${\cal U}_{p,s}$ is a candidate matroid for the initial matroidal error correcting network $\cal G.$

In the forthcoming steps, both the network $\cal G$ and the matroid $\cal M$ are together made to evolve so as to preserve the matroidal nature of $\cal G$ in association with $\cal M.$~\\
~\\
\textit{\textbf{Step 3: Extending the network}}~\\
Let ${\cal G}_{temp}={\cal G},$ ${\cal M}_{temp} = {\cal M},$ $B_{temp}=B,$ ${\cal E}_{temp}={\cal E},$ ${\cal X}_{temp}={\cal X},$ and $In_{temp}(t)=In(t),~ \forall~ t\in {\cal T}.$ Let $f_{temp}:{\cal E}_{temp}\cup \mu \rightarrow {\cal E}({\cal M}_{temp})$ be the function defined as $f_{temp}(a) = f(a), \forall a\in\mu \cup {\cal E}_{temp}.$

Choose a random subset ${\cal E}_C\subseteq {\cal E}_{temp}$ of size at least $2.$ Add a new coding node to ${\cal G}_{temp}$ having incoming edges from the forwarding nodes whose incoming edges correspond to those in ${\cal E}_C.$ Add a new forwarding node, which has an incoming edge denoted as $e_{|{\cal E}_{temp}|+1}$ coming from the newly added coding node.

~\\~\\
\textit{\textbf{Step 4: Extending the matroid}}~\\
Let $cl$ be the closure operator in ${\cal M}_{temp}.$ Let ${\cal K}$ be a modular cut which contains $cl(f_{temp}({\cal E}_C))$ but does not contain $cl\left(B_{temp}-f_{temp}(\mu)\right).$  If such a modular cut does not exist, the algorithm goes back to \textbf{\textit{Step 3}} and proceeds with a different choice for ${\cal E}_C.$ If such a modular cut does not exist for any choice of ${\cal E}_C,$ then the algorithm ends without producing the appropriate output network. 

Let $r$ being the rank function in ${\cal M}_{temp}+_{_{\cal K}}x,$ the single-element extension of ${\cal M}_{temp}$ corresponding to the modular cut $\cal K.$ Then, in the matroid ${\cal M}_{temp}+_{_{\cal K}}x,$ the set $f_{temp}({\cal E}_C)\cup x$ contains a circuit with $x$, as $r(cl(f_{temp}({\cal E}_C))\cup x)=r(cl(f_{temp}({\cal E}_C)))$ by definition of a single-element extension. 
\begin{remark}
If ${\cal M}_{temp}+_{_{\cal K}}x$ is a representable extension, it has a representation of the form
\[
(I_{n+|{\cal E}_{temp}|}~~~{\cal X}'~~~\boldsymbol{x}),
\]
over some finite field such that the following hold.
\begin{itemize}
\item The submatrix ${\cal X}'$ is such that the matrix $(I_{n+|{\cal E}_{temp}|}~~~{\cal X}')$ is also a representation for ${\cal M}_{temp},$ as  
\[
({\cal M}_{temp}+_{_{\cal K}}x)\backslash x = {\cal M}_{temp}.
\]
\item The vector $\boldsymbol{x}$ is a column vector of size $n+|{\cal E}_{temp}|$ and can be obtained as a linear combination of the column vectors of ${\cal X}'$ corresponding to $f_{temp}({\cal E}_C).$  
\item Moreover, the first $n$ components of $\boldsymbol{x}$ are not all zero because $x\notin cl\left(B_{temp}-f_{temp}(\mu)\right),$ as $cl\left(B_{temp}-f_{temp}(\mu)\right) \notin {\cal K}.$
\end{itemize}
\end{remark}

\begin{figure}[htbp]
\centering
\includegraphics[totalheight=2.7in]{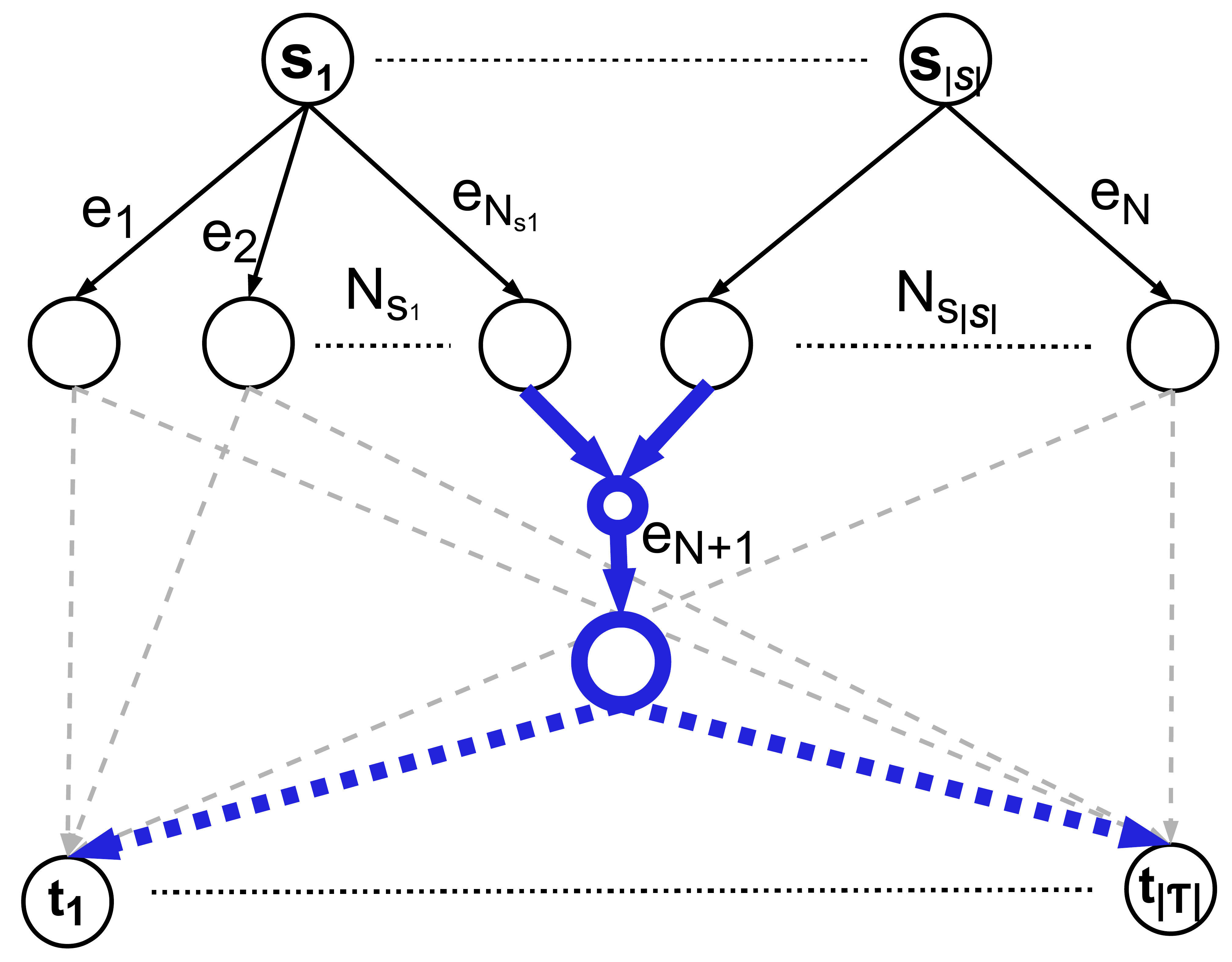}
\caption{Example of an extension of the network in  Fig. \ref{fig:multicastinit} with ${\cal E}_C=\left\{e_{N_{s_1}},e_{R_{|{\cal S}|-1}+1}\right\}.$ The newly added nodes and edges are indicated in blue and in bold. The unremoved incoming edges to the sinks are dimmed as they criss-cross with the newly added nodes and edges.}	
\label{fig:multicastextended}	
\end{figure}

We now add element $y$ in series with element $x$ to get the matroid $\left({\cal M}_{temp}+_{_{\cal K}} x\right) +^s_x y.$ Now the updates to the temporary variables are made as follows.
\begin{itemize}
\item[(a)]${\cal M}_{temp}=\left({\cal M}_{temp}+_{_{\cal K}} x\right) +^s_x y.$
\item[(b)]$B_{temp}= B_{temp} \cup b_{n+|{\cal E}_{temp}|+1},~\text{where}~  b_{n+|{\cal E}_{temp}|+1} = y.$
\item[(c)]$f_{temp}(e_{|{\cal E}_{temp}|+1})=x \in E({\cal M}_{temp}).$
\item[(d)]Let ${\cal G}_{temp}$ be updated by adding the two new nodes (coding node and forwarding node) to the node set, and with ${\cal E}_{temp}={\cal E}_{temp}\cup e_{|{\cal E}_{temp}|+1}.$  Thus the edge $e_{|{\cal E}_{temp}|+1}$ is now referred to as $e_{|{\cal E}_{temp}|}.$
\end{itemize}
\begin{remark}
If ${\cal M}_{temp}+_{_{\cal K}}x$ is representable, then by Lemma \ref{seriesrepresentation}, so is $\left({\cal M}_{temp}+_{_{\cal K}} x\right)+^s_x y$, with the corresponding representation
\begin{equation}
\label{eqn8}
\left(
\begin{array}{cccc}
I_{n+|{\cal E}_{temp}|} & \boldsymbol{0} & {\cal X}' & \boldsymbol{x}\\
\boldsymbol{0} & 1 & \boldsymbol{0} & 1
\end{array}
\right),
\end{equation}
where the $\boldsymbol{0}$s represent zero row and column vectors of the appropriate sizes. The column corresponding to the new element $y$ is then $\left(
\begin{array}{c}
\boldsymbol{0}\\
1
\end{array}
\right).
$ We also make the following update 
\[
{\cal X}_{temp}=
\left(
\begin{array}{cc}
{\cal X}' & \boldsymbol{x}\\
\boldsymbol{0} & 1
\end{array}
\right).
\]
\end{remark}
~\\
\textit{\textbf{Step 5: Updating the incoming edges at the sinks}} ~\\
For each sink $t,$ we update the set $In_{temp}(t)$ at most once as follows. 
\begin{itemize}
\item For some $e_i \in In_{temp}(t),$ if there is some circuit ${\cal C}_{e_i} \subseteq \left(f_{temp}({\cal E}_C) \cup x \cup y\right)$ such that $\left(x\cup f_{temp}(e_i) \subseteq {\cal C}_{e_i}\right),$ then let $In_{temp}(t)= (In_{temp}(t)-e_i)\cup e_{|{\cal E}_{temp}|}.$
\end{itemize}
The update is based on the rationale that if the flow on $e_i$ has been encoded into the flow in the newly added edge $e_{|{\cal E}_{temp}|},$ then in any sink which has $e_i$ as an incoming edge, the edge $e_i$ can be replaced by $e_{|{\cal E}_{temp}|}$ in the set of incoming edges. Such an update is only the most natural one possible. It is possible to update the incoming edges at the sinks more interestingly, however requiring more computations (such an optional update is described in \textbf{\textit{Step 6}} of this algorithm). An example instance of the extended network (from Fig. \ref{fig:multicastinit}), along with the updated incoming edges at the sinks is shown in Fig. \ref{fig:multicastextended}.
~\\~\\
\textit{\textbf{Step 6: Checking the conditions of Definition \ref{matroidalerrornetworkdefinition}}}~\\
The matroid ${\cal M}_{temp}$ along with function $f_{temp}$ and basis $B_{temp}$ satisfies the conditions (A) and (B) of Definition \ref{matroidalerrornetworkdefinition} with respect to the network ${\cal G}_{temp}$ for the following reasons.
\begin{itemize}
\item Condition (A) is satisfied because $f_{temp}(\mu) = \left\{b_1,b_2,...,b_n\right\}\in B_{temp}.$
\item Condition (B1) is satisfied because $f_{temp}(e_{|{\cal E}_{temp}|})=x\notin cl\left(B_{temp}-f_{temp}(\mu)\right),$ as $cl\left(B_{temp}-f_{temp}(\mu)\right)\notin {\cal K}.$
\item We know that ${\cal M}_{temp}$ is the series extension of the matroid ${\cal M}_{temp}/y$ at $x.$ Using this fact, and by applying Lemma \ref{seriesextensionlemma2} (with ${\cal N}$ being the updated matroid ${\cal M}_{temp}$, and with $e_1=x$ and $e_2=y$), we have that any circuit containing $x$ in ${\cal M}_{temp}$ also contains $y$. Therefore, we have, 
\[
x\in cl(f_{temp}({\cal E}_C)\cup y) \text{~but~} x\notin cl(f_{temp}({\cal E}_C)),
\]
where $cl$ is the closure operator in ${\cal M}_{temp}.$ Thus it is seen that Condition (B2) is satisfied as  $f_{temp}(e_{|{\cal E}_{temp}|})=x$ and $y=b_{n+|{\cal E}_{temp}|}.$
\end{itemize}

Condition (C) of Definition \ref{matroidalerrornetworkdefinition} is not ensured by \textbf{\textit{Step 4}} and therefore has to be checked independently. 
\begin{remark}
Suppose ${\cal M}_{temp}$ is representable before extension, and we also wish to obtain a representable extension. This corresponds to a scalar linear network-error correcting code for ${\cal G}_{temp}$. In other words, the vector $\boldsymbol{x}$ of (\ref{eqn8}), which corresponds to a linear combination of the global encoding vectors from existing nodes, has to be designed such that the error correcting capability of the scalar linear network-error correcting code is maintained. Using the techniques of \cite{YeC1,YeC2,Zha,Mat,YaY}, this can always be done as long as the field size is large enough (discussed in Section \ref{seccomplexity}). Once the vector $\boldsymbol{x}$ is found, the matroid is also updated as the vector matroid of the matrix in (\ref{eqn8}). Thus, we can find a suitable extension of the initial matroid such that the updated ${\cal M}_{temp}$ is a representable matroid that maintains Condition (C). However, in this case the field size demanded by the algorithms in \cite{YeC1,YeC2,Zha,Mat,YaY} is in general quite high, and therefore the scalar linear network-error correcting code obtained operates over such a large field.
\end{remark}

In general, ${\cal M}_{temp}$ need not be representable. Therefore we simply check Condition (C) by brute-force. If Condition (C) does not hold, then the algorithm returns to \textit{\textbf{Step 4}} to search for an extension of the matroid which satisfies all the conditions of Definition \ref{matroidalerrornetworkdefinition}.

If Condition (C) of Definition \ref{matroidalerrornetworkdefinition} holds for all sinks and for all error patterns on the incoming edges of the forwarding nodes, then all the concerned variables are updated as follows.
\begin{itemize}
\item[(a)]$In(t)=In_{temp}(t), ~\forall~ t\in {\cal T} .$
\item[(b)]\textit{Optional Update:} Optionally, for any sink $t,$ the set $In(t)$ can be updated as the set $I\cup e_{|{\cal E}_{temp}|},$ where $I$ is the smallest subset of $(In_{temp}(t)-e_{|{\cal E}_{temp}|})$ such that upon fixing $In(t)=I\cup e_{|{\cal E}_{temp}|},$ Condition (C) is still satisfied. This involves further brute-force checking of Condition (C) for each such subset of $In_{temp}(t).$ However, it can generate networks where there are no unnecessary incoming edges at any sink. The implementation of this optional update in our MATLAB program is illustrated in Example \ref{multicastex} of Subsection \ref{sketchandillustrexamples} in the transition between Fig. \ref{fig:multicastconstructionexample2}	and Fig. \ref{fig:multicastconstructionexample3}, and also in Example \ref{unicastncexample} in Section \ref{sec6}.
\item[(c)]${\cal M}={\cal M}_{temp}.$
\item[(d)]${\cal B}={\cal B}_{temp}.$
\item[(e)]If ${\cal M}_{temp}$ is representable, let ${\cal X}={\cal X}_{temp}.$ (Thus the matroid $\cal M$ is again the vector matroid of the matrix of the form $(I_{n+|{\cal E}|}~~~{\cal X})$.)
\item[(f)]${\cal G}={\cal G}_{temp}.$
\item[(g)]${\cal E}={\cal E}_{temp}.$
\item[(h)]$f(a) = f_{temp}(a)~\forall a\in \mu\cup{\cal E}.$

\end{itemize}

If $N_C$ coding nodes have already been added, then the algorithm ends with the output of all the above variables. Otherwise, the algorithm returns back to \textit{\textbf{Step 3}}, to find a new extension to the graph and the matroid. Note that as the network $\cal G$ is maintained to be a matroidal $\alpha$-error correcting network over the matroid $\cal M$ at each addition of a coding node, the resultant network after the final extension is also a matroidal $\alpha$-error correcting network in association with the matroid $\cal M.$ If ${\cal M}$ is a representable matroid, then a scalar linear network-error correcting code is obtained according to the proof of Theorem \ref{matroidalerrornetworkthm}. 

\subsection{Multiple-Unicast Construction}
\label{subsec5c}
We now present a similar algorithm as that of multicast for the construction of multiple-unicast network instances. As this algorithm follows the same pattern as that of the multicast algorithm, we only point out the differences between the two. ~\\
~\\
\textit{\textbf{Step 1: Initializing the multiple-unicast network}}~\\
The network is initialized by creating $n$ source nodes (each of which generate one message), and $1+2\alpha$ forwarding nodes corresponding to each source node, each with one incoming edge from the corresponding source. Let these edges be $\left\{e_1, e_2, ... , e_{n(1+2\alpha)}\right\}={\cal E}.$ Let ${\cal T}$ be the collection of $n$ sink nodes created. 

\begin{figure}[htbp]
\centering
\includegraphics[totalheight=1.9in,width=3.5in]{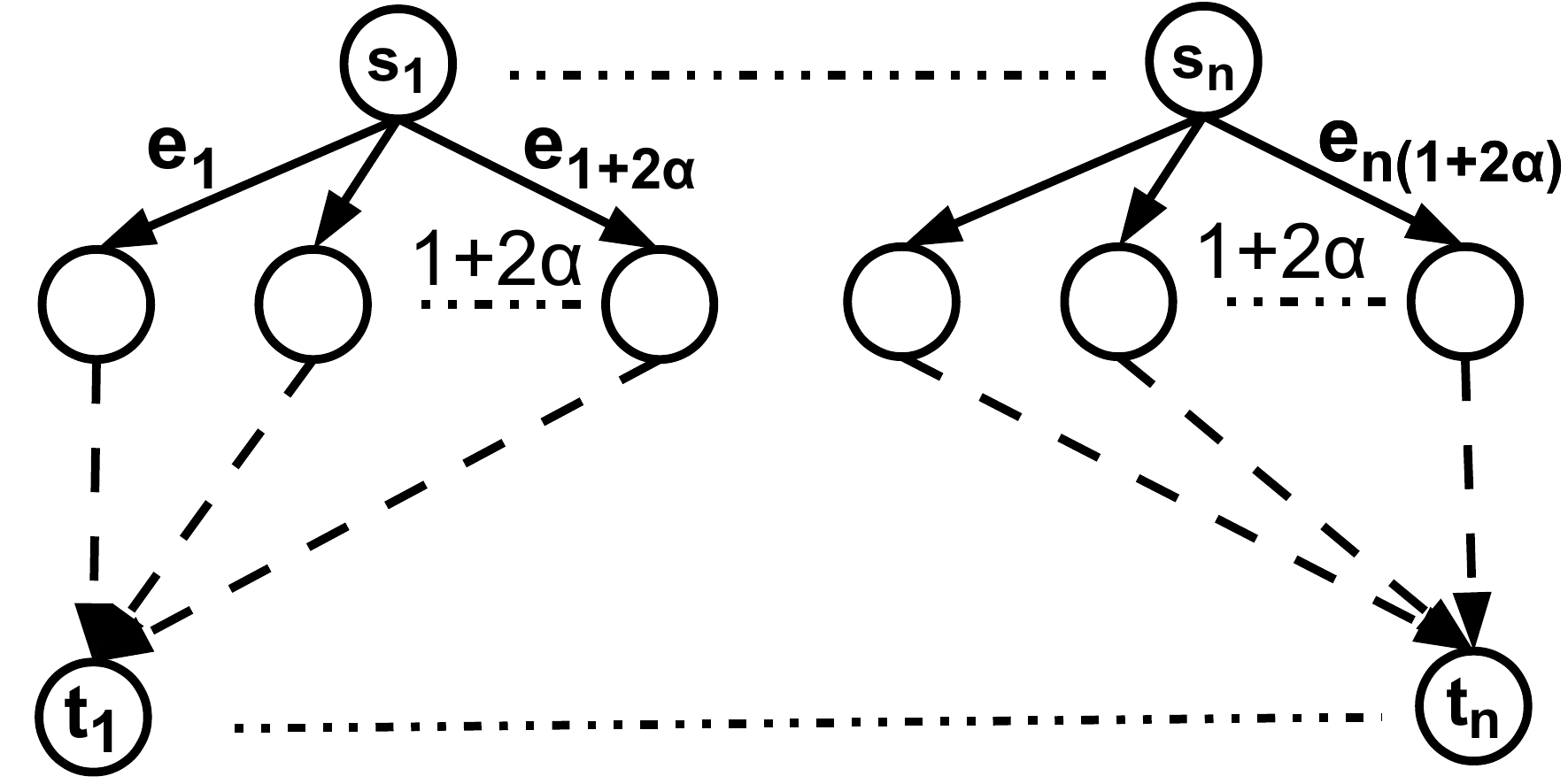}
\caption{Initial network of the multiple-unicast algorithm}	
\label{fig:unicastinit}	
\end{figure}

For the sink $t_i$ which demands the message from source $s_i,$ $1+2\alpha$ imaginary incoming edges are drawn from the forwarding nodes corresponding to that particular source. Again, we abuse our notation and denote by $In(t_i)$ the incoming edges of these forwarding nodes. This initialized network is represented in Fig. \ref{fig:unicastinit}.
~\\
~\\
\textit{\textbf{Step 2: Initializing the matroid}}
~\\
As before, we obtain a matroid $\cal M$ such that the network $\cal G$ is a matroidal $\alpha$-error correcting network with respect to this matroid $\cal M.$ Let $A$ be the $n \times n(1+2\alpha)$ matrix
\[
\left(\begin{array}{cccc}
\boldsymbol{1_{1+2\alpha}} & \boldsymbol{0_{1+2\alpha}} & ... & \boldsymbol{0_{1+2\alpha}}\\
\boldsymbol{0_{1+2\alpha}} & \boldsymbol{1_{1+2\alpha}} & ... & \boldsymbol{0_{1+2\alpha}}\\
. & . & ... & .\\
. & . & ... & . \\
\boldsymbol{0_{1+2\alpha}} & \boldsymbol{0_{1+2\alpha}} & ... & \boldsymbol{1_{1+2\alpha}}
\end{array}
\right),
\]
where $\boldsymbol{1_{1+2\alpha}}$ and  $\boldsymbol{0_{1+2\alpha}}$ represent the all-ones and all-zeros row vectors of size $1+2\alpha$ over some finite field. Let $\cal M$ be the vector matroid of the following matrix,
\[
\left(\begin{array}{cccc}
I_n & \boldsymbol{0} & A \\
\boldsymbol{0} & I_{n(1+2\alpha)} & I_{n(1+2\alpha)}
\end{array}
\right),
\]
where the $\boldsymbol{0}$s represent zero matrices of appropriate sizes. Note that the above matrix is of the form $(I_{n+|{\cal E}|}~~~{\cal X})$ with $|{\cal E}| = n(1+2\alpha).$

%
%

Let $B=\left\{1,2,3,...,n+|{\cal E}|\right\}$ be the basis of $\cal M$ considered. Let 
$
f:{\cal E}\cup\mu\rightarrow E({\cal M})
$
be the function defined as follows. 
\begin{align*}
& f(m_i) = i,~~m_i \in \mu, i=1,2,...,n. \\
& f(e_i) = n+|{\cal E}|+i,~\forall~e_i \in {\cal E}.
\end{align*}
Then it can be seen that this matroid $\cal M$ with the basis $B$ and function $f$ satisfy the conditions of Definition \ref{matroidalerrornetworkdefinition}, as each source is simply employing a repetition code of length $1+2\alpha$.~\\
~\\
\textit{\textbf{Step 3}(extending the network)} and \textit{\textbf{Step 4}(extending the matroid)} are the same as the multicast construction. Therefore we proceed to \textit{\textbf{Step 5}}.~\\
~\\
\textit{\textbf{Step 5: Updating the incoming edges at the sinks}}~\\
In multiple-unicast (or more generally, in the networks with arbitrary demands), there arises the issue of interference from other undesired source symbols with the desired symbols at any sink, thereby necessitating the presence of side information besides the sufficient error correction capability in order to decode correctly. Therefore, unlike the multicast case, simply replacing the encoded edge with the newly formed edge will not suffice to update $In_{temp}(t),$ as the newly formed edge can include additional interference not present in the encoded edge. 

\begin{figure}[htbp]
\centering
\includegraphics[totalheight=2.9in]{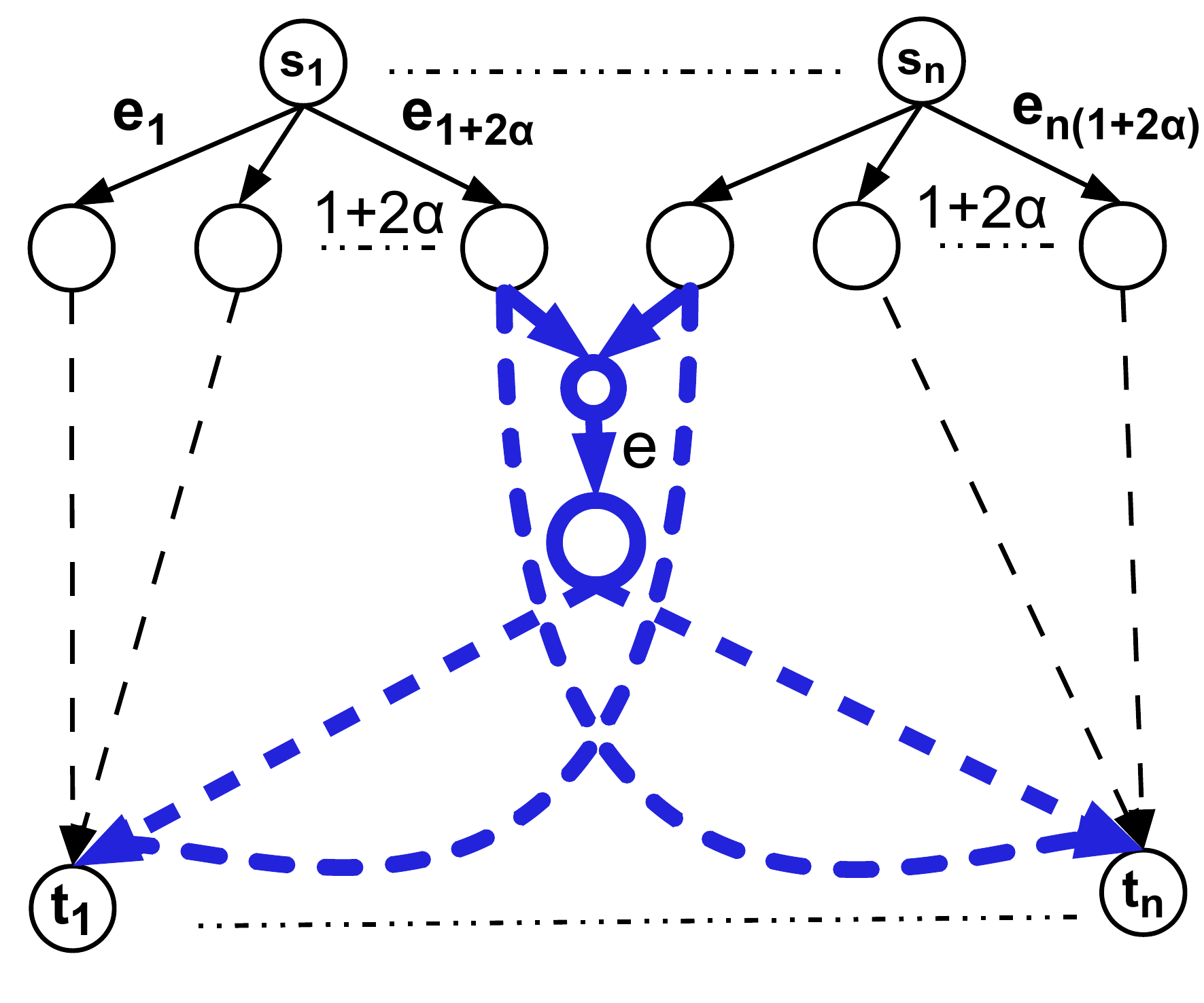}
\caption{Example of an extension of the network in  Fig. \ref{fig:unicastinit}. The newly added nodes and edges are indicated in blue and in bold.}	
\label{fig:unicastextended}	
\end{figure}

The following procedure is therefore adopted to update the incoming edges at each of the sinks.
\begin{enumerate}
\item This is the same as in multicast and done at most once for a sink. For some $e_i \in In_{temp}(t),$ if there is some circuit ${\cal C}_{e_i} \subseteq f_{temp}({\cal E}_C) \cup x\cup y$ such that $x\cup f_{temp}(e_i) \subseteq {\cal C}_{e_i},$ then let $In_{temp}(t)= (In_{temp}(t)-e_i)\cup e_{|{\cal E}_{temp}|+1}.$ If no such $e_i$ exists, there is no need to update $In_{temp}(t)$ and this entire step can be skipped.
\item Let $e_i$ be the element that is replaced in $In_{temp}(t).$ Let $e_j \in {\cal E}_{temp}$ such that the following conditions hold. 
\begin{itemize} 
\item $e_j \notin In_{temp}(t)$ but $f_{temp}(e_j)\in ({\cal C}_{e_i}-f_{temp}(e_i)).$~\\
\item ~\vspace{-0.5cm}
\begin{align*}
r_{{\cal M}_{temp}}\left(\hspace{-0.15cm}\right.&\left.f\left(In_{temp}(t)-e_{|{\cal E}_{temp}|+1}\right)\cup f(e_j)\right) \\
& > r_{{\cal M}_{temp}}\left(f\left(In_{temp}(t)-e_{|{\cal E}_{temp}|+1}\right)\right).
\end{align*}
\end{itemize}
This means that the flow in $e_j$ has been encoded as additional new interference into the flow in the newly added edge $e_{|{\cal E}_{temp}|+1},$ thus creating the necessity of additional side information at the sink $t$ to cancel out this interfering flow. We thus update $In_{temp}(t)$ as $In_{temp}(t)=In_{temp}(t)\cup e_j.$ Thus for each $e_j$ such that the above two conditions hold at sink $t,$ $e_j$ is included in $In_{temp}(t)$ so that sufficient side information is available at the sink to decouple any newly introduced interference and decode the necessary information. This is also to be repeated at each sink.
\end{enumerate}
An example instance of an extension of the network of Fig. \ref{fig:unicastinit}, along with the updated incoming edges at the sinks is shown in Fig. \ref{fig:unicastextended}. As with the multicast algorithm, it is possible to update the sink incoming edges after Condition (C) has been checked. Thus such an update can be optionally included at the end of \textbf{\textit{Step 6}}.
~\\
~\\
\textit{\textbf{Step 6}(checking the conditions of Definition \ref{matroidalerrornetworkdefinition})} is the same as that of the multicast construction, therefore we don't elaborate further. The optional update to the incoming edges to the sinks can also be done in a similar fashion as in \textbf{\textit{Step 6}} of the multicast construction. 

As in the multicast construction, at each step the matroidal property of the network is preserved, thus the output of the algorithm is a matroidal $\alpha$-error correcting network which unicasts the set of messages in the presence of at  most $\alpha$ network-errors.

\subsection{On constructing matroidal error correcting networks associated with nonrepresentable matroids} 
\label{mecnnonrepresentable}
One of the major results of \cite{DFZ} was that nonrepresentable matroids can be used to construct matroidal networks for which Shannon-type information inequalities (the most widely used collection of information inequalities in information theory) cannot bound their capacity as tightly as the non-Shannon-type information inequalities do. In other words, networks connected with nonrepresentable matroids can prove to be very useful in obtaining insights on the general theory of network coding. It can therefore be expected that matroidal error correcting and detecting networks associated with nonrepresentable matroids will be useful in obtaining similar insights for network-error detection and correction. It was already mentioned in the beginning of Section \ref{sec5} that it is not straightforward to obtain representable or nonrepresentable matroids from which we can construct matroidal network-error correcting or detecting networks directly. The difficulty is that, unlike \cite{DFZ}, Definitions \ref{matroidalerrornetworkdefinition} and \ref{matroidalerrorcorrectionnetworkdefinition} for matroidal error detecting and correcting networks require matroids whose contractions have to satisfy specific properties which enable the decoding of the demanded symbols at sinks. Since this is a fundamental requirement of error detecting and correcting networks, it is clear that such a requirement cannot avoided. This motivated the method used in our algorithms to construct such networks, i.e., starting with simple matroids and their counterpart networks and then extending them together while keeping the conditions of error correction intact. The chief reasons for the inability of using our algorithms to obtain  example networks which are associated with nonrepresentable matroids are as follows.
\begin{itemize}
\item Descriptions of nonrepresentable matroids with many elements in its groundset is not an easy task, even on a computing  device. More importantly, computing the extensions (in particular single-element extensions, which involve computation of the flats and the modular cuts) of such nonrepresentable matroids with many elements is computationally intensive. Furthermore, there are a large number of possible single-element extensions for any matroid with many elements in its groundset. Checking the representability or nonrepresentability of such extensions is not easy. 
\item Evaluating the error correcting property of a given linear network-error correcting code involves going through all possible error patterns and checking if the error correction holds for each of them. To the best of the authors' knowledge, such a brute-force technique is used in all available coherent linear network-error correction literature (see \cite{YeC1,Zha,Mat,YaY}, for example) to construct linear network-error correction codes. Thus checking Condition (C) of Definitions \ref{matroidalerrornetworkdefinition} and \ref{matroidalerrorcorrectionnetworkdefinition} demands brute-force analysis of all the contractions corresponding to all possible error patterns. Compared to representable matroids, computing the contractions of nonrepresentable matroids is  computationally intensive. 
\item In \cite{DFZ}, Shannon and non-Shannon information inequalities were used to capture the uniqueness of the \textit{Vamos network}  obtained from the nonrepresentable Vamos matroid (see \cite{DFZ} for more details). In our case, even if we suppose that a matroidal error detecting (or correcting) network associated on a nonrepresentable matroid is obtained through our algorithm, such an analysis seems rather complicated, again the issue being the number of possible error patterns. Verifying that the best possible linear error correction schemes have rates of information transmission less than the best possible nonlinear schemes once again implies going through each of the error-patterns and evaluating the maximum possible rates of transmission. The number of these calculations grows linearly with the number of possible error-patterns and can quickly become unwieldy.
\end{itemize}
Though we do not present examples of networks obtained using our algorithms  which are associated with nonrepresentable matroids from our algorithms because of the above reasons, we present a proposition in this subsection as a first step towards reducing the search-space of matroidal extensions in order to obtain nonrepresentable matroids which satisfy the properties in Definition \ref{matroidalerrornetworkdefinition}. Also, in Section \ref{secinsufficiency}, using ideas from \cite{DFZ3}, we present an example network which is a matroidal 1-error detecting network associated with a nonrepresentable matroid, using which we show that linear network-error detection and correction schemes are not always sufficient to satisfy network demands in the presence of network-errors.
 
Proposition \ref{propprincipal} below shows that if we are to use the constructions of Section \ref{sec5} to obtain matroidal error correcting or detecting networks associated with nonrepresentable matroids, then the extension of the matroid considered in \textbf{\textit{Step 4}} of the multicast and the multiple-unicast constructions must necessarily be a non-principal extension, i.e., the modular cut corresponding to the extension must not be a principal modular cut. The proof of the following proposition is given for the sake of completeness as to the best of the authors' knowledge it seems to be unavailable in matroid theory literature.

\begin{proposition}
\label{propprincipal}
Let $A$ be a matrix of size $k\times m$ ($k\leq m$) with elements from some field $\mathbb{F}_q,$ and let ${\cal M}={\cal M}[A].$ Let ${\cal K}_F$ be the principal modular cut of $\cal M$ generated by flat $F$ of ${\cal M}.$ Then the principal extension ${\cal M}+_{_{{\cal K}_F}} e$ of the matroid $\cal M$ is representable over an extension of $\mathbb{F}_q.$
\end{proposition}
\begin{IEEEproof}
Let $X = A^{F},$ the submatrix of $A$ with respect the column indices given by $F.$ Let $\langle X\rangle_q$ denote the space spanned by the columns of $X$ over $\mathbb{F}_q.$ Let $X_{(0)}, X_{(1)},..,X_{{(M-1)}}$ be the submatrices corresponding to all the flats $F_0, F_1,..., F_{M-1}$ of $\cal M$ which do not contain $F.$ Thus for each $i=0,1,2,...,M-1,$ there exists at least one non-zero vector $v_i \in \mathbb{F}_q^k$ such that $v_i \in \langle X\rangle_q$ but $v_i \notin \langle X_{(i)}\rangle_q.$

Consider the extension field $\mathbb{F}_{Q}, Q=q^M.$ Let $\beta$ be the primitive element of $\mathbb{F}_Q,$ with respect to $\mathbb{F}_q$ as the base field. Thus any element of $\mathbb{F}_Q$ can be uniquely expressed as a polynomial of degree at most $M-1$ in $\beta.$ 

Let 
\[
v=\sum_{i=0}^{M-1}v_i\beta^{i} \in \mathbb{F}_Q^k.
\]
Let $\tilde{A} = \left(A~|~v\right)$ be the matrix over $\mathbb{F}_Q$ where the elements of the submatrix $A$ are viewed as elements from the basefield $\mathbb{F}_{q}$ embedded in $\mathbb{F}_Q.$ We claim that $\tilde{A}$ is the required representation for the matroid extension ${\cal M}+_{_{{\cal K}_F}} e.$ Let $\langle X\rangle_Q$ denote the vector space spanned by the columns of $X$ over $\mathbb{F}_Q.$ According to Definition \ref{singleelement}, to show that $\tilde{A}$ is the required representation, it is enough to show that $v \in	\langle X\rangle_Q$ but $v \notin \langle X_{(i)}\rangle_Q, i=0,1,2,..,M-1.$ 

For each $i=0,1,...,M-1,$ as $v_i \in \langle X\rangle_q$ it is clear that $v_i \in \langle X\rangle_Q,$ also. Thus $v \in	\langle X\rangle_Q.$ Now, for some $r$ such that $0\leq r \leq M-1,$ consider a $\mathbb{F}_Q$ linear combination of the column vectors in $X_{(r)}$ as follows.
\begin{align}
\nonumber
\sum_{j}g_jX_{(r)}^j & = \sum_{j}\left(\sum_{j'=0}^{M-1}g_{j,j'}\beta^{j'}\right)X_{(r)}^j \\
& = \sum_{j'=0}^{M-1}\left(\sum_jg_{j,j'}X_{(r)}^j\right)\beta^{j'},
\end{align}
where $g_j=\sum_{j'=0}^{M-1}g_{j,j'}\beta^{j'} \in \mathbb{F}_Q$ with $g_{j,j'} \in \mathbb{F}_q,~\forall j'.$ As $v_r \notin \langle X_{(r)}\rangle_q,$ we must have that for any $j'=0,1,2,...,M-1,$ 
\[
\sum_jg_{j,j'}X_{(r)}^j \neq v_r.
\]
For the same reason, we must have
\[
\sum_{j}g_jX_{(r)}^j= \sum_{j'=0}^{M-1}\left(\sum_jg_{j,j'}X_{(r)}^j\right)\beta^{j'} \neq \sum_{i=0}^{M-1}v_i\beta^{i} = v,
\]
for any $r = 0,1,2,...,M-1$ and for any linear coefficients $g_j \in \mathbb{F}_Q~\forall j.$ 

Thus $v \notin \langle X_{(i)}\rangle_Q,~\forall i=0,1,2,...,M-1.$ Thus $\tilde{A}$ satisfies the conditions to be a representation for ${\cal M}+_{_{{\cal K}_F}} e.$ This proves the proposition. 
\end{IEEEproof}
\section{Complexity}
\label{seccomplexity}
We now calculate upper bounds on the complexity of the algorithms for the case of scalar linear network-error correcting codes (i.e., representable matroids). These calculations are for the implementation of our algorithms without the execution of the optional update to the incoming edges to the sinks in \textbf{\textit{Step 6}}. Including this optional update step will certainly increase the complexity of the algorithms. However, the calculations that follow capture the essential running time of our algorithms in the representable case. In the case of nonrepresentable matroids, the complexity of our algorithms will depend heavily on the matroidal operations involved to obtain the extensions, computing the contractions and checking the ranks of subsets in the computed contractions in order to verify the error correcting properties of the matroidal network so formed. As such matroidal operations are involved, it is not clear how to proceed in this direction. Hence we take up on computing the complexity of our algorithms in generating networks associated only with representable matroids. In any case, constructing network associated with nonrepresentable matroids using our algorithm can be expected to be at least as difficult as the representable case, since in the representable case all the matroids have matrix representations and all matroid operations are implementable as operations based on linear algebra. 

For obtaining the complexity of our multisource multicast algorithm, we shall directly use the complexity of the construction algorithm for single source multicast scalar linear network-error correcting codes given in \cite{Mat}. Further, we shall also show that our multiple-unicast algorithm (in the case of representable matroids) is equivalent to a variant of the algorithm in \cite{Mat} and therefore the complexity of the algorithm of \cite{Mat} can be used to obtain that of our multiple-unicast algorithm also.
\subsection{Network-Error Correcting Codes - Algorithm of \cite{Mat}}
Algorithm \ref{alg:necc} is a brief version of the algorithm given in \cite{Mat} for constructing an scalar linear $\alpha$-network-error correcting code for a given single source, acyclic network that meets the network Singleton bound given in \cite{YeC1}. The construction of \cite{Mat} is based on the network code construction algorithm of \cite{JSCEEJT}. The algorithm constructs a network code such that all network-errors in upto $2\alpha$ edges will be corrected as long as the sinks know where the errors have occurred. Such a network code is then shown \cite{Mat} to be equivalent to an $\alpha$-network-error correcting code. Other equivalent (in terms of complexity) network-error correction algorithms can be found in \cite{Zha} \cite{YaY}.

\begin{algorithm}
\KwIn{An acyclic network ${\cal G}({\cal V},{\cal E})$ with mincut $N$ from the source $s$ to the set of sinks $\cal T.$} 
\KwOut{An $\alpha$-network-error correcting code for $\cal G$ that meets the network Singleton bound}
\vspace{0.1cm}
\hrule
\vspace{0.1cm}
$(1)$ Let $\cal F$ be the set of all subsets of $\cal E$ of size $2\alpha.$ Add an imaginary source $s'$ and draw $n=N-2\alpha$ edges from $s'$ to $s.$

$(2)$ \ForEach{$F \in {\cal F}$}
{
$(i)$ Starting from the original network, add an imaginary node $v$ at the midpoint of each edge $e\in F$ and add an edge of unit capacity from $s'$ to each $v.$ 

$(ii)$ \ForEach{sink $t\in \cal T$} 
{
Draw as many edge disjoint paths from $s'$ to $t$ passing through the imaginary edges added at Step $(i)$ as possible. Let $m_{t}^F (\leq 2\alpha)$ be the number of such paths. 

Draw $n$ edge disjoint paths passing through $s$ that are also edge disjoint from the $m_{t}^F$ paths drawn in the previous step.
}

$(iii)$ Use the algorithm from \cite{JSCEEJT} using the identified edge disjoint paths such that it ultimately gives a network code with the following property. Let $B_t(F)$ be the $\left(n+2\alpha\right) \times \left(n+m_{t}^F\right)$ matrix, the columns of which are the $N$ length global encoding vectors (representing the linear combination of the $n$ input symbols and $2\alpha$ error symbols) of the incoming edges at sink $t$ corresponding to the $n+m_{t}^F$ edge disjoint paths. Then $B_t(F)$ must be full rank. As proved in \cite{Mat}, this ensures that the network code thus obtained is $\alpha$-network-error correcting and meets the network Singleton bound.
}
\caption{\vspace{0.1cm}Algorithm of \cite{Mat} for constructing a network-error correcting code that meets the network Singleton bound.\vspace{0.1cm}}
\label{alg:necc}
\end{algorithm}
It is shown in \cite{Mat} that Algorithm \ref{alg:necc} results in a network code which is a $\alpha$-network-error correcting code meeting the network Singleton bound, as long as the field size 
\begin{equation}
\label{eqn15}
q > |{\cal T}||{\cal F}|= |{\cal T}|
\left(
\begin{array}{c}
|{\cal E}| \\
2\alpha 
\end{array}
\right).
\end{equation}
The complexity of the algorithm is then $O\left(|{\cal F}||{\cal T}|N\left(|{\cal E}||{\cal F}||{\cal T}|+|{\cal E}|+N+2\alpha\right)\right).$

\subsection{Multicast}
We use the complexity of Algorithm \ref{alg:necc} to calculate the complexity of our multisource multicast algorithm. This requires converting the multisource multicast network to the single source multicast network, as Algorithm \ref{alg:necc} works only on a single source multicast network. This can be done after \textbf{\textit{Step 1}} of the algorithm, where we can add a super-source to the network from which edges flow into the actual set of sources ${\cal S}.$ After \textbf{\textit{Step 1}}, the network is clearly matroidal $\alpha$-error correcting with respect to the direct sum of the uniform matroids. And thus the network after \textbf{\textit{Step 1}} has a multicast scalar linear $\alpha$-network-error correcting code if the direct sum is representable. Constructing the $N_C$ nodes and their global encoding vectors while preserving the error correcting property, i.e. generating the network and appropriate matroid extensions, can be done using Algorithm \ref{alg:necc}, once all the variables have been initialized and the super source has been added. 

We consider errors only at the incoming edges of the forwarding nodes, and there are at most $|{\cal E}|=N+N_C$ such edges at any iteration of our algorithm. Let $\eta = \left(\begin{array}{c} |{\cal E}|\\ 2\alpha \end{array} \right).$ If the field size of operation assumed is greater than $|{\cal T}|\eta$, then by Algorithm \ref{alg:necc}, a suitable extension to the representable matroid (i.e., a suitable global encoding vector to the edge of the newly added incoming node) exists at each iteration of our algorithm, and the total complexity of obtaining the network and the representable matroid (equivalently, the linear network-error correcting code) will be 
$O\left(\eta|{\cal T}|N\left(|{\cal E}|\eta|{\cal T}|+|{\cal E}|+N+2\alpha\right)\right),$
assuming that the other steps in the algorithm can be done in constant time or with negligible complexity compared to \textbf{Step 4} and \textbf{Step 6}. With a smaller field size, the complexity of obtaining the network and the matroid will continue to be bounded similarly, provided the suitable vectors exist at all iterations. At the end of using Algorithm \ref{alg:necc} to obtain the coding nodes and the linear network-error correction code, the super-source and the outgoing edges from the super-source can be removed to give our required network.

\subsection{Multiple Unicast}
Unlike multicast, there exist no known algorithms to construct network-error correcting codes for multiple unicast networks which we can use to compute the complexity according to the requirements of our algorithm. Therefore, we take an indirect approach. At each iteration in our multiple unicast algorithm (omitting the optional update in \textbf{\textit{Step 6}}), we show that the construction of a suitable global encoding vector (for the current edge under processing) for satisfying the multiple-unicast conditions is equivalent to the construction of a suitable global encoding vector such that certain matrices are full-rank as in Step $2(iii)$ of Algorithm \ref{alg:necc} for each error pattern in $\cal F$. Thus, the complexity of our multiple-unicast algorithm can be obtained from the complexity of Algorithm \ref{alg:necc} after suitable changes.

Let ${\cal G}(i)$ be the state of the multiple unicast network at the iteration $i$ ($i=0$ representing the initial state and $i=N_C$ representing the final iteration) of our multiple-unicast algorithm. That is, in the network ${\cal G}(i),$ $i-1$ coding nodes have already been added and the global encoding vectors corresponding to their incoming edges have been fixed. Also, a particular subset of the forwarding nodes have been picked and the $i^{th}$ coding and the corresponding forwarding node have been added according to \textbf{Step 3} of the algorithm. We also update the incoming edges at the sinks according to \textbf{Step 5} even before fixing the global encoding vector of the newly added edge by simply adding edges containing all possible interfering flows as the new side information for the sinks. So the steps that remain to be executed are \textbf{Step 4} and \textbf{Step 6}, i.e., picking a suitable global encoding vector for the newly added edge $e_{n(2\alpha+1)+i}$ (from the newly added coding node) so that the error correction capability and decoding continue to hold at the sinks. After achieving this goal, those edges which carry side information that are not used for the decoding process at the sinks can be removed. 

Let $n_t(i)$ be the number of incoming edges at sink $t$ and $\boldsymbol{F_{{\cal S},t}(i)}$ be the transfer matrix of size $n\times n_t(i)$ from the sources to sink $t$ at the end of iteration $i$ of our multiple-unicast algorithm (i.e., after fixing a suitable global encoding vector for $e_{n(2\alpha+1)+i}$). Towards obtaining a bound on the complexity of our algorithm, we first prove the following lemma.

\begin{lemma}
\label{lemma4}
For each sink $t$ in ${\cal G}(i),$ there exists some full rank square matrix $A_t(i)$ of size $n_t(i)$ such that 
\[
\boldsymbol{F_{{\cal S},t}(i)}A_t(i)=\left(I^j~I^j~..~I^j~|~C(i)\right),
\]
where $I^j$ is the $j^{th}$ basis vector corresponding to the input $x_j$ demanded by sink $t$ and is repeated $2\alpha+1$ times in the above matrix.
\end{lemma}
\begin{IEEEproof}
The claim holds for ${\cal G}(0)$ with $C(0)$ being an empty matrix. We assume that the claim holds for ${\cal G}(i)$ and will prove that it holds for ${\cal G}(i+1)$ as well. Because of the network code and the way the incoming edges at the sinks are updated, we have for some nonsingular square matrix $L$ of size $n_t(i+1),$
\[
\boldsymbol{F_{{\cal S},t}(i+1)}=\left(\boldsymbol{F_{{\cal S},t}(i)}~|~V\right)L,
\]
where $V$ is a matrix with $n$ rows, consisting of the global encoding vectors of the newly added incoming edges (at iteration $i+1$) with interfering flows. Because the claim holds for ${\cal G}(i),$ we must have
\begin{align*}
\boldsymbol{F_{{\cal S},t}}&\boldsymbol{(i+1)}\\
&=\left(\left(I^j~I^j~..~I^j~|~C(i)\right)A_t(i)^{-1}~|~V\right)L\\
&=\left(I^j~I^j~..~I^j~|~C(i)~|~V\right)\left(\begin{array}{cc}A_t(i)^{-1} & \boldsymbol{0}\\ \boldsymbol{0} & I_{V}\end{array}\right)L,
\end{align*}
where the $\boldsymbol{0}$s represent zero matrices of appropriate sizes, and $I_{V}$ is the identity matrix such that $V=VI_{V}.$

The matrix 
\[
B=\left(\begin{array}{cc}A_t(i)^{-1} & \boldsymbol{0} \\ \boldsymbol{0} & I_{V}\end{array}\right)L
\]
is invertible. Let $C(i+1)=\left(C(i)~|~V\right).$ Let $A_t(i+1)=B^{-1}.$ 
\[
\boldsymbol{F_{{\cal S},t}(i+1)}A_t(i+1)=\left(I^j~I^j~..~I^j~|~C(i+1)\right).
\]
By induction on $i$ ($i=1,2,...,N_C$) the lemma is proved. 
\end{IEEEproof}

Let $\boldsymbol{F_{t}(i)}$ denote the matrix $\boldsymbol{F_{t}}$ at the end of the $i^{th}$ iteration. Let $\boldsymbol{F_{supp(\boldsymbol{z}),{t}}(i)}$ denote the submatrix of $\boldsymbol{F_{t}(i)}$ consisting of those rows of $\boldsymbol{F_{t}}$ which are indexed by $supp(\boldsymbol{z}),$ for some error vector $\boldsymbol{z}.$ 

The following lemma is now a direct consequence of Lemma \ref{lemma4} and Lemma \ref{lemmadecoding} and will help us to connect our multiple-unicast algorithm to Algorithm \ref{alg:necc}. 
\begin{lemma}
\label{multipleunicastequivalence}
Let $\bar{A}_t(i)$ be the matrix consisting of the first $2\alpha+1$ columns of $A_t(i).$ The sink $t$ can successfully decode its demanded $j^{th}$ information symbol ($\boldsymbol{{\cal D}_{t}}=j$) in ${\cal G}(i)$ if the square matrix
\[
\left(
\begin{array}{c}1~~1~~.~.~.~~1\\
\hline\vspace{-0.3cm}\\
\boldsymbol{F_{supp(\boldsymbol{z}),{t}}(i)}\bar{A}_t(i)
\end{array}\right)
\]
is full-rank for each error vector $\boldsymbol{z}$ such that $supp(\boldsymbol{z}) \in {\cal F},$ the set of all possible error patterns. 
\end{lemma}
\begin{IEEEproof}
If the given matrix is full-rank for all possible errors, then we must have for any such error vector $\boldsymbol{z}$
\[
cols(I_{\boldsymbol{{\cal D}_{t}}}) \subseteq \left\langle\left(
\begin{array}{c}I^j~~I^j~~.~.~.~~I^j\\
\hline\vspace{-0.3cm}\\
\boldsymbol{F_{supp(\boldsymbol{z}),{t}}(i)}\bar{A}_t(i)
\end{array}\right)\right\rangle,
\]
as $I_{\boldsymbol{{\cal D}_{t}}}=
\left(
\begin{array}{c}I^j\\
\hline\vspace{-0.3cm}\\
\boldsymbol{0}
\end{array}\right)$ and as $\left(
\begin{array}{c}I^j~~I^j~~.~.~.~~I^j\\
\hline\vspace{-0.3cm}\\
\boldsymbol{F_{supp(\boldsymbol{z}),{t}}(i)}\bar{A}_t(i)
\end{array}\right)$ has exactly $2\alpha+1$ non-zero rows. But then, this means
\begin{align}
\nonumber
cols(I_{\boldsymbol{{\cal D}_{t}}}) & \subseteq \left\langle\left(
\begin{array}{c}I^j~~I^j~~.~.~.~~I^j~|~C(i)\\
\hline\vspace{-0.3cm}\\
\boldsymbol{F_{supp(\boldsymbol{z}),{t}}(i)}A_t(i)
\end{array}\right)\right\rangle\\
\label{eqn103}
& \subseteq \left\langle\left(\left(
\begin{array}{c}\boldsymbol{F_{{\cal S},t}}\\
\boldsymbol{F_{supp(\boldsymbol{z}),{t}}(i)}
\end{array}\right)A_t(i)\right)\right\rangle\\
\label{eqn104}
& \subseteq \left\langle\left(
\begin{array}{c}\boldsymbol{F_{{\cal S},t}}\\
\boldsymbol{F_{supp(\boldsymbol{z}),{t}}(i)}
\end{array}\right)\right\rangle,
\end{align}
where (\ref{eqn103}) is because of Lemma \ref{lemma4} and (\ref{eqn104}) is because $A_t(i)$ is full-rank. By Lemma \ref{lemmadecoding}, this means that the demand $\boldsymbol{{\cal D}_{t}}=j$ can be successfully decoded by the sink $t.$
\end{IEEEproof}

Lemma \ref{multipleunicastequivalence} connects the problem of designing a multiple-unicast network-error correcting code for ${\cal G}(i)$ with maintaining the full-rankness of a set of matrices as in Algorithm \ref{alg:necc}. 
Thus, Algorithm \ref{alg:necc} can be used to design a multiple-unicast network-error correcting code for ${\cal G}(i)$ by modifying Step $2(iii)$ to fix the local encoding kernels at the new coding node such that the following condition is satisfied. 
\begin{itemize}
\item The matrix $\left(
\begin{array}{c}1~~1~~.~.~.~~1\\
\hline\vspace{-0.3cm}\\
\boldsymbol{F_{supp(\boldsymbol{z}),{t}}(i)}\bar{A}_t(i)
\end{array}\right)$ is full-rank for each sink $t$ and for each error pattern $supp(\boldsymbol{z}) \in {\cal F},$ at each iteration $i=1,2,...,N_C.$ 
\end{itemize}

As in the multicast case, we have that the maximum number of edges at any particular iteration is less than $|{\cal E}|=N+N_C.$ With $\eta=\left(\begin{array}{c}|{\cal E}| \\ 2\alpha \end{array}\right),$ we invoke the result from \cite{Mat} to note that a suitable choice of the local encoding kernels is possible if $q \geq |{\cal T}|\eta=n\eta.$ The complexity of our multiple-unicast algorithm is $O\left(nN\eta\left(|{\cal E}|n\eta+|{\cal E}|+N+2\alpha\right)\right),$ again assuming that the other steps in the algorithm can be done in constant time or with negligible complexity compared to \textbf{Step 4} and \textbf{Step 6}. 
\section{Insufficiency of Linear Network-Error Detecting and Correcting Codes}
\label{secinsufficiency}

In \cite{DFZ3}, it was shown that there exist networks for which linear network codes (linearity in a very general sense) are insufficient to achieve the maximum rate of information transmission to the sinks, when compared to general network coding (including nonlinear schemes). In other words, the \textit{network coding capacity} of a network could be strictly greater than the \textit{linear network coding capacity} of the network. A network for which linear network coding cannot achieve network coding capacity was explicitly constructed in \cite{DFZ3}. The network in \cite{DFZ3} was constructed by `conjoining' two subnetworks, of which one is linearly solvable over fields of characteristic two, and the other is linearly solvable over fields of odd characteristic. The two subnetworks were constructed based on results from matroid theory, in particular the Fano and the non-Fano matroids \cite{Oxl}. The matrix $A$ shown below considered over any field of characteristic two (for example, $\mathbb{F}_2$)  is a representation for the Fano matroid. 
\begin{equation}
\label{fanononfanorepresentation}
A=\left(
\begin{array}{ccccccc}
1 & 0 & 0 & 1 & 0 & 1 & 1 \\
0 & 1 & 0 & 1 & 1 & 0 & 1 \\
0 & 0 & 1 & 0 & 1 & 1 & 1 
\end{array}
\right).
\end{equation}
The matrix $A$ is also a representation for the non-Fano matroid except that it is over a field with characteristic not equal to two (for example, $\mathbb{F}_3$). Combining the two subnetworks, the conjoined network is shown to be linearly unsolvable. We refer the reader to \cite{DFZ3} for more details. 

Because of the fact that network coding is a special case of network-error correction (or equivalently network-error detection), it is to be expected that linear network-error correcting (detecting) codes must be insufficient for solving network-error correction (detection) problems on general networks. In Subsection \ref{insuffnetwork}, we present an explicit example network for which linear network-error detection (for the case of single edge errors) is not sufficient, using simple extensions of the networks shown in \cite{DFZ3}. The reason for choosing such simple extensions is two fold. Firstly, the networks chosen are sufficient to prove the insufficiency claim. The second reason, as the verification of the linear nonsolvability of the chosen networks will make it clear, is that rigorously proving that linear network-error correcting codes are not sufficient for a particular network can require many times the computations necessary for showing linear network coding is insufficient. Choosing extensions of the networks shown in \cite{DFZ3} to demonstrate the insufficiency of linear network coding makes our job easier. For these two reasons, we work with the chosen networks which are simple extensions of those from \cite{DFZ3}. Nevertheless, it is certainly possible to construct more complicated networks for which linear network-error correction and detection are insufficient. 

In the following subsections, we construct the network for which linear network-error detection is insufficient, while a nonlinear scheme is shown to provide the required error detection. We combine simple extensions of the networks shown in \cite{DFZ3} to create the network that we are looking for. 

\subsection{A network solvable only on alphabets of characteristic two}
\begin{figure}
\centering
\includegraphics[width=3.4in]{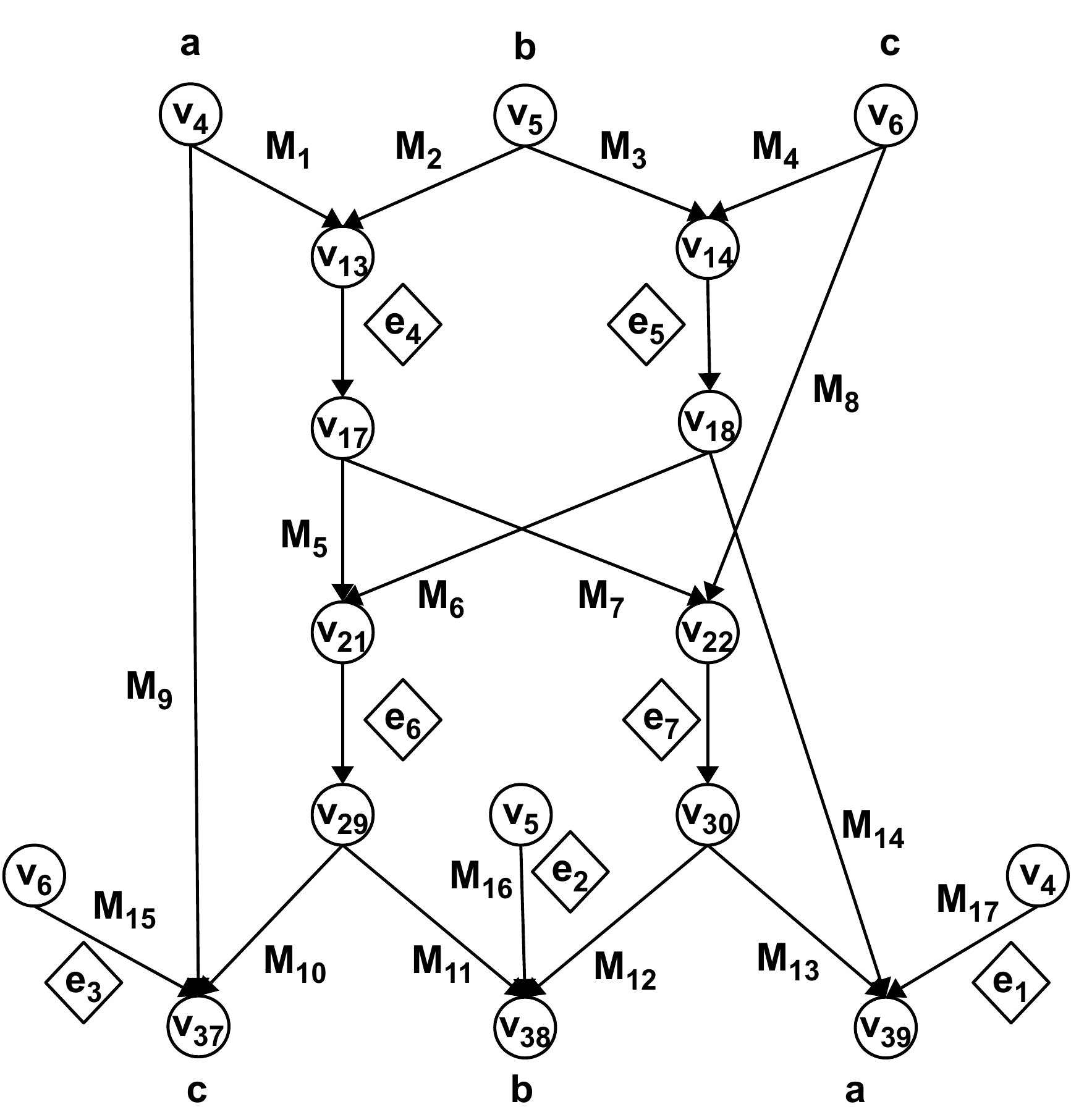}
\caption{The network-error detection network $\tilde{\cal N}_1$ which is solvable only over fields of characteristic two. It is a matroidal $1$-error detecting network associated with the matroid ${\cal M}_{\tilde{\cal N}_1}$ whose representation is shown in (\ref{representationn1}).}	
\label{fig:subnet1}	
\end{figure}

Consider the network $\tilde{\cal N}_1$ shown in Fig. \ref{fig:subnet1}. 
%
The nodes $v_4,$ $v_5$ and $v_6$ generate the messages $a,$ $b$ and $c$ (over some finite field) respectively. The sinks $v_{37},$ $v_{38},$ and $v_{39}$ demand the symbols $c,$ $b,$ and $a$ respectively. Some of the edges in the network are marked by the values $M_i$ which are coefficients of some arbitrary scalar linear network code for the network. Any edge which is not marked by a coefficient is assumed to have the identity element as its coefficient, meaning it just forwards the information from its tail node to the head node. It can be easily seen that these $M_i$s are sufficient to characterise any scalar linear network code for $\tilde{\cal N}_1.$ Each of the sinks have a direct edge from the corresponding node generating their demands, indicated by a duplicate node along with the edges $e_1,$ $e_2$ and $e_3$. The network ${\cal N}_1$ from \cite{DFZ3} is simply the network obtained from $\tilde{\cal N}_1$ by the deletion of the edges $e_1, e_2,$ and $e_3.$ Thus $\tilde{\cal N}_1$ is a simple extension of the network ${\cal N}_1$ from \cite{DFZ3}. We now prove the following lemma.
\begin{lemma}
\label{lemmanetworkn1}
A single edge network-error detection code over a finite field exists for $\tilde{\cal N}_1$ if and only if the finite field used has characteristic two. 
\end{lemma}

\begin{IEEEproof}
%

\textit{Only if part:}

Let the network coding coefficients $M_i$s define a single edge network-error detecting code over some field $\mathbb{F}.$ Note that there are exactly two paths from any source to the corresponding sink, one through the network coded portion of the network and the other through the direct edges $e_1, e_2,$ and $e_3.$ Therefore it is clear that for detecting single-edge errors, we require $M_{15}, M_{16}, M_{17}$ to be nonzero. Thus, we see that the sinks $v_{37}, v_{38}$ and $v_{39}$ can decode the required symbols by observing the symbols on the direct edges $e_3, e_2$ and $e_1$ from $v_6, v_5$ and $v_4$ respectively, as long as these edges are not in error. 

In order to show that the characteristic of the field used should be two for the network code defined using $M_i$s to be a single edge network-error detecting code, we consider the single edge errors at the edges  $e_1, e_2$ and $e_3.$

Consider that the only error in the network occurs in edge $e_3.$ Then the matrix $\left(
\begin{array}{c}
\boldsymbol{F_{{\cal S},t}} \\
\boldsymbol{F_{supp(\boldsymbol{z}),{t}}}
\end{array}
\right)
$ corresponding to $supp(\boldsymbol{z})=e_3$ at the sink $t=v_{37}$ is 
\[
\boldsymbol{F_{v_{37}}^{e_3}} = 
\left(
\begin{array}{ccc}
M_9 & M_1M_5M_{10} & 0 \\
0 & (M_2M_5+M_3M_6)M_{10} & 0 \\
0 & M_4M_6M_{10} & M_{15} \\
\hline
0 & 0 & 1 
\end{array}
\right),
\]
where the ordering of the columns adopted in the above matrix corresponds to the incoming edges at the sink given as follows.
\[
In(v_{37})=\{v_4\rightarrow v_{37},v_{29}\rightarrow v_{37},e_3\}.
\]
By Lemma \ref{lemmadecoding}, for some $x_1,x_2,$ and $x_3$ belonging to the finite field, we must have
\[
\boldsymbol{F_{v_{37}}^{e_3}}(x_1~x_2~x_3)^T = 
\left(
\begin{array}{c}
0 \\
0 \\
1 \\
0
\end{array}
\right).
\]
Thus we must have 
\begin{align*}
M_9x_1 + M_1M_5M_{10}x_2 &= 0. \\
M_2M_5M_{10}x_2+M_3M_6M_{10}x_2 & = 0. \\
M_4M_6M_{10}x_2+M_{15}x_3 &=1.  \\
x_3 & = 0.
\end{align*}
Let $M_9x_1=M_9'$, and $M_{10}x_2=M_{10}'.$ Then we have
\begin{align}
\label{eqnno1}
M_9' + M_1M_5M_{10}' &= 0. \\
\label{eqnno2}
M_2M_5M_{10}'+M_3M_6M_{10}' & = 0. \\
\label{eqnno3}
M_4M_6M_{10}'&=1. 
\end{align}

The transfer matrix corresponding to error at $e_2$ at the sink $t=v_{38}$ ($In(t)=\{v_{29}\rightarrow v_{38},v_{30}\rightarrow v_{38},e_2\}$) is
\[
\boldsymbol{F_{v_{38}}^{e_2}} = 
\left(
\begin{array}{ccc}
M_1M_5M_{11} & M_1M_7M_{12} & 0 \\
(M_2M_5+M_3M_6)M_{11} & M_2M_7M_{12} & M_{16} \\
M_4M_6M_{11} & M_8M_{12} & 0 \\
\hline
0 & 0 & 1 
\end{array}
\right).
\]
As before, by Lemma \ref{lemmadecoding}, for some finite field coefficients $y_1,y_2,$ and $y_3,$ we must have
\begin{align*}
M_1M_5M_{11}y_1 + M_1M_7M_{12}y_2 & = 0. \\
(M_2M_5+M_3M_6)M_{11}y_1 + M_2M_7M_{12}y_2+M_{16}y_3 & = 1.\\
M_4M_6M_{11}y_1+M_8M_{12}y_2 & = 0.\\
y_3& = 0.
\end{align*}
Letting $M_{11}y_1 = M_{11}'$ and $M_{12}y_2=M_{12}',$ we have
\begin{align}
\label{eqnno4}
M_1M_5M_{11}' + M_1M_7M_{12}' & = 0. \\
\label{eqnno5}
M_2M_5M_{11}'+M_3M_6M_{11}'+M_2M_7M_{12}' & = 1. \\
\label{eqnno6}
M_4M_6M_{11}'+M_8M_{12}' &=0.
\end{align}
The transfer matrix corresponding to error at $e_1$ at the sink $t=v_{39}$ ($In(t)=\{v_{30}\rightarrow v_{39},v_{18}\rightarrow v_{39},e_1\}$) is
\[
\boldsymbol{F_{v_{39}}^{e_1}} = 
\left(
\begin{array}{ccc}
M_1M_7M_{13} & 0 & M_{17} \\
M_2M_7M_{13} & M_3M_{14} & 0 \\
M_8M_{13} & M_4M_{14} & 0 \\
\hline
0 & 0 & 1 
\end{array}
\right).
\]
Again, by Lemma \ref{lemmadecoding}, for some finite field coefficients $z_1,z_2,z_3,$ we must have
\begin{align*}
M_1M_7M_{13}z_1 + M_{17}z_3 & = 1. \\
M_2M_7M_{13}z_1 + M_3M_{14}z_2 & = 0. \\
M_8M_{13}z_1 + M_4M_{14}z_2 &= 0. \\
z_3&=0.
\end{align*}
Letting $M_{13}z_1 = M_{13}'$ and $M_{14}z_2=M_{14}',$ we have
\begin{align}
\label{eqnno7}
M_1M_7M_{13}'& = 1. \\
\label{eqnno8}
M_2M_7M_{13}' + M_3M_{14}' & = 0.\\
\label{eqnno9}
M_8M_{13}' + M_4M_{14}' &= 0.
\end{align}
Equations similar to (\ref{eqnno1})-(\ref{eqnno9}) were derived in \cite{DFZ3} for the network ${\cal N}_1.$ Mimicking the arguments in \cite{DFZ3}, we now show that the characteristic of the finite field used must be two.

From (\ref{eqnno3}) and (\ref{eqnno7}), we must have that the matrices $M_1, M_4, M_6, M_7, M_{10}',$ and $M_{13}'$ are all invertible. By (\ref{eqnno2}), we must then have $M_2M_5+M_3M_6=0.$ Thus by (\ref{eqnno5}), we must have 
\begin{equation}
\label{eqnno10}
M_2M_7M_{12}'=1.
\end{equation}
and therefore $M_2$ and $M_{12}'$ are invertible. By (\ref{eqnno4}), $M_5M_{11}'=-M_7M_{12}'$ and thus $M_5$ and $M_{11}'$ are invertible. Furthermore, $M_3M_{14}'=-M_2M_7M_{13}'$ by (\ref{eqnno8}), and $M_9'=-M_1M_5M_{10}'$ by (\ref{eqnno1}). Thus $M_3, M_{14}',$ and $M_9'$ are invertible. As $M_8=-M_4M_{14}'M_{13}'^{-1}$ by (\ref{eqnno9}), the matrix $M_8$ is invertible too. Thus all the matrices in the equations (\ref{eqnno1})-(\ref{eqnno9}) are invertible. 

From (\ref{eqnno4}), we have 
\begin{align*}
0  &= M_5M_{11}'+M_7M_{12}'\\
&= M_5M_{11}'+M_2^{-1}(M_2M_7M_{12}')\\
& = M_5M_{11}'+M_2^{-1}.
\end{align*}
where the last equality follows from (\ref{eqnno10}).

Thus we have
\begin{align}
\label{eqnno11}
M_2M_5M_{11}' = -1.
\end{align}

From (\ref{eqnno6}), we have
\begin{align}
\nonumber
0 & = M_4M_6M_{11}'+M_8M_{12}' \\
& = M_4M_3^{-1}M_3M_6M_{11}'-M_4M_{14}'M_{13}'^{-1}M_{12}', 
\end{align}
where the last equality follows from (\ref{eqnno9}). Now, using (\ref{eqnno2}) and (\ref{eqnno8}), we have
\begin{align}
\nonumber
0& = -M_4M_3^{-1}M_2M_5M_{11}'+M_4M_3^{-1}M_2M_7M_{13}'M_{13}'^{-1}M_{12}' \\
\nonumber
& = M_4M_3^{-1}(M_2M_7M_{12}'-M_2M_5M_{11}') \\
\nonumber
&= M_4M_3^{-1}(1-M_2M_5M_{11}'), 
\end{align}
where the last equality follows from (\ref{eqnno10}). Thus we must have
\begin{align}
\label{eqnno12}
M_2M_5M_{11}'=1.
\end{align}
Thus, from (\ref{eqnno11}) and (\ref{eqnno12}), we see that we require $1=-1.$ This is true only in a field of characteristic two.

\textit{If part:}

It is easy to verify that using $M_i=1 \in \mathbb{F}_{2^m}$ (for any $m$) for all $i$ results in a single edge network-error detecting code for $\tilde{\cal N}_1.$ 
\end{IEEEproof}

In the case of a network code with all $M_i=1 \in \mathbb{F}_{2^m},~\forall i,$ we now argue that the network $\tilde{\cal N}_1$ is a matroidal $1$-error detecting network with respect to the vector matroid ${\cal M}_{\tilde{\cal N}_1}$ of the matrix over $\mathbb{F}_{2^m}$ shown below. 
\begin{equation}
\label{representationn1}
\left(
\begin{array}{cccccccc}
& 1 & 0 & 0 & 1 & 0 & 1 & 1 \\
& 0 & 1 & 0 & 1 & 1 & 0 & 1 \\
& 0 & 0 & 1 & 0 & 1 & 1 & 1 \\
&  &  &  &  &  &  &  \\
& 1 & 0 & 0 & 0 & 0 & 0 & 0 \\
I_{10} & 0 & 1 & 0 & 0 & 0 & 0 & 0 \\
& 0 & 0 & 1 & 0 & 0 & 0 & 0 \\
& 0 & 0 & 0 & 1 & 0 & 1 & 1 \\
& 0 & 0 & 0 & 0 & 1 & 1 & 0 \\
& 0 & 0 & 0 & 0 & 0 & 1 & 0 \\
& 0 & 0 & 0 & 0 & 0 & 0 & 1 
\end{array}
\right)
\end{equation}
Let the function with respect to which the matroid ${\cal M}_{\tilde{\cal N}_1}$ is associated be
\begin{align}
\label{f1}
f_1:\mu_{\tilde{\cal N}_1}\cup{\cal E}_{{\tilde{\cal N}_1}}\rightarrow E({\cal M}_{\tilde{\cal N}_1}).
\end{align}
The function $f_1$ maps the input symbols ($\mu_{\tilde{\cal N}_1}=\{a,b,c\}$) and the edges of $\tilde{\cal N}_1$ to the elements of the groundset $E({\cal M}_{\tilde{\cal N}_1}).$ The labeling on the columns (i.e., the mapping given by $f_1$) of the matrix given in (\ref{representationn1}) is as follows. The first three columns correspond to the inputs $\mu_{\tilde{\cal N}_1}$. The next seven columns constitute the basis elements of the errors at $\{e_i:i=1,2,..,7\}$ as shown in Fig. \ref{fig:subnet1}. The last seven columns correspond to the linear combination of the input symbols and the errors flowing on these edges. Though there are a total of $21$ edges in $\tilde{\cal N}_1,$ these seven edges are sufficient to characterise the matroid associated with the single edge network-error detecting code on $\tilde{\cal N}_1.$ It is easy to verify that the function $f_1$ and the matroid $\tilde{\cal M}_{\tilde{\cal N}_1}$ satisfy all the requirements of Definition \ref{matroidalerrornetworkdefinition} for a single edge network-error detecting code. We list the elements of the ground set of ${\cal M}_{\tilde{\cal N}_1}$ in the ordering of the columns shown in (\ref{representationn1}) as follows. 
\begin{align}
\nonumber
E({\cal M}_{\tilde{\cal N}_1}) = \left\{x_i:i=1,2,3\right\}&\cup\left\{y_i:i=1,2,...,7\right\} \\ 
\label{groundsetn1}
&\cup\left\{y_i':i=1,2,...,7\right\}.
\end{align}

Finally, we have the following lemma which follows from Lemma \ref{lemmanetworkn1} and the discussion above. 
\begin{lemma}
\label{n1representation}
The network $\tilde{\cal N}_1$ is a matroidal 1-error detecting network associated with a $\mathbb{F}_2$-representable matroid.
\end{lemma}
\subsection{A network not solvable on alphabets of characteristic two}
Consider the network $\tilde{\cal N}_2$ shown in Fig. \ref{fig:subnet2}. The network has five sources $v_7, v_8, v_3, v_{11}$ and $v_{12}$ generating the information symbols $a, b, c, d,$ and $e$ respectively. There are seven sinks $v_{40}, v_{41}, v_{42}, v_{43}, v_{44}, v_{45}$ and $v_{46}$ demanding the symbols $c, b, a, c, e, d,$ and $c$ respectively. The network ${\cal N}_2$ of \cite{DFZ3} is the subnetwork of $\tilde{\cal N}_2$ consisting of all nodes and edges except the direct edges from $v_3, v_8, v_7, v_3, v_{12}, v_{11},$ and $v_3$ to the sinks. We seek the conditions to be satisfied by the finite field over which a single edge network-error detection code can be designed for $\tilde{\cal N}_2.$ 

Again, it is easy to verify that assuming all $1$s from a finite field with characteristic not equal to two as the network coding coefficients of $\tilde{\cal N}_2$ results in a single edge network-error detection code. The network $\tilde{\cal N}_2$ is then a matroidal $1$-error detecting network associated with the matroid ${\cal M}_{\tilde{\cal N}_2}$ whose representation (over any field with characteristic not equal to two) is shown in (\ref{representationn2}) at the top of the next page. 
\begin{figure*}
\begin{equation}
\label{representationn2}
\left(
\begin{array}{c|c|ccccccccccccccc}
& & 1 & 0 & 0 & 0 & 0 & 0 & 0 & 1 & 1 & 1 & 0 & 0 & 0 & 0 & 0 \\
& & 0 & 1 & 0 & 0 & 0 & 0 & 0 & 1 & 1 & 0 & 1 & 0 & 0 & 0 & 0 \\
I_5 & \boldsymbol{0} & 0 & 0 & 1 & 1 & 1 & 0 & 0 & 1 & 0 & 1 & 1 & 1 & 1 & 1 & 0 \\
& & 0 & 0 & 0 & 0 & 0 & 1 & 0 & 0 & 0 & 0 & 0 & 1 & 1 & 0 & 1 \\
& & 0 & 0 & 0 & 0 & 0 & 0 & 1 & 0 & 0 & 0 & 0 & 1 & 0 & 1 & 1 \\
\hline
\boldsymbol{0} & I _{15} & & & & & & & I_{15} & & & & & & & & 
\end{array}
\right)
\end{equation}
\hrule
\end{figure*}
The corresponding function $f_2$ is given as
\begin{align}
\label{f2}
f_2:\mu_{\tilde{\cal N}_2}\cup{\cal E}_{{\tilde{\cal N}_2}}\rightarrow E({\cal M}_{\tilde{\cal N}_2}),
\end{align}
where $\mu_{\tilde{\cal N}_2}=\{a,b,c,d,e\}$ is the collection of the input symbols.
As with $\tilde{\cal N}_1,$ not all the edges of $\tilde{\cal N}_2$ are considered in the representation of ${\cal M}_{\tilde{\cal N}_2}.$ The columns of the matrix shown in (\ref{representationn2}) (and therefore the mappings of the function $f_2$) are indexed as follows. The first five columns correspond to the five input symbols. The next $15$ columns correspond to the error basis elements at the edges $\{e_i:i=1,2,..,15\}$ as shown in Fig. \ref{fig:subnet2}. The final $15$ columns correspond to the linear combination of the inputs and error symbols flowing at these $15$ edges. We list the elements of the ground set of ${\cal M}_{\tilde{\cal N}_2}$ in the ordering of the columns shown in (\ref{representationn2}) as follows. 
\begin{align}
\nonumber
E({\cal M}_{\tilde{\cal N}_2}) = \left\{x_i:i=1,2,..,5\right\}&\cup\left\{z_i:i=1,2,...,15\right\} \\ 
\label{groundsetn2}
&\cup\left\{z_i':i=1,2,...,15\right\}.
\end{align}

As with $\tilde{\cal N}_1,$ it can be seen that in the absence of errors in the additional direct edges to the sinks (those not in ${\cal N}_2$), the sinks of $\tilde{\cal N}_2$ can straight away decode their required demands. Assuming single edge network-errors on these additional edges and using arguments equivalent to those in \cite{DFZ3} (as was done in the proof of Lemma \ref{lemmanetworkn1}), we have the following lemma, which we state without proof.

\begin{lemma}
\label{lemmanetworkn2}
The network $\tilde{\cal N}_2$ has a single edge network-error detecting code if and only if the finite field used has characteristic not equal to two. 
\end{lemma}
The following lemma follows directly from Lemma \ref{lemmanetworkn2} and the preceding discussion. 
\begin{lemma}
\label{n2representation}
The network $\tilde{\cal N}_2$ is a matroidal 1-error detecting network associated with a $\mathbb{F}_3$-representable matroid.
\end{lemma}
\begin{figure*}
\centering
\includegraphics[width=6in]{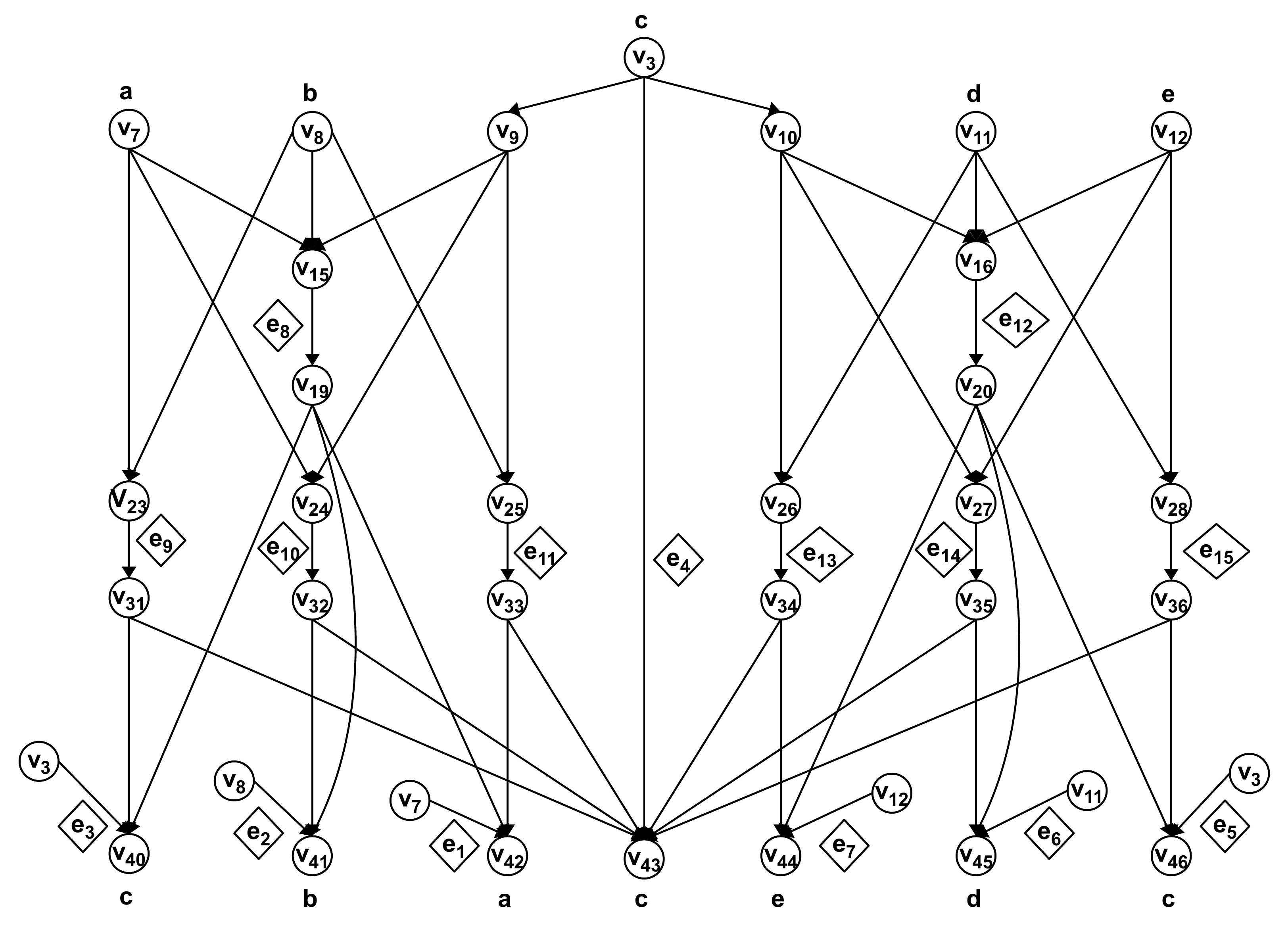}
\caption{The network-error detection network $\tilde{\cal N}_2$ which is not solvable over fields with characteristic two. This network is a matroidal $1$-error detecting network associated with the matroid ${\cal M}_{\tilde{\cal N}_2}$ whose representation is shown in (\ref{representationn2}).}	
\label{fig:subnet2}	
\hrule
\end{figure*}

It can be seen from the proof of Lemma \ref{lemmanetworkn1} that particular error patterns were considered in order to verify whether the linear network code defined over a particular alphabet satisfies the required network-error detection (correction) properties. Given an arbitrary network, it may be necessary to consider all possible error patterns, i.e., $\left(\begin{array}{c}{\cal E} \\ \beta \end{array} \right)$ of them to verify the $\beta$ network-error detection capability. This is why proving insufficiency of linear network coding for network-error correction or detection could be computationally much harder than proving insufficiency of linear network codes for network coding with no errors. 

\subsection{A network for which linear network-error detection is insufficient}
\label{insuffnetwork}
\begin{figure*}
\centering
\includegraphics[width=7in]{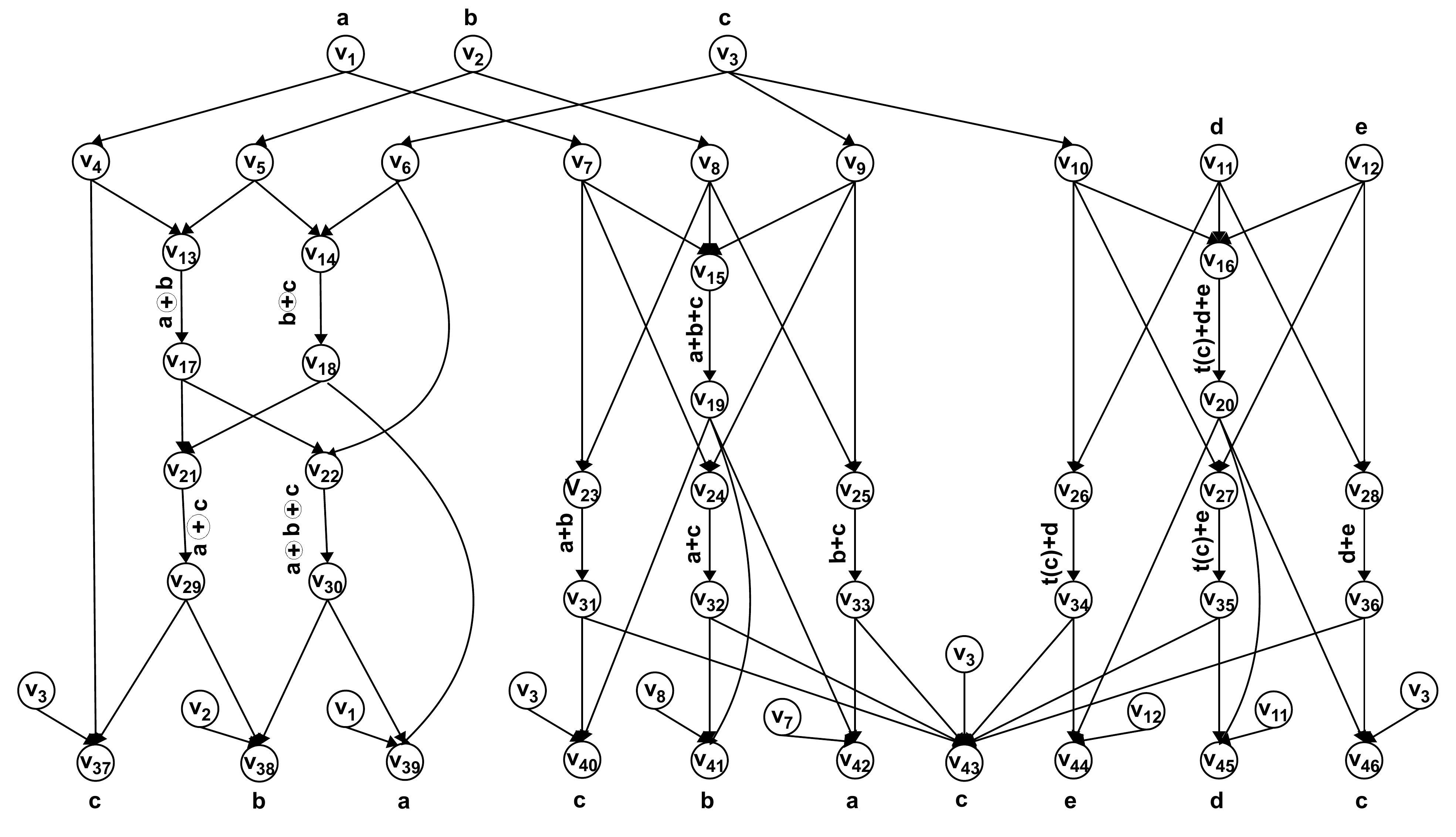}
\caption{The network-error detection network $\tilde{\cal N}_3$ which does not have a linear single edge network-error detecting code. The code shown is a nonlinear single edge network-error detecting code. This network is a matroidal $1$-error detecting network associated with the matroid ${\cal M}_{\tilde{\cal N}_3},$ which is an amalgam of the matroids ${\cal M}_{\tilde{\cal N}_1}$ and ${\cal M}_{\tilde{\cal N}_2}.$}	
\label{fig:nonlinearnet}
~\\	
\hrule
\end{figure*}

We now present the network $\tilde{\cal N}_3$ shown in Fig. \ref{fig:nonlinearnet} for which linear network coding is insufficient to achieve the sinks demands in the presence of network-errors. The network $\tilde{\cal N}_3$ is a conjoining of the network $\tilde{\cal N}_1$ and $\tilde{\cal N}_2$ with the exception of a few additional dummy edges. Thus, we assume ${\cal E}_{\tilde{\cal N}_3}={\cal E}_{\tilde{\cal N}_1}\cup {\cal E}_{\tilde{\cal N}_2}.$ We ignore the dummy edges for the sake of the clarity. The network ${\cal N}_3$ shown in \cite{DFZ3} is equivalent to $\tilde{\cal N}_3$ except for the direct edges to the sinks from the corresponding sources. Because of Lemmas \ref{lemmanetworkn1} and Lemma \ref{lemmanetworkn2}, the network $\tilde{\cal N}_3$ does not have a linear single edge network-error detecting code over any field. 

However, there is a nonlinear single edge network-error detecting code over an alphabet $\cal A$ of size $4$ , the corresponding edge functions of which are shown along the edges of $\tilde{\cal N}_3$ in Fig. \ref{fig:nonlinearnet}. Except for the additional direct edges from the sources to the corresponding sinks, the network coding functions on $\tilde{\cal N}_3$ are adopted from the network code for ${\cal N}_3$ in \cite{DFZ3}. All the missing edge functions are considered to be identity. The symbols $+$ and $-$ indicate the addition and subtraction in the ring ${\mathbb Z}_4,$ while the symbols $\oplus$ indicates the bitwise XOR operation in ${\mathbb Z}_2 \oplus {\mathbb Z}_2.$ In other words, for any two elements $a,b \in {\cal A}$, the element $a+b$ and $a-b$ indicate the sum of $a$ and $b$ and the difference between $a$ and $b$ viewing them as elements from ${\mathbb Z}_4.$ The element $a\oplus b$ indicates the bitwise XOR between $a$ and $b$ viewing them as elements from ${\mathbb Z}_2 \oplus {\mathbb Z}_2.$  For some $a\in {\cal A},$ $t(a)$ is the element of ${\cal A}$ obtained by switching the components of $a$ considered as element of ${\mathbb Z}_2 \oplus {\mathbb Z}_2.$ The nonlinearity of the network-error correction code comes from the nonlinearity of the function $t,$ and because $\oplus$ is linear in ${\mathbb Z}_2 \oplus {\mathbb Z}_2$ but nonlinear in ${\mathbb Z}_4,$ while $+$ and $-$ are linear in ${\mathbb Z}_4$ but nonlinear in ${\mathbb Z}_2 \oplus {\mathbb Z}_2.$ Using the arguments developed in \cite{DFZ3}, it is straightforward to show that these coding functions define a single edge network-error detection code for $\tilde{\cal N}_3.$ 

We can now ask the question - \textit{Is the network $\tilde{\cal N}_3$ a matroidal 1-error detecting network?} If the answer is yes, then it would mean that our definition of a matroidal error detecting network (Definition \ref{matroidalerrornetworkdefinition}) has a wider scope and is not limited to linear network-error detection and representable matroids. Also, an equivalent question can be raised about the network ${\cal N}_3$ shown in \cite{DFZ3} -  \textit{Is the network ${\cal N}_3$ a matroidal network?} This second question is left unanswered in both \cite{DFZ3} (where the insufficiency results for linear network coding in ${\cal N}_3$ was first presented) and in \cite{DFZ} (where the matroidal connections to the construction of ${\cal N}_1$ and ${\cal N}_2$ were discussed). We answer these questions in the affirmative. In the rest of this Subsection, we obtain a matroid ${\cal M}_{\tilde{\cal N}_3}$ associated with which the network $\tilde{\cal N}_3$ is a  matroidal 1-error detecting network. That the network ${\cal N}_3$ of \cite{DFZ3} is matroidal follows easily. 

We first prove the following lemma. 
\begin{lemma}
\label{conditionsn3matroidal}
Let $E({\cal M}_{\tilde{\cal N}_3})=E({\cal M}_{\tilde{\cal N}_1})\cup E({\cal M}_{\tilde{\cal N}_2})$ be the groundset of a matroid ${\cal M}_{\tilde{\cal N}_3}.$ If the matroid ${\cal M}_{\tilde{\cal N}_3}$ satisfies the following two conditions
\begin{align}
\label{cond1}
{\cal M}_{\tilde{\cal N}_3}|_{E({\cal M}_{\tilde{\cal N}_1})} = {\cal M}_{\tilde{\cal N}_1}.\\
\label{cond2}
{\cal M}_{\tilde{\cal N}_3}|_{E({\cal M}_{\tilde{\cal N}_2})} = {\cal M}_{\tilde{\cal N}_2},
\end{align}
then the network $\tilde{\cal N}_3$ is matroidal 1-error detecting associated with ${\cal M}_{\tilde{\cal N}_3}.$
\end{lemma}
\begin{IEEEproof}
Let $\mu_{\tilde{\cal N}_3}=\mu_{\tilde{\cal N}_1}\cup \mu_{\tilde{\cal N}_2}.$ Clearly, $\mu_{\tilde{\cal N}_3}=\mu_{\tilde{\cal N}_2}.$ Let $f_3:\mu_{\tilde{\cal N}_3}\cup {\cal E}_{\tilde{\cal N}_3} \rightarrow E({\cal M}_{\tilde{\cal N}_3})$ be a function such that
\begin{align*}
&f_3(\mu_{\tilde{\cal N}_3})=f_2(\mu_{\tilde{\cal N}_2}),\\
&f_3(e) = f_1(e), \forall e\in {\cal E}_{\tilde{\cal N}_1}, \\
&f_3(e) = f_2(e), \forall e\in {\cal E}_{\tilde{\cal N}_2},
\end{align*}
where $f_1$ and $f_2$ are defined as in (\ref{f1}) and (\ref{f2}) respectively. Since $\tilde{\cal N}_3$ is a conjoining of the networks ${\tilde{\cal N}_1}$ and ${\tilde{\cal N}_2},$ i.e. as ${\cal E}_{\tilde{\cal N}_3}={\cal E}_{\tilde{\cal N}_1}\cup {\cal E}_{\tilde{\cal N}_2},$ it is clear that the function $f_3$ is well defined. 

Now, since the networks $\tilde{\cal N}_1$ and $\tilde{\cal N}_2$ are already matroidal 1-error detecting networks associated to ${\cal M}_{\tilde{\cal N}_1}$ (with respect to $f_1$) and ${\cal M}_{\tilde{\cal N}_2}$ (with respect to $f_2$) respectively, by the definition of $f_3$ it follows that $\tilde{\cal N}_3$ is a matroidal 1-error detecting network associated with $\tilde{\cal N}_3$ with respect to $f_3.$
\end{IEEEproof}

In order to show that $\tilde{\cal N}_3$ is matroidal 1-error detecting, we have to demonstrate a matroid which satisfies the conditions in Lemma \ref{conditionsn3matroidal}. In the rest of this subsection, we show that such a matroid can be obtained. We use Definition \ref{matroiddefnrank} of a matroid based on its rank function to obtain our matroid ${\cal M}_{\tilde{\cal N}_3}.$

Let $r:2^{E({\cal M}_{\tilde{\cal N}_1})\cup E({\cal M}_{\tilde{\cal N}_2})}\rightarrow \mathbb{Z}^+\cup\left\{0\right\}$  be a function defined as
\[
r(X) = r_{{\cal M}_{\tilde{\cal N}_1}}(X_1)+r_{{\cal M}_{\tilde{\cal N}_2}}(X_2)-r_{{\cal M}_{\tilde{\cal N}_2}}(X_{1,2}),
\]
where $X_1=X\cap E({\cal M}_{\tilde{\cal N}_1}), X_2=X\cap E({\cal M}_{\tilde{\cal N}_2}),$ and $X_{1,2}=X\cap E({\cal M}_{\tilde{\cal N}_1})\cap E({\cal M}_{\tilde{\cal N}_2})=X\cap\{x_1,x_2,x_3\} = X_1\cap X_2.$ The functions $r_{{\cal M}_{\tilde{\cal N}_1}}$ and $r_{{\cal M}_{\tilde{\cal N}_2}}$ are the rank functions of the matroids ${\cal M}_{\tilde{\cal N}_1}$ and ${\cal M}_{\tilde{\cal N}_2}$ respectively. Clearly the function $r$ is well defined. Also, as $r_{{\cal M}_{\tilde{\cal N}_2}}(X_2)\geq r_{{\cal M}_{\tilde{\cal N}_2}}(X_{1,2}),$ we must have $r(X)\geq 0,~\forall X.$ Also, for any $X\subseteq E({\cal M}_{\tilde{\cal N}_1})\cup E({\cal M}_{\tilde{\cal N}_2}),$ we note that
\begin{align}
\label{simplifyeqn}
r_{{\cal M}_{\tilde{\cal N}_2}}(X_{1,2})=r_{{\cal M}_{\tilde{\cal N}_1}}(X_{1,2})=|X_{1,2}|.
\end{align}

Now, suppose there is a matroid with the above function $r$ as its rank function. Then it can be seen that from the definition of the function $r$ that such a matroid satisfies the requirements of Lemma \ref{conditionsn3matroidal}. This is because for any $X\subseteq E({\cal M}_{\tilde{\cal N}_1}), r(X)=r_{{\cal M}_{\tilde{\cal N}_1}}(X),$ and for any $X\subseteq E({\cal M}_{\tilde{\cal N}_2}), r(X)=r_{{\cal M}_{\tilde{\cal N}_2}}(X).$ Thus the network $\tilde{\cal N}_3$ would be a matroidal $1$-error detecting network associated with such a matroid. We now prove the following lemma which shows that the function $r$ defines a matroid.
\begin{lemma}
\label{functionrmatroid}
The function $r$ is the rank function of a matroid.
\end{lemma}
\begin{IEEEproof}
We have to show that the function $r$ satisfies the properties \textbf{R1}, \textbf{R2}, and \textbf{R3} of Definition \ref{matroiddefnrank}. 

We first consider the condition \textbf{R1}. We have by the definition of $r,$ for $X\subseteq E({\cal M}_{\tilde{\cal N}_1})\cup E({\cal M}_{\tilde{\cal N}_2}),$
\[
r(X) = r_{{\cal M}_{\tilde{\cal N}_1}}(X_1)+r_{{\cal M}_{\tilde{\cal N}_2}}(X_2)-r_{{\cal M}_{\tilde{\cal N}_2}}(X_{1,2}),
\]
where $X_1=X\cap E({\cal M}_{\tilde{\cal N}_1}), X_2=X\cap E({\cal M}_{\tilde{\cal N}_2}),$ and $X_{1,2}=X\cap E({\cal M}_{\tilde{\cal N}_1})\cap E({\cal M}_{\tilde{\cal N}_2})=X\cap\{x_1,x_2,x_3\}.$ Because $r_{{\cal M}_{\tilde{\cal N}_1}}$ and $r_{{\cal M}_{\tilde{\cal N}_2}}$ are rank functions and by (\ref{simplifyeqn}), we must have
\begin{align}
\nonumber
r(X) &\leq |X_1|+|X_2|-|X_{1,2}|\\
\nonumber
&\leq |X_1|+|(X_2-X_{1,2})\uplus X_{1,2}|-|X_{1,2}|\\
\label{no9eqn}
r(X) &\leq |X_1|+|X_2-X_{1,2}|=|X|.
\end{align}
We have already seen that $r(X)\geq 0, \forall X.$ Along with (\ref{no9eqn}), this means that the function $r$ satisfies \textbf{R1}. Now we prove that \textbf{R2} holds. 

Let $X\subseteq Y\subseteq E({\cal M}_{\tilde{\cal N}_1})\cup E({\cal M}_{\tilde{\cal N}_2}).$ Then $X_1=X\cap E({\cal M}_{\tilde{\cal N}_1})\subseteq Y_1 = Y\cap E({\cal M}_{\tilde{\cal N}_1}).$ Similarly, $X_2\subseteq Y_2,$ and $X_{1,2}\subseteq Y_{1,2}.$

Let $B_{X_1}$ be a subset of $X_1$ of the largest size which is independent in ${\cal M}_{\tilde{\cal N}_1}$. Similarly let $B_{X_2}\subseteq X_2, B_{X_{1,2}}\subseteq X_{1,2}, B_{Y_1}\subseteq Y_1, B_{Y_2}\subseteq Y_2, B_{Y_{1,2}}\subseteq Y_{1,2}$ be some maximal independent subsets in the appropriate matroids. Because $X_{1,2}\subseteq X_1\subseteq Y_1,$ we can always find $B_{X_{1,2}}, B_{X_1}, B_{Y_1}$ such that $B_{X_{1,2}}\subseteq B_{X_1}\subseteq B_{Y_1},$ by repeated application of \textbf{I3} in Definition \ref{matroiddefnindp}. Similarly, we assume $B_{X_{1,2}}\subseteq B_{X_2}\subseteq B_{Y_2}$ and $B_{X_{1,2}}\subseteq B_{Y_{1,2}}.$

By the definition of $r,$ we have
\begin{align}
\nonumber
r(X)&=|B_{X_1}|+|B_{X_2}|-|B_{X_{1,2}}|\\
\nonumber
&=|B_{X_{1,2}}\uplus (B_{X_1}-B_{X_{1,2}})|+|B_{X_2}|-|B_{X_{1,2}}|\\
\nonumber
&=|B_{X_{1,2}}|+|B_{X_1}-B_{X_{1,2}}|+|B_{X_2}|-|B_{X_{1,2}}|\\
\label{no2eqn}
r(X)&=|B_{X_1}-B_{X_{1,2}}|+|B_{X_2}|.
\end{align}
As in (\ref{no2eqn}), we have 
\begin{align}
\nonumber
r(Y)&=|B_{Y_1}-B_{Y_{1,2}}|+|B_{Y_2}|\\
\nonumber
&\geq |B_{X_1}-B_{Y_{1,2}}|+|B_{Y_2}|~~~~~(\text{as}~B_{X_1}\subseteq B_{Y_1})  \\
\nonumber
&\geq |B_{X_1}-(B_{X_{1,2}}\uplus (B_{Y_{1,2}}-B_{X_{1,2}}))|+|B_{Y_2}|\\
\label{no3eqn}
r(Y)&\geq |B_{X_1}-B_{X_{1,2}}|-|B_{Y_{1,2}}-B_{X_{1,2}}|+|B_{Y_2}|.
\end{align}
We also have the following equations.
\begin{align}
\nonumber
|B_{Y_2}|&=|B_{X_2}\uplus (B_{Y_2}-B_{X_2})|\\
\nonumber
&=|B_{X_2}|+|B_{Y_2}-B_{X_2}|\\
\nonumber
&\geq |B_{X_2}|+|(B_{Y_2}-B_{X_2})\cap E({\cal M}_{\tilde{\cal N}_1})|\\
\label{no4eqn}
|B_{Y_2}|&\geq |B_{X_2}|+|B_{Y_{1,2}}-B_{X_{1,2}}|.
\end{align}
By (\ref{no3eqn}) and (\ref{no4eqn}), we have
\begin{align}
\label{no5eqn}
r(Y)\geq |B_{X_1}-B_{X_{1,2}}|+|B_{X_2}|.
\end{align}
Comparing (\ref{no2eqn}) and (\ref{no5eqn}), we have $r(X)\leq r(Y).$ Hence \textbf{R2} holds. Finally, we prove the condition \textbf{R3} also holds. 

Let $X,Y \subseteq E({\cal M}_{\tilde{\cal N}_1})\cup E({\cal M}_{\tilde{\cal N}_2}).$ By the definition of $r$ and (\ref{simplifyeqn}), we have
\begin{align}
\nonumber
r&(X)+r(Y)-r(X\cap Y)\\
\nonumber
&=r_{{\cal M}_{\tilde{\cal N}_1}}(X_1)+r_{{\cal M}_{\tilde{\cal N}_2}}(X_2)-|X_{1,2}|\\
\nonumber
&~~+r_{{\cal M}_{\tilde{\cal N}_1}}(Y_1)+r_{{\cal M}_{\tilde{\cal N}_2}}(Y_2)-|Y_{1,2}|\\
\label{no6eqn}
&~~-r_{{\cal M}_{\tilde{\cal N}_1}}(X_1\cap Y_1)-r_{{\cal M}_{\tilde{\cal N}_2}}(X_2\cap Y_2)+|X_{1,2}\cap Y_{1,2}|.
\end{align}
Also, we have
\begin{align}
\nonumber
r&(X\cup Y)\\
\nonumber
&=r_{{\cal M}_{\tilde{\cal N}_1}}(X_1\cup Y_1)+r_{{\cal M}_{\tilde{\cal N}_2}}(X_2\cup Y_2)-|X_{1,2}\cup Y_{1,2}|\\
\nonumber
&\leq r_{{\cal M}_{\tilde{\cal N}_1}}(X_1)+r_{{\cal M}_{\tilde{\cal N}_1}}(Y_1)-r_{{\cal M}_{\tilde{\cal N}_1}}(X_1\cap Y_1)\\
\nonumber
&~~+r_{{\cal M}_{\tilde{\cal N}_2}}(X_2)+r_{{\cal M}_{\tilde{\cal N}_2}}(Y_2)-r_{{\cal M}_{\tilde{\cal N}_2}}(X_2\cap Y_2)\\
\label{no7eqn}
&~~-|X_{1,2}\cup Y_{1,2}|,
\end{align}
where the last inequality follows from the fact that $r_{{\cal M}_{\tilde{\cal N}_1}}$ and $r_{{\cal M}_{\tilde{\cal N}_2}}$ are rank functions.

From (\ref{no6eqn}) and (\ref{no7eqn}), to show that $r(X\cup Y)\leq r(X)+r(Y)-r(X\cap Y),$ we must prove
\begin{align}
\label{no8eqn}
|X_{1,2}\cup Y_{1,2}|\geq |X_{1,2}|+|Y_{1,2}|-|X_{1,2}\cap Y_{1,2}|.
\end{align}
But (\ref{no8eqn}) holds with equality by the law of unions of sets, and thus the condition \textbf{R3} holds for the function $r$. 
\end{IEEEproof}

Thus from Lemma \ref{functionrmatroid}, the function $r$ defines a matroid. Let this matroid be the candidate matroid ${\cal M}_{\tilde{\cal N}_3}$ as in Lemma \ref{conditionsn3matroidal}. Note that ${\cal M}_{\tilde{\cal N}_3}$ satisfies the conditions of Lemma \ref{conditionsn3matroidal}, as explained in the discussion preceding Lemma \ref{functionrmatroid}. Thus, if ${\cal M}_{\tilde{\cal N}_3}$ is representable over some field $\mathbb{F},$ then the matroids ${\cal M}_{\tilde{\cal N}_1}$ and ${\cal M}_{\tilde{\cal N}_2}$ must also be $\mathbb{F}$-representable, as restrictions of $\mathbb{F}$-representable matroids are $\mathbb{F}$-representable. However, the matroids ${\cal M}_{\tilde{\cal N}_1}$ and ${\cal M}_{\tilde{\cal N}_2}$ can never have representations over the same field because of Lemma \ref{lemmanetworkn1} and Lemma \ref{lemmanetworkn2}. Thus ${\cal M}_{\tilde{\cal N}_3}$ is nonrepresentable. We thus have the following lemma.
\begin{lemma}
\label{n3representation}
The network $\tilde{\cal N}_3$ is a matroidal 1-error detecting network associated with the nonrepresentable matroid ${\cal M}_{\tilde{\cal N}_3}.$
\end{lemma}

Thus Definition \ref{matroidalerrornetworkdefinition} applies to error detecting networks associated with nonrepresentable matroids  also. A similar argument can be given for Definition \ref{matroidalerrorcorrectionnetworkdefinition} also.
\begin{remark}
\label{amalgam}
A matroid ${\cal M}$ on the groundset $E=E_1\cup E_2$ is said to be an \textit{amalgam} of the matroids ${\cal M}_1={\cal M}|E_1$ and ${\cal M}_2={\cal M}|E_2$ (the reader is referred to \cite{Oxl} for more details). Thus the matroid ${\cal M}_{\tilde{\cal N}_3}$ is an amalgam of ${\cal M}_{\tilde{\cal N}_1}$ and ${\cal M}_{\tilde{\cal N}_2}.$ 
\end{remark}

By Lemma \ref{n3representation} and because of the connection between the network $\tilde{\cal N}_3$ and the network ${\cal N}_3$ shown in \cite{DFZ3}, it is easy to prove that ${\cal N}_3$ is a matroidal network associated with a nonrepresentable matroid, one which is constructed as an amalgam of the matroids $M_{\tilde{\cal N}_1}/\left\{y_i:i=1,...,7\right\}$ and $M_{\tilde{\cal N}_2}/\left\{z_i:i=1,...,15\right\}.$ We leave the details of this proof to the reader.
\section{More Examples}
\label{sec6}
In this section, we present some examples of networks with scalar linear network codes and network-error correcting codes to illustrate our construction algorithms. Each example shown in this section is obtained by running an instance of the corresponding algorithm fixing the number of sources ($|{\cal S}|$), number of messages ($n$), number of correctable errors ($\alpha$), number of coding nodes to be added ($N_C$), number of sinks $|{\cal T}|$ (necessary for multicast) and the finite field used. Furthermore, for ease of computation, we also fix the number of edges whose symbols are to be encoded at any iteration in the construction algorithm to the new coding node, i.e., $|{\cal E}_C|$ is fixed. These examples are obtained by randomly picking existing forwarding nodes at any iteration in the algorithm to combine their information flows, and then checking if the resultant network code (or the equivalent matroid) satisfies the necessary properties. The MATLAB codes that generate these examples will be provided by the authors on request. All the figures and the corresponding matroid representations (or network coding coefficients) are shown at the end of the manuscript.
\subsection{Multicast}
\begin{figure*}
\centering
\includegraphics[totalheight=8in]{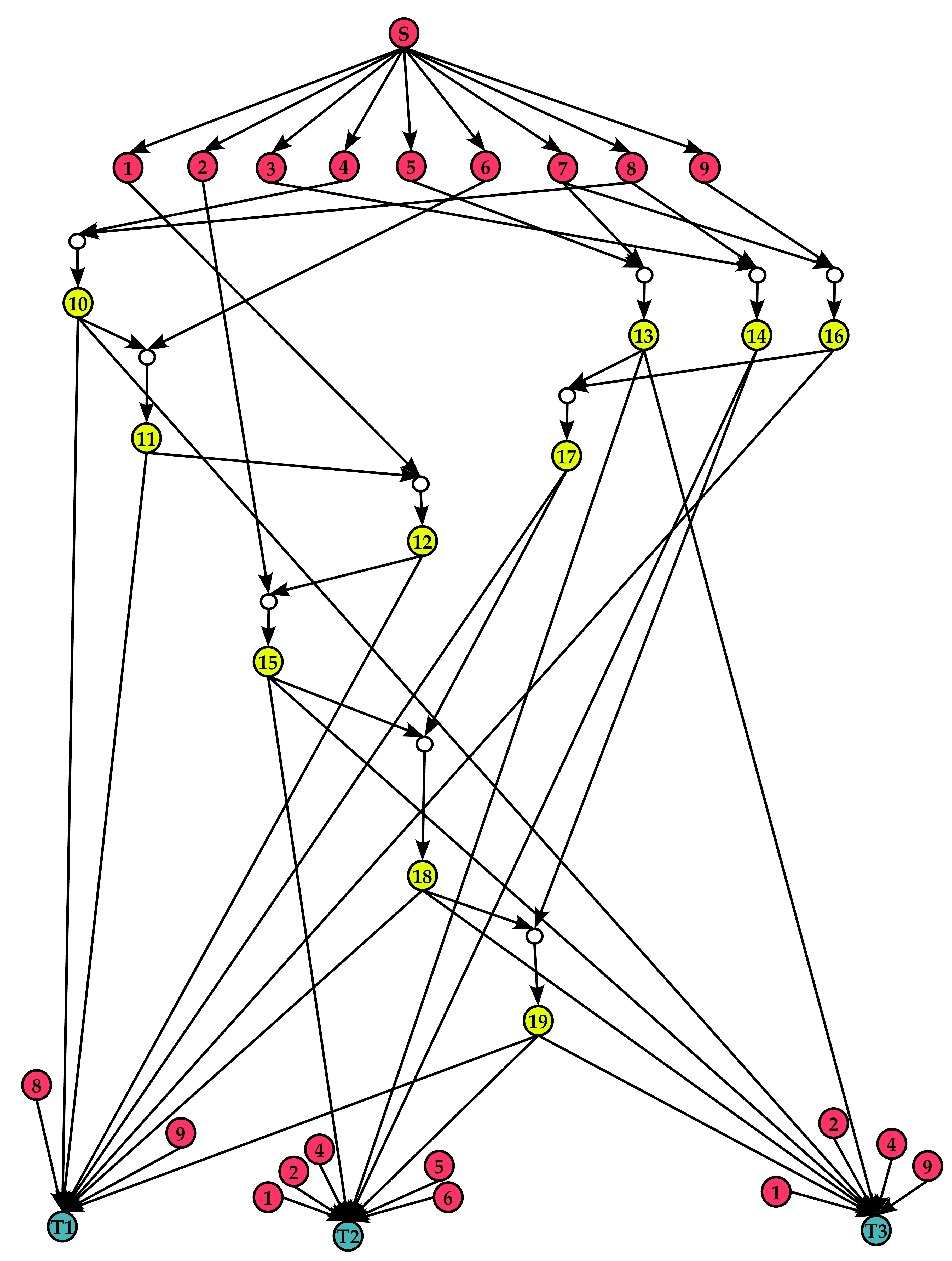}
\caption{A network with a $3$-network-error correcting code multicasting $3$ information symbols. The direct links from the different sources to the sinks are indicated by incoming edges from the corresponding duplicate nodes. The corresponding network coding coefficients are shown in Table \ref{tab1}.} 	
\label{fig:bigmulticastexm}	
\end{figure*}

\begin{example}
Fig. \ref{fig:bigmulticastexm} shows a single source multicast network with a scalar linear $3$-error correcting network code and $N_C=10$. Table \ref{tab1} shows all the relevant parameters using which the algorithm designs the network and the linear network coding coefficients obtained as outputs of the algorithm. The global encoding vectors of the $N$ outgoing edges from the source in the network correspond to the columns of a generator matrix of an MDS code with minimum distance $2\alpha+1=7$ and length $N=n+2\alpha=9.$ The values in the last column of Table \ref{tab1} represent the particular linear combination using which the information flows from the existing forwarding nodes (specified by the first column in Table \ref{tab1}) are combined at the new coding node formed (the corresponding forwarding node is given by the second column of Table \ref{tab1}). These linear encoding coefficients are represented by the decimal equivalents of the polynomial representations of the respective finite field elements. Also in Fig. \ref{fig:bigmulticastexm}, the direct links from the source to the sinks are indicated by incoming edges from the corresponding duplicate nodes (which are unconnected to the rest of the network).

This example also illustrates the ability of our multicast algorithm to construct scalar linear network-error correcting codes for multicast networks over smaller fields when compared with existing algorithms in \cite{Zha,Mat,YaY}. To see this, suppose that the network shown in Fig. \ref{fig:bigmulticastexm} was given as the input network to the algorithms in \cite{Zha,Mat,YaY} in order to design a multicast $3$-network-error correcting code. These algorithms require a field size $q$ such that
\[
q \geq 
\sum_{t\in{\cal T}}\left(
\begin{array}{c}
|{\cal E}| \\
2\alpha
\end{array}
\right) \geq \sum_{t\in{\cal T}}\left(
\begin{array}{c}
N_C \\
2\alpha
\end{array}
\right) = 3\left(
\begin{array}{c}
10 \\
6
\end{array}
\right) = 630
\]
to design a multicast linear network-error correcting code that can correct any $3$ network-errors in the given network. Thus only if $q\geq 630,$ the algorithms in \cite{Zha,Mat,YaY} guarantee the construction of a suitable network-error correcting code for our final network. However, our algorithm obtains a network-error correcting code for this network over $\mathbb{F}_{16}$ because it designs the network and the associated matroids together and representations of these associated matroids can be given over $\mathbb{F}_{16}$. The topology of the network is controlled by our algorithm. This is in contrast with the algorithms in  \cite{Zha,Mat,YaY}, which take a given network as the input and design the network-error correcting code for that network. The field size demands of \cite{Zha,Mat,YaY} are less dependent on the actual topology of the network and depend more on its size.
\end{example}
\subsection{Multiple-Unicast}
\begin{figure*}
\centering
\subfigure[Unicast Network with $3$ information symbols with no error correction at initial stage of multiple-unicast construction]{
\includegraphics[width=3.4in]{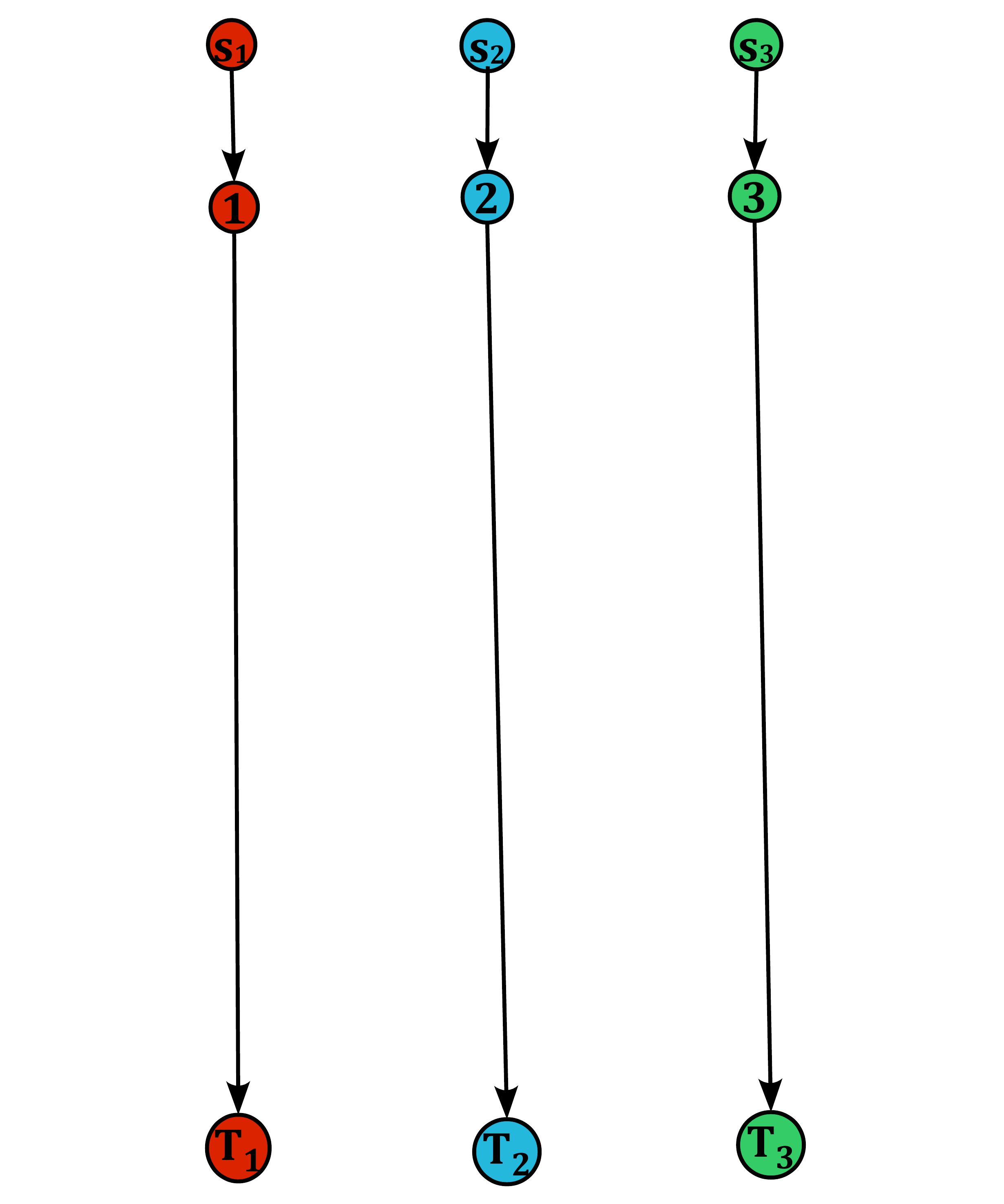}
\label{fig:unicastncnetwork0}	
}
\subfigure[Multiple-unicast network after first iteration]{
\includegraphics[width=3.4in]{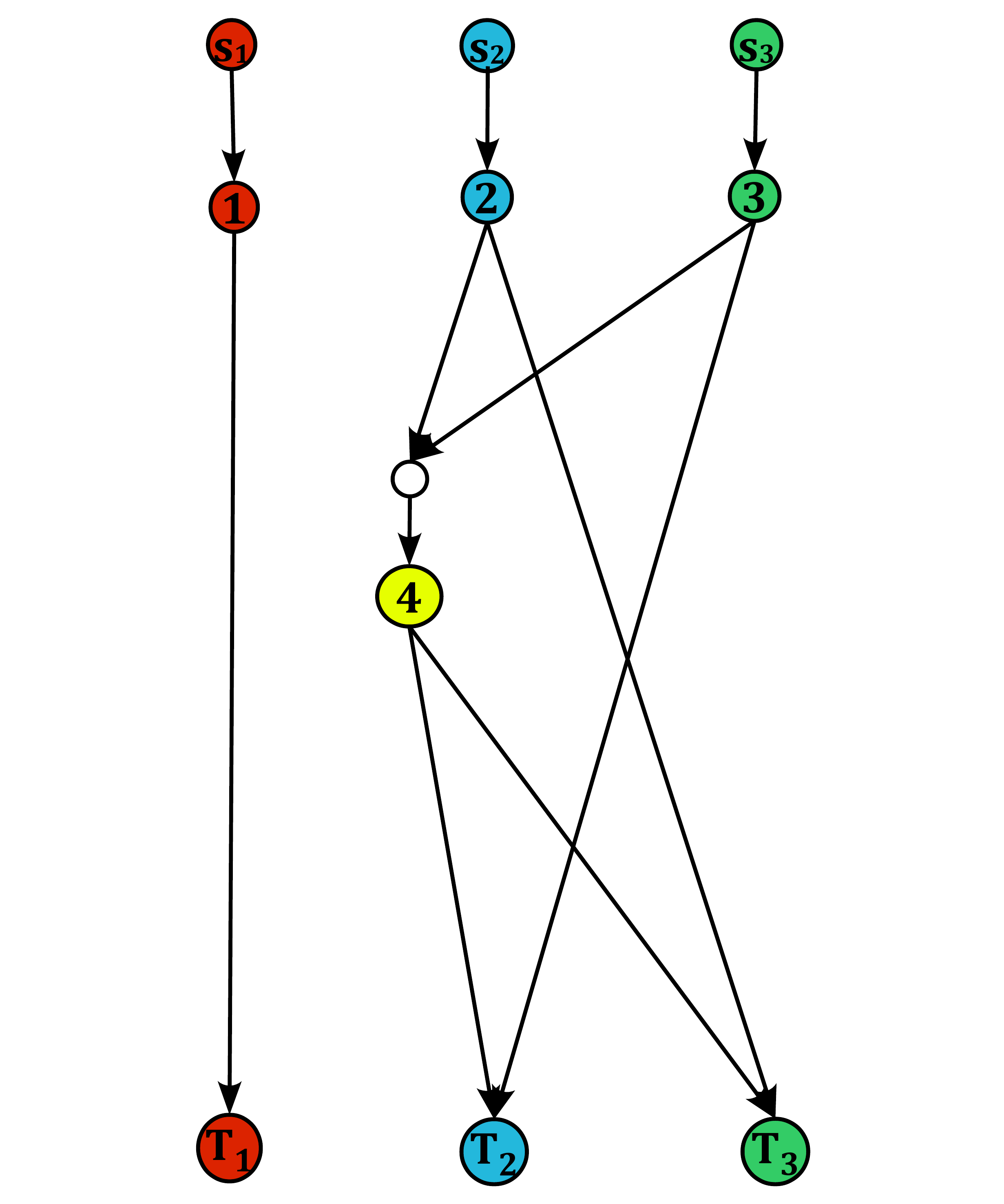}
\label{fig:unicastncnetwork1}	
}
\subfigure[Multiple-unicast network after fourth iteration. The direct links from the different sources to the sinks are indicated by incoming edges from the corresponding duplicate nodes. ]{
\includegraphics[width=3.4in]{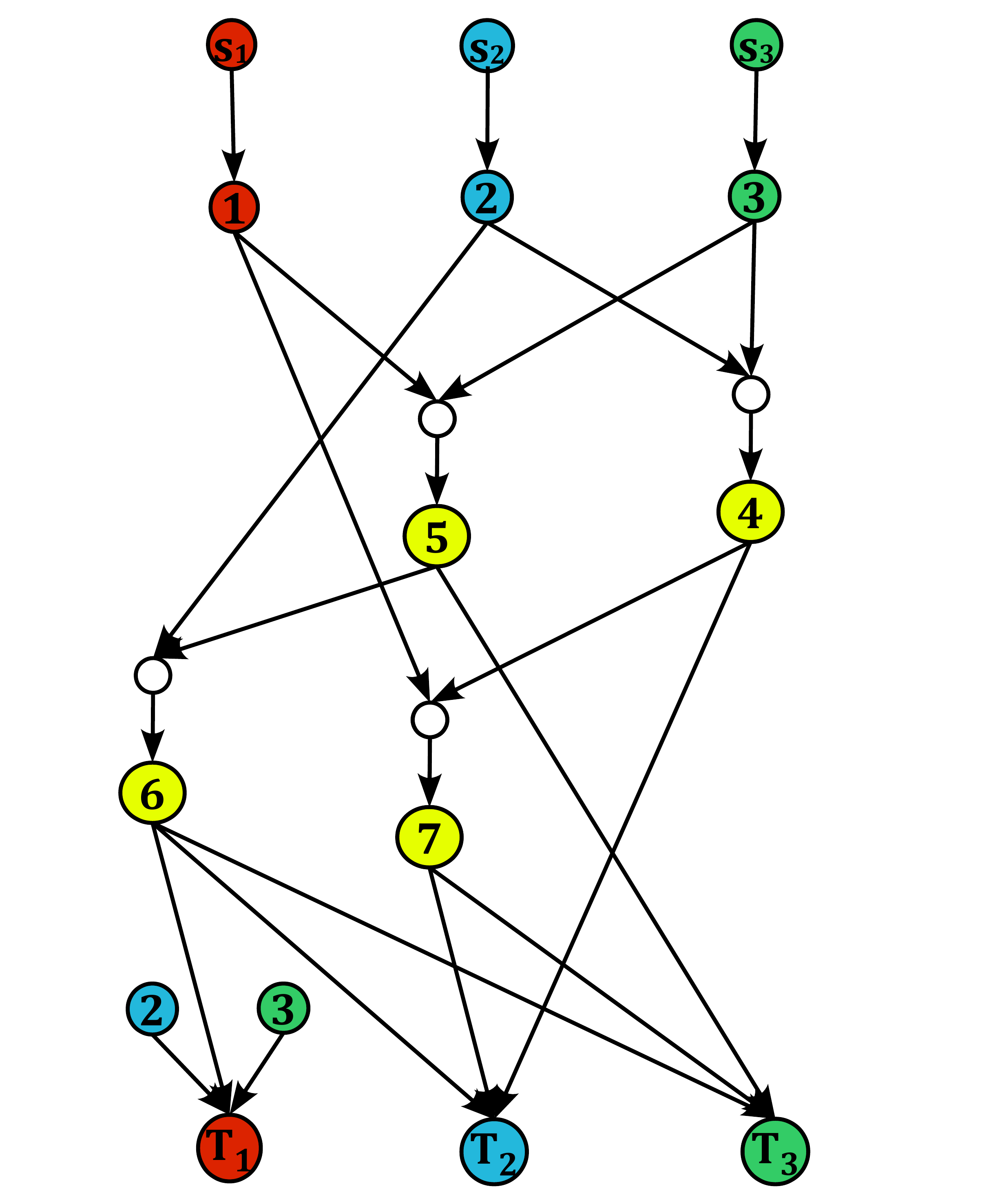}
\label{fig:unicastncnetwork4}	
}
\subfigure[Multiple-unicast network after fifth(final) iteration. Note the reduction in the number of incoming edges at Sink $T_2.$ This is a result of using the optional update in \textbf{\textit{Step 6}} of our multiple-unicast algorithm. In this case the interference from sources $s_1$ and $s_3$ to sink $T_2$ is aligned by the network itself rather than canceled.]{
\includegraphics[width=3.4in]{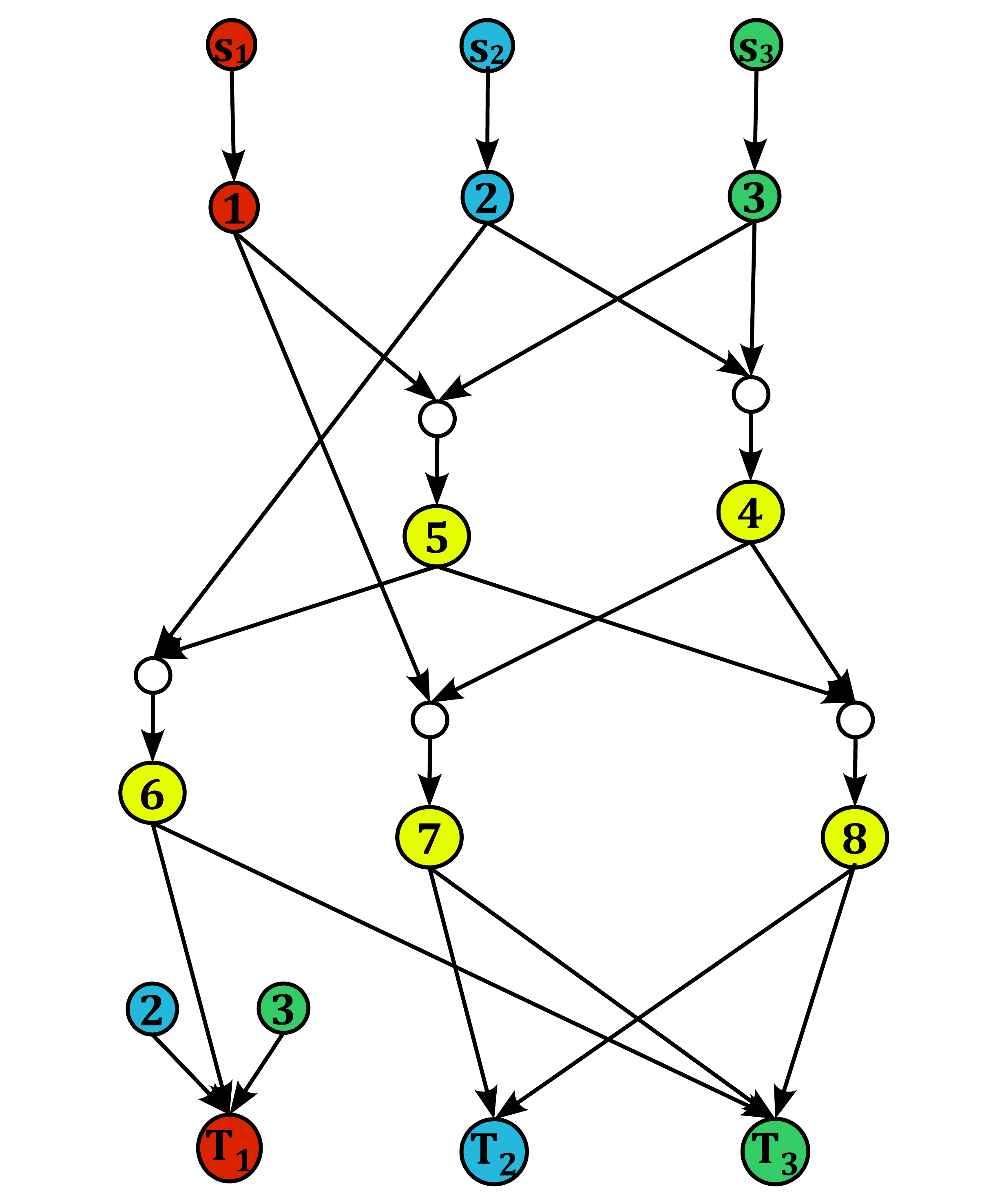}
\label{fig:unicastncnetwork5}	
}	
\caption{The stages of network evolution in the construction of a multiple-unicast network with no error correction, i.e., using only network coding. Only those networks corresponding to the initial stage and the first, fourth, and fifth iterations are given here. The representations of the associated matroids is given in Fig. \ref{fig:unicastnceqns}.}
\label{unicastnwstages}
\end{figure*}

\begin{figure*}
\centering
\includegraphics[totalheight=8in]{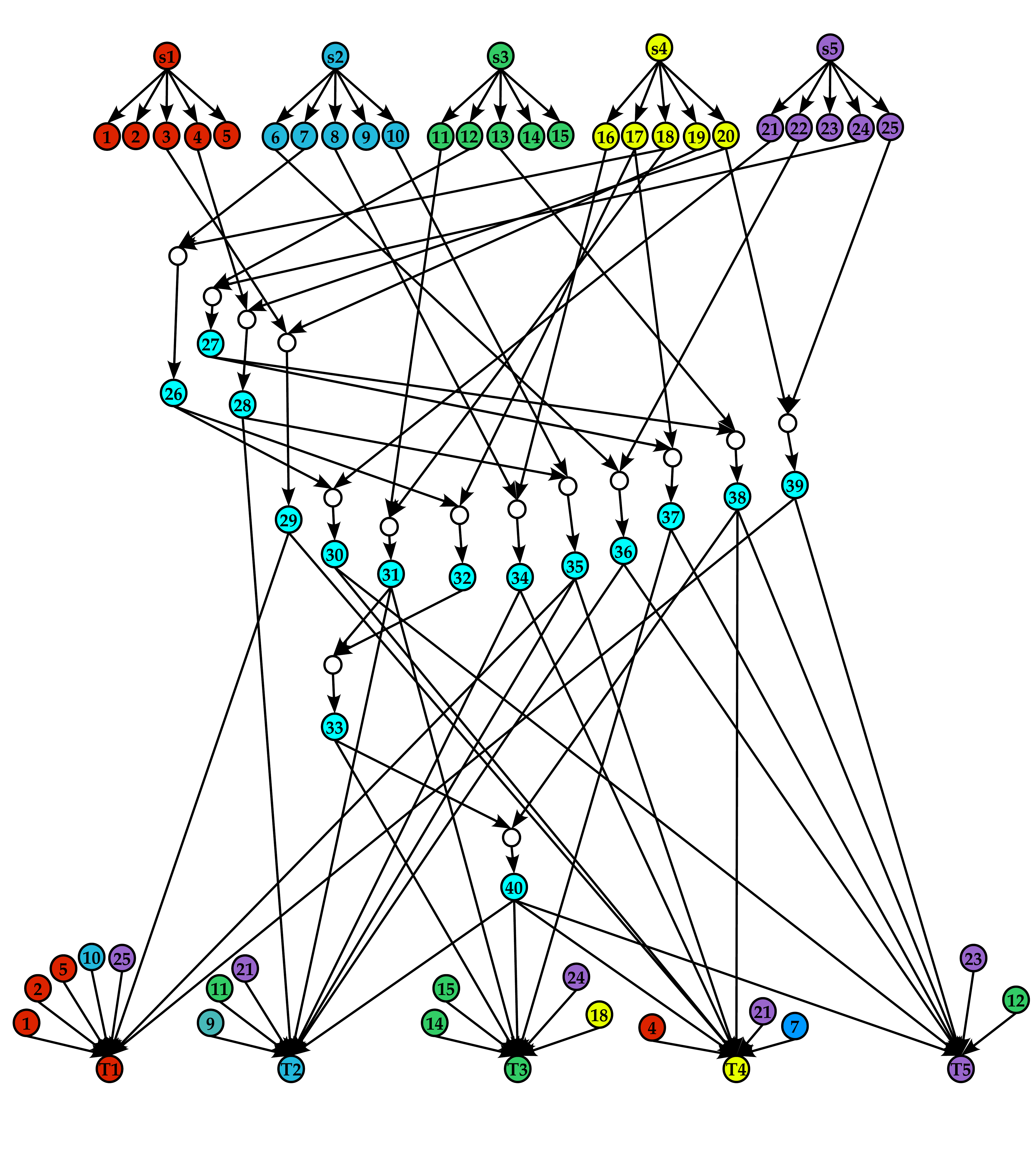}
\caption{A network with a $2$-network-error correcting $5$-unicast code. The direct links from the different sources to the sinks are indicated by incoming edges from the corresponding duplicate nodes. Table \ref{tab2} gives the corresponding network coding coefficients.}	
\label{fig:bigmultipleunicastexm}	
\end{figure*}

\begin{example}
\label{unicastncexample}
Fig. \ref{fig:unicastncnetwork0}-\ref{fig:unicastncnetwork5} show the stages of the network evolution of a multiple-unicast network with parameters $n=3, \alpha=0$ (no error correction) and $N_C=5$. The direct links from the different sources to the sinks are indicated by incoming edges from the corresponding duplicate nodes. Every sink demands the information symbol generated by the corresponding source. The representative matrices of the corresponding matroids are shown in (\ref{eqnstage0unicastnc})-(\ref{eqnstage3unicastnc}) in Fig. \ref{fig:unicastnceqns}. Every network is a matroidal $0$-error correcting network with the corresponding matroid and function $f,$ as defined in Example \ref{multicastex}. Note the reduction in the number of incoming edges at Sink $T_2$ from three in Fig. \ref{fig:unicastncnetwork4} to two in Fig. \ref{fig:unicastncnetwork5}. This is a result of using the optional update in \textbf{\textit{Step 6}} of our multiple-unicast algorithm. The transfer matrix from the sources to sink $T_2$ at the end of the final iteration is 
\[
\boldsymbol{F_{{\cal S},T_2}}=
\left(
\begin{array}{cc}
1 & 2\\
4 & 1\\
3 & 6
\end{array}
\right),
\]
where the matrix is over $\mathbb{F}_8$ (with modulo polynomial $x^3+x+1$), with the entries being the decimal equivalents of the polynomial representations of elements from $\mathbb{F}_8.$ The demanded symbol at $T_2$ is generated by $s_2,$ and corresponds to the second row above. The interference from source $s_3$, corresponding to the third row, is seen to be a scaled version of the interference from $s_1$, corresponding to the first row. Thus in this case, our multiple-unicast algorithm generates a network for which the interference is aligned by the network rather than canceled. However, the sink itself is enabled to cancel the interference. It is easily seen that a linear combination of the two columns of $\boldsymbol{F_{{\cal S},T_2}}$ generates the basis vector $(0~1~0~0)^T,$ enabling the sink $T_2$ to decode the demanded information symbol generated by source $s_2.$ \end{example}
\begin{example}
Fig. \ref{fig:bigmultipleunicastexm} shows a multiple-unicast network with a $2$-error correcting code, with all relevant parameters of which are shown in Table \ref{tab2}. The $i^{th}$ sink demands the information symbol generated by the $i^{th}$ source. Each source employs a repetition code of length $2\alpha+1=5$ on its outgoing edges. As in Table \ref{tab1}, the values in the last column of Table \ref{tab2} represent the decimal equivalents of the field elements in their polynomial representation. The direct links from the different sources to the sinks are indicated by incoming edges from the corresponding duplicate nodes. 
\end{example}
\section{Concluding remarks and Discussion}
\label{sec7}
The matroidal connections to network-error correction and detection have been analysed in this paper. It was shown that networks with scalar linear network-error correcting and detecting codes correspond to representable matroids with certain special properties. We also presented algorithms which can construct matroidal error correcting networks. The same algorithms can also be used to construct matroidal error detecting networks. By restricting ourselves to the class of representable matroids, we can therefore obtain a large number of networks with scalar linear network-error correcting and detecting codes, some of which were presented as examples. Further restricting ourselves to the matroids which are representable over particular fields, we can obtain networks which have scalar linear network-error correcting codes over those particular fields. This may facilitate some intuition towards finding the minimum field size requirement for scalar linear network-error correcting codes to exist, which is known to be a hard problem. Also, running our algorithms along with the optional update of sink incoming edges in \textbf{\textit{Step 6}} may provide insight on the solvability and capacity of general multisource multicast and multiple-unicast networks in the presence of errors. In particular, the multiple-unicast algorithm can then be used to generate multiple-unicast networks where interference from other sources is not always canceled by the network nodes, as shown by Example \ref{unicastncexample}. Following techniques similar to \cite{DFZ3}, it was also shown that linear network codes prove are not always sufficient to provide the demanded error correction.

It is known \cite{Oxl} that characterising all possible modular cuts of a matroid, and therefore all possible extensions of a matroid is in general a difficult task. Moreover, we require extensions which satisfy certain constraints for the resultant network to be matroidal, and have to satisfy even more constraints if they have to be associated with representable matroids. Characterising such extensions could be a particularly rewarding exercise. As a first step towards characterising such extensions and also towards obtaining matroidal error correcting networks associated with nonrepresentable matroids, we proved Proposition \ref{propprincipal} regarding the principal extensions of a representable matroid. It can be expected that deeper theoretical insights on the theory of network coding and error correction can be gained with more powerful machinery from matroid theory.


\begin{table*}
\centering
\normalsize
\caption{Multicast example (Fig. \ref{fig:bigmulticastexm}): $n=3,~\alpha=3,~N_C=10,~|{\cal E}_C|=2,~{|{\cal T}|}=3,$ Finite field used=$\mathbb{F}_{16}$ (modulo $x^4+x+1$)}
\label{tab1}
\begin{tabular}{|c|c|c|}
\hline
\textbf{Nodes used to form new}& \textbf{New forwarding node} & $\boldsymbol{\mathbb{F}}$ \textbf{linear combination of nodes of} \\
\textbf{coding node (see figure)} & \textbf{formed (see figure)} & \textbf{column $1$ used to form new node}\\
\hline
(4,8) & 10 & (1,2) \\
\hline
 (6,10) & 11 & (1,5) \\
\hline
(1,11) & 12 & (1,9) \\
\hline
(5,7) & 13 & (1,2) \\
\hline
(3,8) & 14 & (1,3) \\
\hline
(2,12) & 15 & (1,8) \\
\hline
(7,9) & 16 & (1,13) \\
\hline
(13,16) & 17 & (1,1) \\
\hline
(15,17) & 18 & (1,2) \\
\hline
(14,18) & 19 & (1,1) \\
\hline
\end{tabular}~\\~\\~\\
\hrule
\end{table*}
\begin{figure*}
\begin{minipage}[htbp]{.5\textwidth}
\begin{equation}
\label{eqnstage0unicastnc}
\left(
\begin{array}{cccc}
&1&0&0\\
&0&1&0\\
&0&0&1\\
&&&\\
I_{6}&1&0&0\\
&0&1&0\\
&0&0&1\\
\end{array}
\right)
\end{equation}
\end{minipage}
\begin{minipage}[htbp]{.5\textwidth}
\begin{equation}
\label{eqnstage1unicastnc}
\left(
\begin{array}{ccccc}
&1&0&0&0\\
&0&1&0&1\\
&0&0&1&2\\
&&&&\\
I_{7}&1&0&0&0\\
&0&1&0&1\\
&0&0&1&2\\
&0&0&0&1\\
\end{array}
\right)
\end{equation}
\end{minipage}
\begin{minipage}[htbp]{.5\textwidth}
\begin{equation}
\label{eqnstage2unicastnc}
\left(
\begin{array}{cccccccc}
&1&0&0&0&1&6&1\\
&0&1&0&1&0&1&4\\
&0&0&1&2&2&7&3\\
&&&&&&&\\
&1&0&0&0&1&6&1\\
&0&1&0&1&0&1&4\\
I_{10}&0&0&1&2&2&7&3\\
&0&0&0&1&0&0&4\\
&0&0&0&0&1&6&0\\
&0&0&0&0&0&1&0\\
&0&0&0&0&0&0&1\\
\end{array}
\right)
\end{equation}
\end{minipage}
\begin{minipage}[htbp]{.5\textwidth}
\begin{equation}
\label{eqnstage3unicastnc}
\left(
\begin{array}{ccccccccc}
&1&0&0&0&1&6&1&2\\
&0&1&0&1&0&1&4&1\\
&0&0&1&2&2&7&3&6\\
&&&&&&&&\\
&1&0&0&0&1&6&1&2\\
&0&1&0&1&0&1&4&1\\
I_{11}&0&0&1&2&2&7&3&6\\
&0&0&0&1&0&0&4&1\\
&0&0&0&0&1&6&0&2\\
&0&0&0&0&0&1&0&0\\
&0&0&0&0&0&0&1&0\\
&0&0&0&0&0&0&0&1\\
\end{array}
\right)
\end{equation}
\end{minipage}
\caption{The stages of evolution in the representable matroid in the construction of a multiple-unicast network (shown in Fig. \ref{unicastnwstages}) with only network coding and no network-error correction. Only those representations corresponding to the initial stage and the first, fourth, and fifth iterations are given here. All matrices are over $\mathbb{F}_8$ (with modulo polynomial $x^3+x+1$) and the entries are the decimal equivalents of the polynomial representations of elements from $\mathbb{F}_8.$}
\label{fig:unicastnceqns}	
\end{figure*}
\begin{table*}
\centering
\normalsize
\caption{Multiple-unicast example (Fig. \ref{fig:bigmultipleunicastexm}): $n=5,~\alpha=2,~N_C=15,~|{\cal E}_C|=2,$ Finite field used=$\mathbb{F}_{8}$ (modulo $x^3+x+1$)}
\label{tab2}
\begin{tabular}{|c|c|c|}
\hline
\textbf{Nodes used to form}& \textbf{New forwarding node formed} & $\boldsymbol{\mathbb{F}}$ \textbf{linear combination of nodes of} \\
\textbf{new coding node} & & \textbf{column $1$ used to form new node}\\
\hline
(7,8) & 26 & (1,4) \\
\hline
(12,24) & 27 & (1,2) \\
\hline
(4,20) & 28 & (1,5) \\
\hline
(3,19) & 29 & (1,7) \\
\hline
(21,26) & 30 & (1,1) \\
\hline
(11,18) & 31 & (1,3) \\
\hline
(17,26) & 32 & (1,2) \\
\hline
(30,32) & 33 & (1,3) \\
\hline
(8,16) & 34 & (1,6) \\
\hline
(10,28) & 35 & (1,3) \\
\hline
(6,22) & 36 & (1,3) \\
\hline
(17,27) & 37 & (1,2) \\
\hline
(13,27) & 38 & (1,2) \\
\hline
(20,25) & 39 & (1,6) \\
\hline
(33,38) & 40 & (1,6) \\
\hline
\end{tabular}
\end{table*}
\end{document}